\documentclass[a4paper,11pt]{article}
\pdfoutput=1 % if your are submitting a pdflatex (i.e. if you have
\usepackage{jcappub} % for details on the use of the package, please
\usepackage[T1]{fontenc} % if needed
\usepackage[caption=false]{subfig}
\usepackage{graphicx}
\usepackage{epstopdf}
\usepackage{mathrsfs}
\usepackage{amssymb}
\usepackage{verbatim}
\usepackage{color}
\usepackage{amsmath}
\usepackage{subfig}
\usepackage{hyperref}
\usepackage[utf8]{inputenc} 
\newcommand{\beq}{\begin{equation}}
\newcommand{\eeq}{\end{equation}}
\def\be{\begin{equation}}
\def\ee{\end{equation}}
\def\bea{\begin{eqnarray}}
\def\eea{\end{eqnarray}}
\def\mrm{\mathrm}
\def\lsim{\raise0.3ex\hbox{$\;<$\kern-0.75em\raise-1.1ex\hbox{$\sim\;$}}}
\def\gsim{\raise0.3ex\hbox{$\;>$\kern-0.75em\raise-1.1ex\hbox{$\sim\;$}}}

\subheader{LPT-ORSAY-16-17, CPHT-RR013-032016 }

\title{\boldmath Re-opening dark matter windows compatible with a diphoton excess}
\author[a]{Giorgio Arcadi}
\author[a,b]{Pradipta Ghosh,}
\author[a]{Yann Mambrini}
\author[a]{and Mathias Pierre}

\affiliation[a]{Laboratoire de Physique Th\'eorique, CNRS, Univ. Paris-Sud,\\
Universit\'e Paris-Saclay, 91405 Orsay, France}
\affiliation[b]{Centre de Physique Th\'eorique, Ecole polytechnique, CNRS, \\ 
Universit\'e Paris-Saclay, 91128 Palaiseau Cedex, France }

\emailAdd{giorgio.arcadi@th.u-psud.fr}
\emailAdd{pradipta.ghosh@th.u-psud.fr}
\emailAdd{yann.mambrini@th.u-psud.fr}
\emailAdd{mathias.pierre@th.u-psud.fr}

\abstract{We investigate a simple setup in which an excess in the di-photon invariant mass distribution 
around $750$ GeV, as seen by the ATLAS and CMS collaborations, is originated
through a pair of collimated photon pairs. In this framework a scalar state 
$s$ decays into two light pseudo-Goldstone bosons $a$, each of which subsequently decays into 
a pair of collimated photons which are misidentified as a single photon. In a minimal context of spontaneous 
symmetry breaking, we show that coupling a complex scalar field $\Phi=(s+ia)/\sqrt{2}$ 
to a fermionic dark matter candidate $\chi$, 
also responsible for generating its mass, allows for the correct relic density in a large region of the 
parameter space, while not being excluded by the direct or indirect detection experiments. Moreover, 
the correct relic abundance can naturally co-exist with a relatively large width for the 
resonant field $s$.}


\arxivnumber{1603.05601}

\begin{document}
\maketitle
\flushbottom

%%%%%%%%%%%%%%%%%%%%%%%%%%%%%%%%%%%%%%%%%%%%%%%%%%%%%%%%%%%%%%%%%
%%%%%%%%%%%%%%%%%%%%%%%%%%%%%%%%%%%%%%%%%%%%%%%%%%%%%%%%%%%%%%%%%
%%%%%%%%%%%%%%%%%%%%%%%%%%%%%%%%%%%%%%%%%%%%%%%%%%%%%%%%%%%%%%%%%
\setcounter{equation}{0}

%%%%%%%%%%%%%%%%%%%%%%%%%%%%%%%%%%%%%%%%%%%%%%%%%%%%%%%%%%%%%%%%%%%%%%

%%%%%%%%%%%%%%%%%%%%%%%%%%%%%%%%%%%%%%%%%%%%%%%%%%%%%%%%%%%%%
\section{Introduction}
\label{sec:intro}
%%%%%%%%%%%%%%%%%%%%%%%%%%%%%%%%%%%%%%%%%%%%%%%%%%%%%%%%%%%%%

Recently the ATLAS \cite{ATLASdiph} and CMS \cite{CMS:2015dxe} collaborations have reported an excess in the 
di-photon invariant mass distribution in the vicinity of
750 GeV in their 13 TeV data at the level of $10\pm3$ fb \cite{ATLASdiph} and $6\pm3$ fb \cite{CMS:2015dxe}, 
respectively which can be interpreted as a (pseudo)scalar resonance with a (relatively) large width. 
The ATLAS has analyzed 3.2 $\mrm{fb^{-1}}$ of data and observed a local significance of 
3.9$\sigma$ for an excess peaked at 750 GeV whereas the CMS, using 3.3 $\mrm{fb^{-1}}$ of data, has reported 
a local significance of $2.8\sigma$ - $2.9\sigma$ \cite{CMS:2016owr} for an excess peaked at\footnote{
Combining with 19.7 $\mrm{fb^{-1}}$ of data at 8 TeV, the largest CMS excess, assuming
a narrow width, i.e., $\sim\mathcal{O}(0.1~{\rm GeV})$, 
appears at 750 GeV with a local and global significance of $3.4\sigma$
and $1.6\sigma$, respectively \cite{CMS:2016owr}.} 760 
GeV. The global significance is reduced to 2.0$\sigma$ \cite{Moriondatlas} 
and $<1\sigma$ \cite{CMS:2016owr}  for the ATLAS and CMS, respectively. 
These announcements have initiated a large amount of works on the subject 
\cite{Huong:2016kpa,DeRomeri:2016xpb,Ahriche:2016mcx,Tsai:2016lfg,
Li:2016tqf,Lazarides:2016ofd,Ren:2016gyg,Molinaro:2016oix,
Redi:2016kip,Ko:2016sxg,Baek:2016uqf,Staub:2016dxq,Mantilla:2016sew,
Hamada:2016vwk,Gross:2016ioi,Bae:2016xni,Ge:2016xcq,Li:2016xcj,
Ding:2016udc,Hektor:2016uth,Bertuzzo:2016fmv,Kawamura:2016idj,
Geng:2016xin,Nomura:2016rjf,King:2016wep,Abel:2016pyc,Aydemir:2016qqj,
Chiang:2016ydx,Cao:2016cok,Han:2016bvl,Chao:2016aer,Nomura:2016seu,
Davis:2016hlw,Ding:2016ldt,Yu:2016lof,Cao:2016udb,Ko:2016wce,
Hati:2016thk,Stolarski:2016dpa,Borah:2016uoi,Fichet:2016pvq,
Sahin:2016lda,D'Eramo:2016mgv,Bhattacharya:2016lyg,Zhang:2016pip,
Ito:2016zkz,Deppisch:2016scs,Dutta:2016jqn,Hernandez:2016rbi,
Karozas:2016hcp,Chao:2016mtn,Danielsson:2016nyy,Ghorbani:2016jdq,
Ko:2016lai,Han:2016bus,Nomura:2016fzs,Palti:2016kew,Dasgupta:2015pbr,
Kaneta:2015qpf,Jiang:2015oms,Hernandez:2015hrt,
Kanemura:2015vcb,Kanemura:2015bli,Huang:2015svl,Hamada:2015skp,
Kang:2015roj,Ibanez:2015uok,Bizot:2015qqo,Dev:2015vjd,Goertz:2015nkp,
Chao:2015nac,Gao:2015igz,Kim:2015xyn,Cao:2015scs,Cai:2015hzc,
Wang:2015omi,Cao:2015apa,An:2015cgp,Tang:2015eko,Li:2015jwd,
Salvio:2015jgu,Park:2015ysf,Han:2015yjk,Hall:2015xds,Zhang:2015uuo,
Liu:2015yec,Cheung:2015cug,Das:2015enc,Davoudiasl:2015cuo,
Allanach:2015ixl,Gu:2015lxj,Cvetic:2015vit,Altmannshofer:2015xfo,
Cao:2015xjz,Chakraborty:2015gyj,Badziak:2015zez,Patel:2015ulo,
Moretti:2015pbj,Huang:2015rkj,Dev:2015isx,deBlas:2015hlv,
Dey:2015bur,Hernandez:2015ywg,Murphy:2015kag,Boucenna:2015pav,
Bauer:2015boy,Cline:2015msi,Cho:2015nxy,Berthier:2015vbb,
Kim:2015ksf,Bi:2015uqd,Dhuria:2015ufo,Heckman:2015kqk,
Huang:2015evq,Cao:2015twy,Wang:2015kuj,Feng:2015wil,
Bardhan:2015hcr,Chang:2015sdy,Luo:2015yio,Han:2015qqj,
Han:2015dlp,Ding:2015rxx,Chakraborty:2015jvs,Chang:2015bzc,Han:2015cty,
Chao:2015nsm,Bernon:2015abk,Carpenter:2015ucu,Alves:2015jgx,
Kim:2015ron,Benbrik:2015fyz,Gabrielli:2015dhk,Bai:2015nbs,
Csaki:2015vek,Chakrabortty:2015hff,Bian:2015kjt,Curtin:2015jcv,
Chao:2015ttq,Demidov:2015zqn,No:2015bsn,Becirevic:2015fmu,
Cox:2015ckc,Martinez:2015kmn,Kobakhidze:2015ldh,Cao:2015pto,
Molinaro:2015cwg,Petersson:2015mkr,Gupta:2015zzs,Bellazzini:2015nxw,
Higaki:2015jag,DiChiara:2015vdm,Pilaftsis:2015ycr,Buttazzo:2015txu,
Nakai:2015ptz,Harigaya:2015ezk}, trying to fit the data in effective 
frameworks \cite{Bharucha:2016jyr,Djouadi:2016eyy,
Falkowski:2015swt,Dutta:2015wqh,Low:2015qep,Ellis:2015oso,
McDermott:2015sck,Angelescu:2015uiz}, featuring loop induced (typically by 
new vector-like fermions, with respect to the standard model (SM)) 
couplings of the resonance, say $s$ for example, with gluons 
and photons which are described by dimension five operators of the form:
%%%%%%%%%%%%%%%%%%%%%%%%%%%%%%%%%%%%%%%%%%%%%%%%%%%%%%%%%%%%%
\be
-{\cal L}_0 = \frac{c_{BB}}{\Lambda} s B^{\mu \nu} B_{\mu \nu} 
+ \frac{c_{WW}}{\Lambda} s W^{\mu \nu}_i W_{\mu \nu}^i 
+\frac{c_{GG}}{\Lambda} s G^{\mu \nu}_\alpha G_{\mu \nu}^\alpha,
\label{Eq:l0}
\ee
%%%%%%%%%%%%%%%%%%%%%%%%%%%%%%%%%%%%%%%%%%%%%%%%%%%%%%%%%%%%%

where $B_{\mu \nu}$ $W^i_{\mu \nu}$ and $G_{\mu \nu}^\alpha$ are the 
$U(1)_Y$, $SU(2)_L$ and $SU(3)_C$ field strength, respectively.
The parameter $\Lambda$ represents the scale at which new dynamics appears, 
which is typically of the order of the mass of new fermions entering in the loop
to generate an effective Lagrangian as shown in eq.~(\ref{Eq:l0}).
However, it was quickly realized that, in the presence of couplings only with gluons and photons, 
in order to produce the observed production 
cross-section $\sim \mathcal{O}(10 ~\mbox{fb})$ and simultaneously a rather large decay width of the 
resonance, $\Gamma_s$, such that $\Gamma_s/m_s \sim 1-10 \%$ \cite{CMS:2016owr,
Moriondatlas} with $m_s$ giving the mass of the resonance, 
large values of the coupling $c_{GG}$ would be essential. This possibility, however, 
contradicts (see ref.~\cite{Franceschini:2015kwy}) with 
the current observations from di-jet 
searches $gg \rightarrow s \rightarrow gg$ at 
the LHC \cite{CMS:2015neg,Aad:2014aqa}, predicting $c_{GG} \lesssim 0.1$
for $\Lambda\sim 2$ TeV. 
Similarly, high values, $\gsim 0.1$, of the coupling $c_{BB}$ 
are disfavoured\footnote{This scenario
corresponds to $c_{WW}=0$ in eq.~(\ref{Eq:l0}).}, given the 
constraint from the production through photon fusion \cite{Fichet:2015vvy}.

A possible solution to increase $\Gamma_s$ can be achieved by allowing new 
decay channels for the resonance. An intriguing possibility is 
represented by the case in which the additional width is provided by the decay
channel into dark matter pairs. Along this line, the authors of 
refs.~\cite{Mambrini:2015wyu,D'Eramo:2016mgv,Backovic:2015fnp,Barducci:2015gtd} 
(see also ref.~\cite{Chu:2012qy}) have 
considered a coupling of the resonance with a fermionic (Dirac or Majorana) dark matter candidate as: 
%%%%%%%%%%%%%%%%%%%%%%%%%%%%%%%%%%%%%%%%%%%%%%%%%%%%%%%%%%%%% 
\be
-{\cal L}_1 = g_{\chi} s \bar{\chi} \chi + m_\chi \bar \chi \chi,
\ee
%%%%%%%%%%%%%%%%%%%%%%%%%%%%%%%%%%%%%%%%%%%%%%%%%%%%%%%%%%%%% 
 
with $g_\chi$ as the relevant Yukawa coupling
and $m_\chi$ representing the mass of the dark matter $\chi$.

However, all these studies converged to show that in the said scenario, the direct 
detection (in the case of a scalar 
resonance) or the indirect detection (in the case of a pseudoscalar resonance), strongly constraints the 
dark matter mass up to $\sim\mathcal{O}(1~{\rm TeV})$, due to the concerned values of $c_{GG}$ and $c_{BB}$ 
couplings, respectively. 
These associated values, even after respecting the di-jet \cite{Franceschini:2015kwy} and 
photon fusion \cite{Fichet:2015vvy} constraints, appear naturally to accommodate the observed
di-photon production cross-section.
Combining these limits further with the constraints from monojet searches \cite{Khachatryan:2014rra,
Aad:2014nra}, $gg \rightarrow \bar{\chi} \chi g$, the light dark matter region, 
$m_\chi \lesssim 375$ GeV, i.e., $m_\chi \lsim m_s/2$ appears excluded \cite{Barducci:2015gtd}.

An interesting alternative has been proposed in refs.~\cite{Dasgupta:2016wxw,Bi:2015lcf,Knapen:2015dap,
Aparicio:2016iwr}, considering the possibility that the di-photon production process 
$pp \rightarrow s \rightarrow \gamma \gamma$ can be mimicked 
by a process of the form $pp \rightarrow s \rightarrow aa \rightarrow 4\gamma$, where $a$ is 
a light pseudoscalar decaying into two collimated photons.
In this scenario a sizable contribution to $\Gamma_s$ could come from 
the $s \rightarrow aa$ decay process. 
This possibility has been investigated, for instance, in ref.~\cite{Aparicio:2016iwr}, 
where the $saa$ coupling is generated by the spontaneous breaking of a Peccei-Quinn (PQ) symmetry. 
However, a sizable $saa$ coupling to secure $\Gamma_s=40$ GeV 
through $s\to aa$ decay process is rather unnatural since it would require a small PQ 
scale $\sim\mathcal{O}(300~\rm{GeV}$). Thus, the presence of an invisible branching fraction $(Br)$ 
has nevertheless been invoked, although the new final states have not been identified as 
dark matter candidates. The similar and extensive studies have been performed in the case of the 
next-to-minimal supersymmetric standard model (NMSSM) \cite{Badziak:2016cfd,Domingo:2016unq,Ellwanger:2016qax}. 
In this model the LHC signal can be reproduced only for masses 
$\sim\mathcal{O}~(100~{\rm MeV})$ of the lightest CP-odd Higgs\footnote{Collimated
photons from a light CP-odd Higgs in the NMSSM
were previously studied in refs.~\cite{Dobrescu:2000jt,Dobrescu:2000yn}.} since otherwise its 
branching ratio into two photons would be heavily suppressed 
compared to the same into SM fermion pairs.

In this work, we will combine the two preceding approaches in the context of a simple model featuring 
spontaneously broken symmetry. The coupling $saa$ is induced after the spontaneous symmetry breaking 
by a classical potential term $|\Phi|^4 \to |(s+ia)|^4$.
Thus, the derived $saa$ coupling remains proportional to the quartic coupling. 
The same complex scalar field $\Phi$ is also coupled to the dark matter 
$\chi$ and, is responsible for generating $m_\chi$ after the spontaneous
symmetry breaking.

We will show that when $\Gamma_s$ is primarily determined by
$s\to aa$ process, i.e., via $\Gamma(s\to aa)$, the di-photon excess can 
be obtained for values of $c_{GG}$ well below $0.1$, the limit
predicted from the di-jet constraint. 
Such low values of $c_{GG}$ couplings, as an outgrowth, 
strongly relax the constraints from monojet and direct dark matter searches 
while still remain compatible with the WMAP/PLANCK \cite{Hinshaw:2012aka,Ade:2013zuv} 
favoured values of the dark matter relic density.

Regarding the WMAP/PLANCK constraints, three regions of the parameters space are now open:
(i) the electroweak-scale dark matter ($m_\chi \simeq 100$ GeV), 
(ii) a heavier window ($\gtrsim 400$ GeV, i.e., $\gsim m_s/2$) and 
(iii) a very light case (keV dark matter). 
For all of these windows, we have checked that the direct and indirect detection limits
are respected and are compatible with a relatively large width of the scalar, exchanged as a mediator. 
Our results are summarized in the final plot where we have shown points respecting all 
the relevant LHC constraints (e.g., di-jet, monojet, etc.),
observations (di-photon production cross-section
and large $\Gamma_s$) and giving, at the same time, the correct relic abundance for the dark matter.

The paper is organized as follows. After a summary of the model in 
section~\ref{Sec:model}, we will briefly review under which conditions 
the process $pp \rightarrow s \rightarrow aa \rightarrow 4\gamma$ can mimic the observed di-photon signal. 
We will then determine the parameter space allowed by the LHC observations, 
and show that the di-photon signal can remain compatible with a large width of $s$ in section~\ref{Sec:LHC}.
In the next section we introduce the dark matter candidate and compute its relic abundance, 
evaluating the parameter space respecting
at the same time the observed di-photon excess and the WMAP/PLANCK constraints. Finally we conclude.

%%%%%%%%%%%%%%%%%%%%%%%%%%%%%%%%%%%%%%%%%%%%%%%%%%%%%%%%%%%%%
\section{The Model}
\label{Sec:model}
%%%%%%%%%%%%%%%%%%%%%%%%%%%%%%%%%%%%%%%%%%%%%%%%%%%%%%%%%%%%%

The Lagrangian of eq.~(\ref{Eq:l0}) can be expressed
for a complex scalar $\Phi$ in a similar way. Adding 
a SM singlet fermion (Majorana or Dirac like)
further with eq.~(\ref{Eq:l0}) is straightforward.
The complete Lagrangian can then be written as:
%
%%%%%%%%%%%%%%%%%%%%%%%%%%%%%%%%%%%%%%%%%%%%%%%%%%%%%%%%%%%%%
\bea
&&
{\cal L}= {\cal L}_0 + {\cal L}_\Phi + {\cal L}_{\chi},
\eea
%%%%%%%%%%%%%%%%%%%%%%%%%%%%%%%%%%%%%%%%%%%%%%%%%%%%%%%%%%%%%
%
with
%%%%%%%%%%%%%%%%%%%%%%%%%%%%%%%%%%%%%%%%%%%%%%%%%%%%%%%%%%%%%
\bea
-{\cal L}_0&&= 
\frac{c_{BB}}{\Lambda} \Phi B^{\mu \nu} B_{\mu \nu} + \frac{c_{WW}}{\Lambda}\Phi W^{\mu \nu}_i W_{\mu \nu}^i 
+\frac{c_{GG}}{\Lambda} \Phi G^{\mu \nu}_\alpha G^\alpha_{\mu \nu} 
\nonumber
\\
&&
-
\frac{i c_{BB}}{\Lambda} \Phi B^{\mu \nu} \widetilde B_{\mu \nu} - \frac{i c_{WW}}{\Lambda} \Phi W^{\mu \nu}_i 
\widetilde W_{\mu \nu}^i 
-\frac{i c_{GG}}{\Lambda} \Phi G^{\mu \nu}_\alpha \widetilde G^\alpha_{\mu \nu} + \mrm{~h.c.,}  
\label{Eq:lagrangian1}
\eea
%%%%%%%%%%%%%%%%%%%%%%%%%%%%%%%%%%%%%%%%%%%%%%%%%%%%%%%%%%%%%

where $\widetilde B_{\mu \nu}$, $\widetilde W_{\mu \nu}^i$, $\widetilde G_{\mu \nu}^\alpha$ 
denote the dual field strength, e.g. 
$\widetilde{B}_{\mu \nu}=\frac{1}{2}\epsilon_{\mu \nu \rho \sigma}B^{\rho \sigma}$, 
and $\Phi=(s +ia)/\sqrt{2}$ is a complex
scalar field whose Lagrangian can be expressed as:
%
%%%%%%%%%%%%%%%%%%%%%%%%%%%%%%%%%%%%%%%%%%%%%%%%%%%%%%%%%%%%%
\beq
{\cal L}_\Phi = \partial_\mu \Phi \partial^\mu \Phi^* + \mu_\Phi^2 |\Phi|^2 
- \lambda_\Phi |\Phi|^4 + \frac{\epsilon_\Phi^2}{2} (\Phi^2 + \mrm{~h.c.}).
\label{Eq:lagrangian2}
\eeq
%%%%%%%%%%%%%%%%%%%%%%%%%%%%%%%%%%%%%%%%%%%%%%%%%%%%%%%%%%%%%

We can see that eq.~(\ref{Eq:lagrangian2}) 
contains a term that explicitly breaks the $U(1)$ symmetry, giving mass to 
the Goldstone boson $a$, which is then promoted to a pseudo-Goldstone state. 
A dark matter $\chi$ can naturally be introduced in this framework through the 
Lagrangian\footnote{We will work in the 
framework of Majorana dark matter throughout this analysis. 
The extension to the Dirac case is straightforward.}:
%
%%%%%%%%%%%%%%%%%%%%%%%%%%%%%%%%%%%%%%%%%%%%%%%%%%%%%%%%%%%%%
\bea\label{Eq:lagrangian3}
{\cal L}_\chi &&=\frac{1}{2} i \bar \chi \gamma^\mu \partial_\mu \chi - g_\chi \Phi \bar \chi \chi 
+ \mrm{h.c.} \nonumber\\
&&
= \frac{1}{2} i \bar \chi \gamma^\mu \partial_\mu \chi - \frac{g_\chi}{\sqrt{2}} s \bar \chi \chi 
- i \frac{g_\chi}{\sqrt{2}} a \bar \chi \gamma^5 \chi.
\eea
%%%%%%%%%%%%%%%%%%%%%%%%%%%%%%%%%%%%%%%%%%%%%%%%%%%%%%%%%%%%%

Once $\Phi$ acquires a vacuum expectation value ($vev$) through the spontaneous 
symmetry breaking mechanism, $\Phi= \frac{1}{\sqrt{2}}(v_\Phi + s + i a)$ with 
$v^2_\Phi = \frac{\mu_\Phi^2}{\lambda_\Phi} + \frac{\epsilon^2_\Phi}{\lambda_\Phi} 
\simeq \frac{\mu_\Phi^2}{\lambda_\Phi}$,
it generates automatically the mass of $s$ ($m_s=\sqrt{2 \lambda_\Phi} v_\Phi$), the
$saa$ coupling ($\lambda_{saa}= \lambda_\Phi v_\Phi$) and the dark matter mass ($m_\chi=\sqrt{2} g_\chi v_\Phi$).
Moreover, the parameter $\Lambda$ in eq.~(\ref{Eq:lagrangian1}) 
can always be freely chosen up to a redefinition of the couplings $c_{ii}$'s. 
Choosing logically $\Lambda=v_\Phi$, being the scale of new physics, one can
eliminate $v_\Phi$ and re-express the relevant part of the 
Lagrangian (see eq.~(\ref{Eq:lagrangian1})) as the function of masses and couplings as:
%
%%%%%%%%%%%%%%%%%%%%%%%%%%%%%%%%%%%%%%%%%%%%%%%%%%%%%%%%%%%%%
 \bea \label{Eq:lagrangian4}
  -{\cal L}  &&\supset \frac{\sqrt{\lambda_\Phi} C_{GG}}{m_s} s G_{\mu \nu}^\alpha G^{\mu \nu}_\alpha
 + \frac{\sqrt{\lambda_\Phi} C_{GG}}{m_s} a G_{\mu \nu}^\alpha \widetilde G^{\mu \nu}_\alpha \nonumber\\
 &&   + \frac{\sqrt{\lambda_\Phi} C_{BB}c_W^2}{m_s} s F_{\mu \nu} F^{\mu \nu}
 + \frac{\sqrt{\lambda_\Phi} C_{BB}c_W^2}{m_s} a F_{\mu \nu} \widetilde F^{\mu \nu} \nonumber\\
&&  + \frac{m_s^2}{2} s^2 + \frac{m_a^2}{2}a^2 
 + \sqrt{\frac{\lambda_\Phi}{2}} m_s sa^2
+ \sqrt{\frac{\lambda_\Phi}{2}} m_s s^3  + \frac{\lambda_\Phi}{4} (s^2+a^2)^2 \nonumber\\
&& +\frac{1}{2}m_\chi \bar \chi \chi + \sqrt{\frac{\lambda_\Phi}{2}} \frac{m_\chi}{m_s} 
 (s \bar \chi \chi + i a \bar \chi \gamma^5 \chi),
 \eea
%%%%%%%%%%%%%%%%%%%%%%%%%%%%%%%%%%%%%%%%%%%%%%%%%%%%%%%%%%%%% 

where $C_{ii}= 2 c_{ii}$, $c^2_W=cos^2\theta_W$, 
$F_{\mu\nu}(\widetilde F_{\mu\nu})$ is the field(dual field) strength for the photon field,
$m_a=\sqrt{2}\epsilon_\Phi$ is the pseudoscalar mass and we have worked
with\footnote{The coupling $C_{BB}$, from the requirement of gauge invariance,  
also generates interactions like $sZZ,\,sZ\gamma$, etc. We, however, do not 
explore them in detail since they are observed to have a marginal impact in our analysis.} 
$c_{WW}\to 0$. If now one fixes the mass of the 
scalar at $750$ GeV,
all the physics is then completely determined by a set of five parameters 
($C_{GG}$, $C_{BB}$, $\lambda_\Phi$, $m_a$, $m_\chi$). If one further wants to fit
simultaneously, the width of $s$, the di-photon cross-section and the relic abundance, 
one is left with only two free parameters. The model then becomes 
very predictive.

Notice that, sticking to an effective field theory approach, the dimension-5 interaction terms in 
eq.~(\ref{Eq:lagrangian1}), between the field $\Phi$ and the SM gauge bosons, do not 
preserve the $U(1)$ symmetry. As already mentioned, an explicit  breaking of the $U(1)$ 
symmetry has already been introduced in eq.~(\ref{Eq:lagrangian2}) 
in order to have a non-zero mass for the pseudoscalar field $a$ and thus, 
the presence of these dimension-5 operators is rather natural. 
Moreover, in the studied scenario, as will be clarified subsequently,
the coefficients of these operators, e.g., $c_{BB},\,c_{GG}$, required to be much 
suppressed in order to generate an experimentally viable production cross-section 
with collimated photons. 
In this setup, one might also consider the inclusion of dimension-6 operators, 
e.g., $\frac{\widetilde {c}_{GG}}{\Lambda^2}\Phi^\dagger\Phi G^{\mu\nu}_a G^a_{\mu\nu}$, 
which preserve the $U(1)$ symmetry. These terms, with the choice of $\Lambda=v_\Phi$, 
after spontaneous breaking of the $U(1)$ symmetry would produce new 
effective dimension-5 interactions for the scalar, featuring the same $1/v_\Phi$ 
suppression associated with the $U(1)$ violating ones. This would modify 
the trilinear couplings between the scalar component of $\Phi$ and the SM gauge bosons
and thus, will affect the production vertex and decay width of the scalar field.
Now, as will be explained later, in the studied framework the decay width
is dominated by the tree-level $saa$ coupling and hence, the 
impact of including dimension-6 terms would hardly affect our findings.
The relevant interactions for the pseudoscalar, however, remain
the same since only dimension-5 operators can lead to couplings linear in the $a$ field.

The main purpose of this paper is to provide a phenomenological explanation of the di-photon excess 
through the production of collimated photon pairs simultaneously with
the main aspects of dark-matter physics.  For this reason in our analysis we 
will freely vary the couplings of various operators and hence, the effect of  
dimension-6 operators can be straightforwardly encapsulated through a redefinition, 
including the contributions from different dimensionalities, of the effective couplings 
between the scalar field $s$ and the SM gauge bosons. 
Thus, we will not explicitly refer 
any further to the presence of dimension-6 operators.

The existence of dimension-six interactions, nevertheless, would appear effective 
for the pair production of $ss$ or $aa$ which would lead to characteristic 
signals, e.g., $gg\to ss\to 4a \to 8\gamma$. A detail and subsequent discussion of such 
signal, however, is beyond the scopes of the current article.

We note in passing that for this work we consider
the width of the resonance in the span of $4-60$ GeV, the di-photon
production cross-section in the range of $1-10$ fb and finally, the 
relic density $\Omega h^2 \approx 0.12$ \cite{Ade:2013zuv}.

%%%%%%%%%%%%%%%%%%%%%%%%%%%%%%%%%%%%%%%%%%%%%%%%%%%%%%%%%%%%%
\section{The LHC analysis} 
\label{Sec:LHC}
%%%%%%%%%%%%%%%%%%%%%%%%%%%%%%%%%%%%%%%%%%%%%%%%%%%%%%%%%%%%%

%%%%%%%%%%%%%%%%%%%%%%%%%%%%%%%%%%%%%%%%%%%%%%%%%%%%%%%%%%%%%
\subsection{Condition to mimic a di-photon process}
%%%%%%%%%%%%%%%%%%%%%%%%%%%%%%%%%%%%%%%%%%%%%%%%%%%%%%%%%%%%%

The decay of a substantially light pseudoscalar can produce highly collimated photons (dubbed as 
``photon jets'') which can be potentially misidentified as a single photon. 
The minimal condition to realize this kind of scenario is that the opening angle 
$\Delta \phi \sim {2}/{\gamma}$ ($\gamma=$ ${m_s}/{2 m_a}$, being the boost factor) 
of the two photons emitted in the $a \rightarrow \gamma \gamma$ process remains below the angular 
resolution of the LHC detectors, which is $\sim\mathcal{O}(20$ mrad)\footnote{This 
corresponds to $\sim1.15^\circ$.}~\cite{Chatrchyan:2013dga,ATLAS:2012ana}. 
This condition provides an upper bound as $m_a \lesssim 2\,
\mbox{GeV}$~\cite{Agrawal:2015dbf,Chala:2015cev,Dasgupta:2016wxw}. At the same time, a very high boost 
typically generates an enhanced decay length for $a$. In order to
mimic the LHC di-photon signal, the decay should occur before the 
electromagnetic calorimeter (ECAL), which is at a distance of $\sim1\,\mbox{meter}$ 
from the collision point. A minimal requirement is then set on the decay length of 
$a$, $l = \beta \gamma/\Gamma_{a} \ll 1$ meter, where
$\beta\gamma=\sqrt{\gamma^2-1}$ and $\Gamma_{a}$ represents the total
decay width for $a$. This constraint is mostly relevant for values of $m_a
\lsim 3m_{\pi^0}$ ($m_{\pi^0}$ denotes the neutral pion mass), such that only 
the decay process $a\to \gamma\gamma$ remains actually accessible\footnote{One 
should note that in this regime the invisible branching fraction
from $a\to\bar{\chi}\chi$ process remains negligible, given the 
suppression of the $a\bar{\chi}\chi$ coupling (see eq.~(\ref{Eq:lagrangian4})) for $m_\chi\lsim m_a/2$.}. 
As a consequence, the decay length, for $m_a\lsim 3m_{\pi^0}$, 
can be expressed only as the function of $m_a$ and $C_{BB}$ as:  
%%%%%%%%%%%%%%%%%%%%%%%%%%%%%%%%%%%%%%%%%%%%%%%%%%%%%%%%%%%%%
\beq
 l \approx 7.3\,\mbox{cm} {\left(\frac{m_s}{750\,\mbox{GeV}}\right)}^3  
 {\left(\frac{500\,\mbox{MeV}}{m_a}\right)}^4 
 \left(\frac{0.05}{\Gamma_s/m_s}\right) {\left(\frac{0.01}{C_{BB}}\right)}^{2}.
\eeq
%%%%%%%%%%%%%%%%%%%%%%%%%%%%%%%%%%%%%%%%%%%%%%%%%%%%%%%%%%%%%

Now the requirement of $l\lsim 1\,\mbox{meter}$ gives the following lower bound on $C_{BB}$:
%%%%%%%%%%%%%%%%%%%%%%%%%%%%%%%%%%%%%%%%%%%%%%%%%%%%%%%%%%%%%
\beq
\label{eq:CBBlow}
 C_{BB} \gtrsim 2.7\times 10^{-3} {\left(\frac{m_s}{750\,\mbox{GeV}}\right)}^{3/2}
 {\left(\frac{500\,\mbox{MeV}}{m_a}\right)}^2 {\left(\frac{0.05}{\Gamma_s/m_s}\right)}^{1/2}. 
\eeq
%%%%%%%%%%%%%%%%%%%%%%%%%%%%%%%%%%%%%%%%%%%%%%%%%%%%%%%%%%%%%
We remark that eq.~(\ref{eq:CBBlow}) should be regarded as a conservative bound. As argued 
in ref.~\cite{Knapen:2015dap}, for example, displaced photons from $a\to\gamma\gamma$ process 
corresponding to $l\sim 1-10\,\mbox{cm}$, might already be distinguishable from 
a pair of prompt photons originating from $s\to\gamma\gamma$ process.

%%%%%%%%%%%%%%%%%%%%%%%%%%%%%%%%%%%%%%%%%%%%%%%%%%%%%%%%%%%%%
\begin{figure}[tbp]
\centering
\includegraphics[width=8cm]{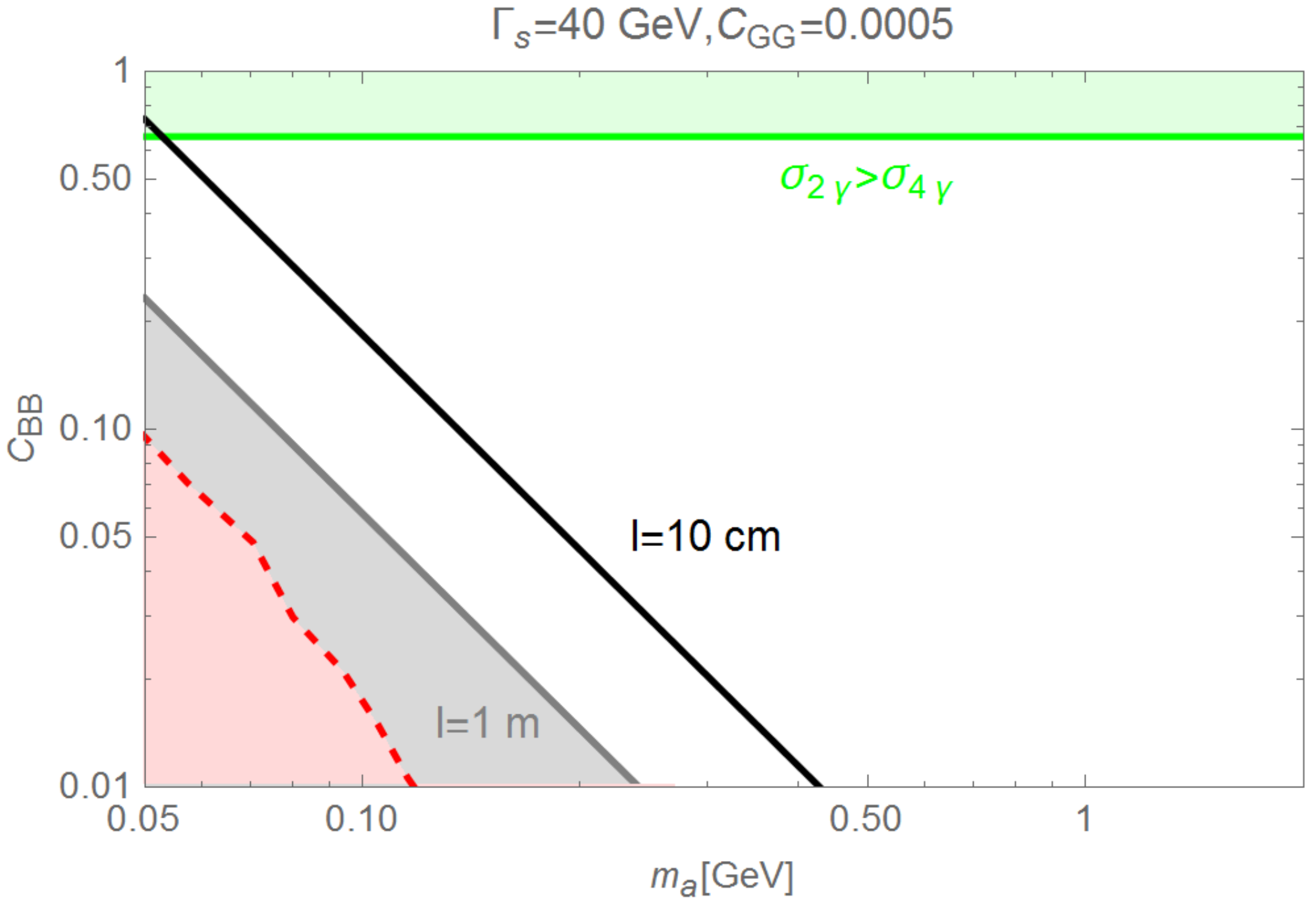}
 \caption{{Allowed region in the $(m_a,C_{BB})$ plane 
 when $a\to \gamma\gamma$ process leads to a collimated photon pairs
 and simultaneously, the associated decay length remains $\lsim 1$ meter 
 (excluding the gray coloured region). The black coloured solid line
 represents a typical displaced decay with $l=10$ cm. The 
 red coloured region is excluded by constraints reported in ref.~\cite{Dobrich:2015jyk}. 
 The green coloured region corresponds to $\sigma_{2\gamma}>\sigma_{4\gamma}$ 
 which is duly explained in the text.}}
\label{Fig:alptraum}
\end{figure}
%%%%%%%%%%%%%%%%%%%%%%%%%%%%%%%%%%%%%%%%%%%%%%%%%%%%%%%%%%%%%

We present in figure.~\ref{Fig:alptraum}, for illustration, 
the allowed $m_a$ values according to the one of $C_{BB}$, 
demanding a pair of collimated photons from $a\to \gamma\gamma$ 
process. This plot is made 
with a fixed value of $C_{GG}=0.0005$ and of $\lambda_\Phi$, corresponding to $\Gamma_s=40$ GeV. 
A further constraint on the associated decay length $l \lsim1\,\mbox{meter}$
disfavours the gray coloured region corresponding to $m_a \lesssim 200\,\mbox{MeV}$.
For reference we have also plotted the line corresponding to $l=10\,\mbox{cm}$
in the $(m_a,C_{BB})$ plane. We also show the $m_a\lesssim 100\,\mbox{MeV}$
region, as reported in ref.~\cite{Dobrich:2015jyk}, (represented by the red coloured region)  
which remains constrained from the direct searches of a light pseudoscalar 
in beam-dump and fixed target experiments.

It is important to note that the process $pp \rightarrow s \rightarrow \gamma \gamma$ 
is also possible in our model since $s$, just like $a$, can couple to two photons through the 
same coupling $C_{BB}$. We have thus represented (in green colour) in figure~\ref{Fig:alptraum} 
the region where\footnote{In reality in figure~\ref{Fig:alptraum} 
the region corresponding to $C_{BB} \gtrsim 0.1$ is excluded by the 8 TeV 
searches of $s\to ZZ,\,Z\gamma$ processes \cite{CMS:2016all,Aad:2015kna,
Khachatryan:2015cwa,Aad:2014fha,CMS:2015lza}. 
One should interpret figure~\ref{Fig:alptraum} just 
as an illustrative plot.} 
$\sigma_{2\gamma}\equiv \sigma(pp \rightarrow s \rightarrow \gamma \gamma) > 
\sigma_{4\gamma}\equiv \sigma(pp \rightarrow s \rightarrow aa \rightarrow 4\gamma)$.

The allowed range, $0.2\,\mbox{GeV} \leq m_a \leq 2\,\mbox{GeV}$, can be constrained 
further by considering suitable 
isolation cuts. Indeed, experimental searches of the somewhat similar 
$h \rightarrow aa \rightarrow 4 \gamma$ process, with $h$ being the SM Higgs boson, 
can discriminate, through suitably defined calorimeter variables, isolated photons from the 
ones coming from $\pi^0$ decays. Given the similarity between the $\pi^0 \rightarrow \gamma \gamma$ 
and $a \rightarrow \gamma \gamma$ processes, it is 
useful to adopt those variables to reduce the probability of 
faking single photon signals~\cite{Draper:2012xt}. 
%
%%%%%%%%%%%%%%%%%%%%%%%%%%%%%%%%%%%%%%%%%%%%%%%%%%%%%%%%%%%%%
\begin{figure}[tbp]
\centering
\begin{center}
 \includegraphics[width=8cm]{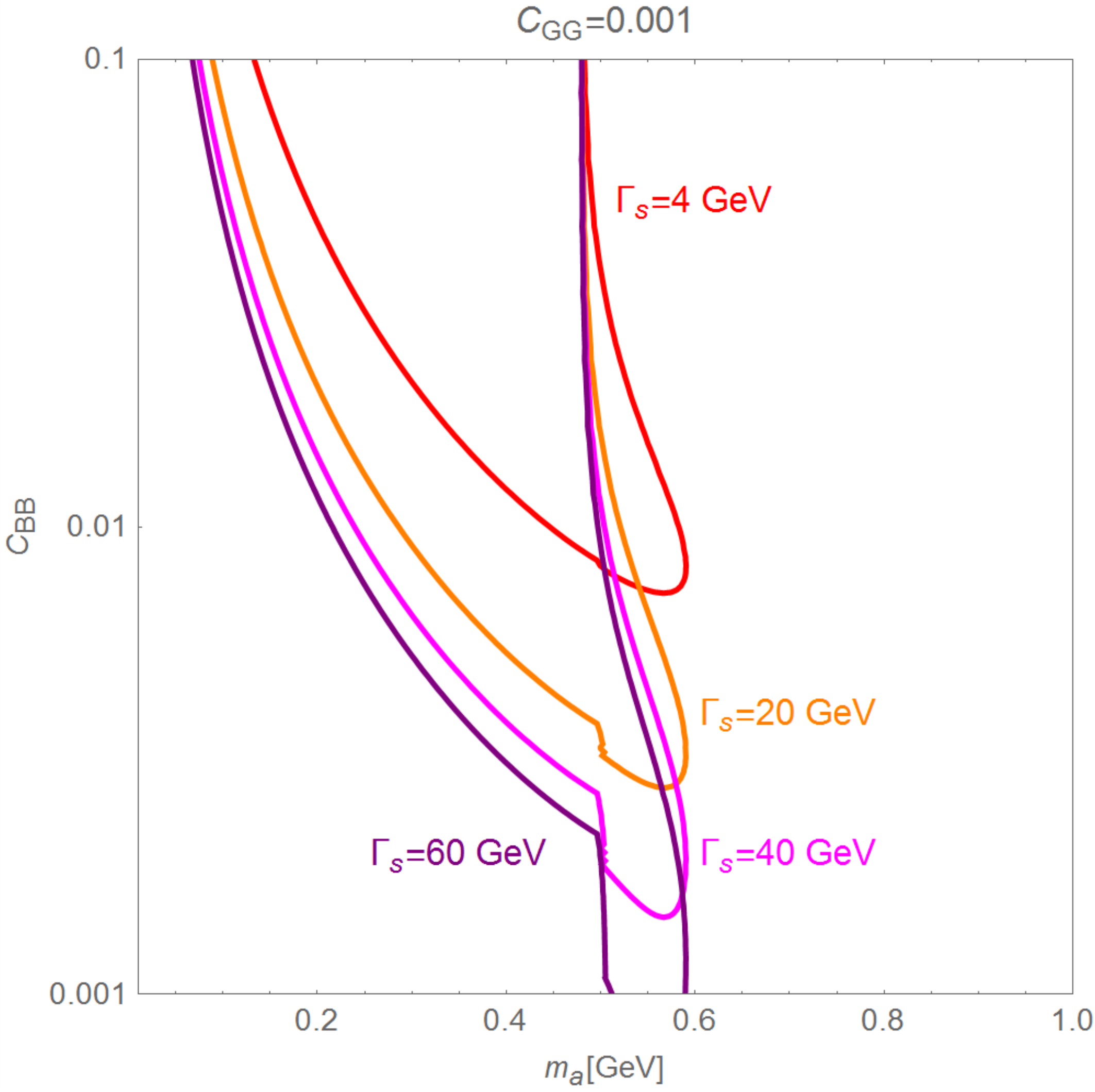}
 \caption{The variations of coupling $C_{BB}$ with $m_a$  
 for the different values of $\Gamma_s$, ensuring that the pseudoscalar
 decays before the ECAL and produces sufficiently collimated photon pairs to mimic 
 the observed di-photon signal. For this plot $C_{GG}=0.001$.}
\label{Fig:ma}
\end{center}
\end{figure}
%%%%%%%%%%%%%%%%%%%%%%%%%%%%%%%%%%%%%%%%%%%%%%%%%%%%%%%%%%%%%

We have then redetermined, in figure~\ref{Fig:ma} the allowed parameter space in the 
$(m_a,C_{BB})$ plane, for $C_{GG}=0.001$ and for four values of $\lambda_\Phi$, ranging from $0.25$ to $4$
giving $\Gamma_s$ in the span of $4$ GeV to $60$ GeV, by using the 
treatment illustrated in ref.~\cite{Aparicio:2016iwr}. 
The upper limit on $m_a$, compared to figure~\ref{Fig:alptraum}, is now lowered to $500-600\,\mbox{MeV}$ 
(based on the similar considerations 
a slightly weaker limit of 800 MeV has been found in ref.~\cite{Knapen:2015dap}). Interestingly, 
most of the allowed parameter space now lies below the threshold of $3m_{\pi^0}$, 
above which the hadronic decays become accessible for the pseudoscalar. 
As a consequence, in this regime, $Br(a \rightarrow \gamma \gamma)=1$ irrespective of the values of 
$C_{BB}$ and $C_{GG}$. 
As will be clarified in the next section, this will lead to a rather predictive scenario. 
To utilize this predictive behaviour, we will use this region of the parameter space for 
most of the analytical estimates presented in this work. In the absence of a dedicated experimental study, 
we have nevertheless included the region 
$m_{3 \pi^0} \leq m_a \leq 2\,\mbox{GeV}$ in the numerical computations.

We finally remark that the allowed parameter space can be constrained 
even further by considering the different photon conversion rate, i.e., the probability of 
the interaction of a photon within the calorimeter to produce a $e^+e^-$ pair, 
between $2\gamma$ and $4\gamma$ events. A detailed investigation of this possibility has 
been performed in ref.~\cite{Dasgupta:2016wxw}.

%%%%%%%%%%%%%%%%%%%%%%%%%%%%%%%%%%%%%%%%%%%%%%%%%%%%%%%%%%%%%
\subsection{Di-jet constraint}
%%%%%%%%%%%%%%%%%%%%%%%%%%%%%%%%%%%%%%%%%%%%%%%%%%%%%%%%%%%%%

As already emphasized, that the couplings $C_{GG}$ and $C_{BB}$ are constrained, 
by the requirement of the relative absence of 
additional signals, e.g., di-jets, $Z\gamma$ or $ZZ$ etc., compared to the di-photon one. 
In our scenario the strongest constraint on $C_{GG}$ comes from the di-jet searches.
For this purpose we have imposed that $\sigma(pp \rightarrow s \rightarrow jj) < 2.5\,\mbox{pb}$,
at $8$ TeV centre-of-mass\footnote{One can, for example, see ref.~\cite{ATLAS:2015nsi,Khachatryan:2015dcf}
for a similar bound with $13$ TeV centre-of-mass energy.}  energy~\cite{CMS:2015neg,Aad:2014aqa}.

In the $Br(a \rightarrow \gamma \gamma)=1$ regime the corresponding cross-section is given by:
%%%%%%%%%%%%%%%%%%%%%%%%%%%%%%%%%%%%%%%%%%%%%%%%%%%%%%%%%%%%%
\bea \label{eq:dijet}
\sigma(pp \rightarrow s \rightarrow jj)&&=\frac{\pi^2}{8 m_s \bf{s}} 
I_{GG, 8 TeV} \frac{2 C_{GG}^2 \lambda_\Phi m_s}
{\pi} Br(s \rightarrow gg)\nonumber\\
&& \approx \frac{\pi^2}{8 m_s \bf{s}} I_{GG, 8 TeV} \frac{2 C_{GG}^2 
\lambda_\Phi m_s}{\pi}\frac{32 C_{GG}^2}{1+32 C_{GG}^2},
\eea
%%%%%%%%%%%%%%%%%%%%%%%%%%%%%%%%%%%%%%%%%%%%%%%%%%%%%%%%%%%%%

where we have used eq.~(\ref{Eq:lagrangian4}) and retained the dependence 
only on the $aa$ and $gg$ channels (see next section) in the determination of $\Gamma_s$. 
The quantities $\bf{s}$ and $I_{GG, 8 TeV}$ 
represent the square of the centre-of-mass energy
and the integral of a function involving dimensionless quantities like
parton distribution functions (PDFs), evaluated at $\sqrt{{\bf s}}=8$ TeV, respectively. 
It can be easily seen from eq.~(\ref{eq:dijet}) that the condition 
$\sigma(pp\to s\to jj)<2.5$ pb, with\footnote{Here we have considered MSTW 2008 NLO 
PDF~\cite{Martin:2010db,Martin:2009bu,Martin:2009iq}
and included the next-to-leading order scaling factor.} $I_{GG, 8 TeV}=280$,
is satisfied for $C_{GG} \lesssim 0.1$.

%%%%%%%%%%%%%%%%%%%%%%%%%%%%%%%%%%%%%%%%%%%%%%%%%%%%%%%%%%%%% 
\subsection{Fitting the di-photon excess}
%%%%%%%%%%%%%%%%%%%%%%%%%%%%%%%%%%%%%%%%%%%%%%%%%%%%%%%%%%%%%

\begin{figure}[tbp]
\centering
 \includegraphics[width=0.5\linewidth]{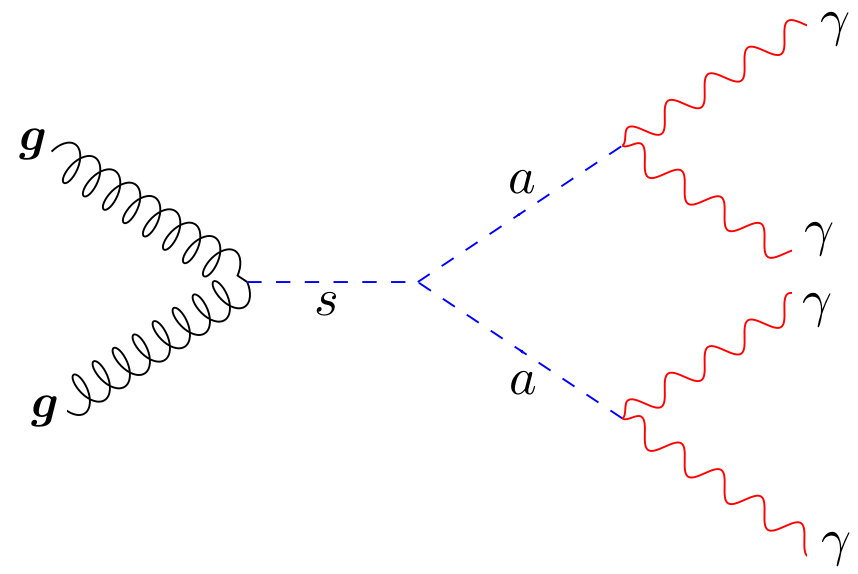}
 \caption{Main channel contributing to the di-photon excess observed at the LHC.}
\label{Fig:feynman1}
\end{figure}

The production cross-section for the studied di-photon signal through the gluon fusion 
process $gg \rightarrow s \rightarrow aa \rightarrow 4 \gamma$ (see 
figure~\ref{Fig:feynman1}) is given by

%%%%%%%%%%%%%%%%%%%%%%%%%%%%%%%%%%%%%%%%%%%%%%%%%%%%%%%%%%%%%
\beq
\sigma_{4 \gamma} = \frac{\pi^2}{8 m_s  \bf{s}} \Gamma(s \rightarrow gg) Br(s \rightarrow aa)
[Br(a \rightarrow \gamma \gamma)]^2~ I_{GG}, 
\label{Eq:siggg}
\eeq
%%%%%%%%%%%%%%%%%%%%%%%%%%%%%%%%%%%%%%%%%%%%%%%%%%%%%%%%%%%%%

with $I_{GG}$ being the integral over the PDFs. This is 
estimated to be $\simeq 3400$ using MSTW 2008 NLO PDF~\cite{Martin:2010db,Martin:2009bu,Martin:2009iq}
and including the next-to-leading order scaling factor in the definition of $I_{GG}$. 
The desired signal, as already mentioned, can also be induced through the photon 
fusion~\cite{Harland-Lang:2016qjy} process $\gamma \gamma \rightarrow s 
\rightarrow aa \rightarrow 4 \gamma$. 
We, however, do not consider this possibility in this article.

%%%%%%%%%%%%%%%%%%%%%%%%%%%%%%%%%%%%%%%%%%%%%%%%%%%%%%%%%%%%%
%%%%%%%%%%%%%%%%%%%%%%%%%%%%%%%%%%%%%%%%%%%%%%%%%%%%%%%%%%%%%

At this point one can use eq.~(\ref{Eq:lagrangian4}) to compute:
%%%%%%%%%%%%%%%%%%%%%%%%%%%%%%%%%%%%%%%%%%%%%%%%%%%%%%%%%%%%%
\be\label{Eq:saasgg}
\Gamma(s \rightarrow aa) = \frac{\lambda_\Phi m_s}{16 \pi}, ~~ 
\Gamma(s \rightarrow gg) = \frac{2 \lambda_\Phi m_s}{\pi} C_{GG}^2. ~~
\ee
%%%%%%%%%%%%%%%%%%%%%%%%%%%%%%%%%%%%%%%%%%%%%%%%%%%%%%%%%%%%%

Now the processes $s\to \gamma\gamma$, $s\to Z Z$, $s\to Z\gamma$
are suppressed compared to $s\to gg$ mode by factors like 
$8/\cos^4 \theta_W,\,8/\sin^4 \theta_W,\,8/\sin^2 \theta_{2W}$, respectively, 
even when $C_{BB}\sim C_{GG}$.
The upper limit of $C_{BB}$ is constrained from the 
photon fusion process. The invisible decay mode $s\to \bar{\chi}\chi$ also suffers suppression,
either from the $s\bar{\chi}\chi$ coupling $\propto {m_\chi}/{m_s}$ 
(for $m_\chi\ll m_s$) or from the phase space factor
(when $m_\chi \sim m_s/2$). Thus, collectively one can use eq.~(\ref{Eq:saasgg}) to get 
%%%%%%%%%%%%%%%%%%%%%%%%%%%%%%%%%%%%%%%%%%%%%%%%%%%%%%%%%%%%%
\be
\label{eq:braa}
Br(s \rightarrow aa) \approx \frac{1}{1+ 32 C_{GG}^2}.
\ee
%%%%%%%%%%%%%%%%%%%%%%%%%%%%%%%%%%%%%%%%%%%%%%%%%%%%%%%%%%%%%

This equation is especially valid for $m_a \lesssim 500$ MeV, i.e., below 
the $3m_{\pi^0}$ threshold when $Br(a \rightarrow \gamma \gamma)=1$. In this region,
combining eq.~(\ref{eq:braa}) with eq.~(\ref{Eq:siggg}) one obtains:
%
%%%%%%%%%%%%%%%%%%%%%%%%%%%%%%%%%%%%%%%%%%%%%%%%%%%%%%%%%%%%%
\bea
\sigma_{4 \gamma}&&\approx \frac{4 \pi^2}{{\bf s}} I_{GG} 
\frac{C_{GG}^2}{(1+ 32 C_{GG}^2)^2} \frac{\Gamma_s}{m_s} \,
\simeq 16~\mrm{fb} \frac{(\Gamma_s/m_s)}{0.05} \frac{\left( \frac{C_{GG}}{0.001} 
\right)^2}{(1+ 32 C_{GG}^2)^2}.
\label{Eq:siggg2}
\eea
%%%%%%%%%%%%%%%%%%%%%%%%%%%%%%%%%%%%%%%%%%%%%%%%%%%%%%%%%%%%%

Interestingly, for $m_a \lsim m_{3\pi^0}$ the $4\gamma$ production cross-section depends only on 
$C_{GG}$ and $\lambda_\Phi$ couplings while remains independent 
of $C_{BB}$, the $a\gamma\gamma$ coupling. 
Indeed, from eq.~(\ref{Eq:siggg2}) one can see the possibility to obtain 
$\sigma_{4\gamma}\sim\mathcal{O}(1-10~{\rm fb})$ together with a large $\Gamma_s$ 
($\sim 5 \%~ m_s$) for low values of $C_{GG}$, consistent with the di-jet constraints. 
This comes trivially from the fact that a large $\lambda_\Phi$ coupling is responsible 
for the large $\Gamma_s$ and a sizable $\sigma_{4\gamma}$ 
without requiring high values of $C_{GG}$ and $C_{BB}$ couplings. 
We finally notice that the di-jet limit, $C_{GG} \lesssim 0.1$
implies that for all practical purposes $\Gamma_s \approx \Gamma(s\to aa)$.
This feature allows us to trade, in the analytical expressions considered in this section, the free 
parameter $\lambda_\Phi$ with the physical observable $\Gamma_s$.

It is important to note that although the coupling $C_{BB}$
does not directly appear in $\sigma_{4\gamma}$, it is not
unconstrained in nature. In fact, the absence of di-photon
signal from photon fusion process and the requirement 
of a pseudoscalar decaying before the ECAL provide
an upper and lower bound for $C_{BB}$, respectively. In reality, however,
there exists one more way to constrain the coupling $C_{BB}$
by demanding a suppressed $s\to \gamma\gamma$ process.
We have already used this to derive eq.~(\ref{eq:braa}).
The condition $\frac{\Gamma(s \rightarrow \gamma \gamma)}{\Gamma(s \rightarrow aa)}\ll 1$, implies
$C_{BB} \ll 0.65$.

A good fit for $\sigma_{4\gamma}$ can also be obtained for $m_a \gtrsim 500\,\mbox{MeV}$ 
with a sizable branching fraction of $a$ into the hadronic channels. 
In such a case the cross-section is obtained by multiplying eq.~(\ref{Eq:siggg2}) 
with $\left(Br(a\to\gamma\gamma)\right)^2 \approx ({C_{BB}^4 c_W^8}/{64 C_{GG}^4})$ 
and is given by:
%%%%%%%%%%%%%%%%%%%%%%%%%%%%%%%%%%%%%%%%%%%%%%%%%%%%%%%%%%%%%
\beq
\sigma_{4 \gamma} \approx 5\,\mbox{fb} \frac{(\Gamma_s/m_s)}{0.05} 
\frac{1}{(1+ 32 C_{GG}^2)^2} {\left(\frac{C_{BB}}{0.005}\right)}^4 {\left(\frac{0.005}{C_{GG}}\right)}^2.
\eeq
%%%%%%%%%%%%%%%%%%%%%%%%%%%%%%%%%%%%%%%%%%%%%%%%%%%%%%%%%%%%%%%

The condition for the dominance of $s\to4\gamma$ channel
compared to $s\to\gamma\gamma$ mode is similarly modified and becomes:
%%%%%%%%%%%%%%%%%%%%%%%%%%%%%%%%%%%%%%%%%%%%%%%%%%%%%%%%%%%%%%%
\beq
C_{GG} \lesssim 0.27 C_{BB}^{1/2}. 
\eeq
%%%%%%%%%%%%%%%%%%%%%%%%%%%%%%%%%%%%%%%%%%%%%%%%%%%%%%%%%%%%%

%%%%%%%%%%%%%%%%%%%%%%%%%%%%%%%%%%%%%%%%%%%%%%%%%%%%%%%%%%%%%
\section{The analysis implementing the dark sector}
\label{Sec:DM}
%%%%%%%%%%%%%%%%%%%%%%%%%%%%%%%%%%%%%%%%%%%%%%%%%%%%%%%%%%%%%

\begin{figure}[tbp]
\centering
\includegraphics[width=0.6\linewidth]{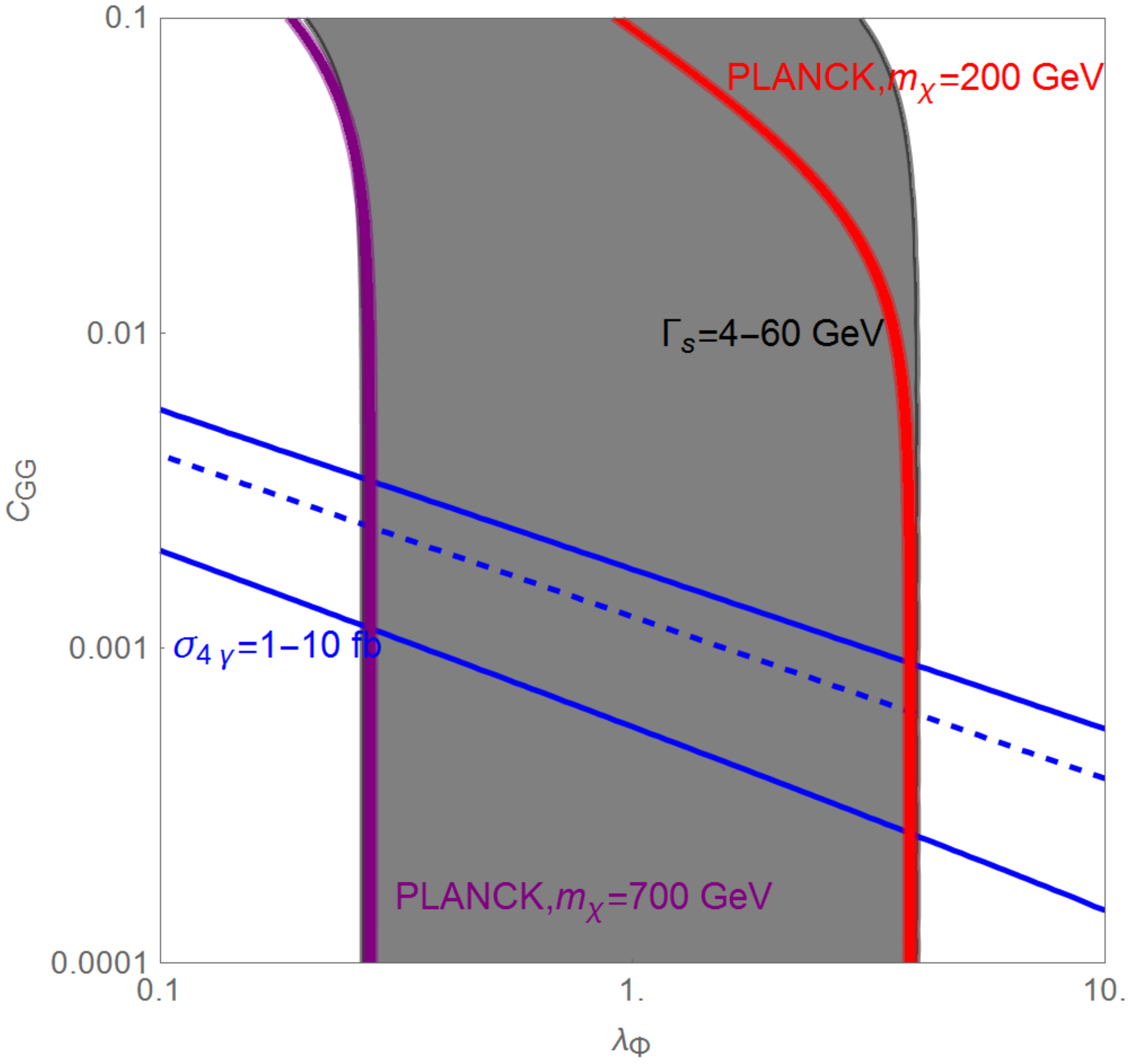}.
\caption{The allowed parameter space in the ($\lambda_\Phi$, $C_{GG}$) plane
consistent with the observations of: (a) correct relic density,
(b) the observed di-photon production cross-section
and (c) a relatively large width of the resonance.
The gray coloured band corresponds to the large width region ($4-60$ GeV) whereas the blue
coloured lines (solid and dashed) represent
a di-photon production cross-section of $1-10$ fb, as observed at the LHC.
The red and magenta coloured lines correspond
to points that respect the PLANCK constraints for $m_\chi= 200$ GeV and $700$ GeV, respectively.
For this plot we consider $C_{BB}=0.005$.}
\label{Fig:cgglphi}
\end{figure}

Once one couples the dark matter to the field $\Phi$ through eq.~(\ref{Eq:lagrangian3}), 
one can compute the relic abundance of $\chi$ as a function of $C_{GG}$ and $\lambda_\Phi$ 
for different $m_\chi$ values and check if there exists
parameter space compatible with the observed di-photon excess.  
The result is presented in figure~\ref{Fig:cgglphi}
where we have shown in the ($\lambda_\phi$, $C_{GG}$) plane 
the parameter space allowed by the three constraints,
namely: (i) $1~{\rm fb}\leq \sigma_{4\gamma}\leq 10~{\rm fb}$, 
(ii) $4~{\rm GeV}\leq \Gamma_s\leq 60~{\rm GeV}$
and (iii) correct relic abundance $\approx 0.12$ \cite{Ade:2013zuv}. 
It is evident from figure~\ref{Fig:cgglphi} that for 
any value of the dark matter mass between 200 GeV
to 700 GeV, there exist values of $C_{GG}$ that give a $\sigma_{4\gamma}$ 
$\sim\mathcal{O}(1-10~\rm{fb})$ together with $4$ GeV $\lsim \Gamma_s\lsim 60$ GeV
while still respecting the PLANCK constraints. We will study the different regimes in detail in 
the following sub-sections. 
%
%%%%%%%%%%%%%%%%%%%%%%%%%%%%%%%%%%%%%%%%%%%%%%%%%%%%%%%%%%%%%
%%%%%%%%%%%%%%%%%%%%%%%%%%%%%%%%%%%%%%%%%%%%%%%%%%%%%%%%%%%%%

\subsection{\boldmath $m_\chi \lesssim 375$ GeV}

For $m_a < m_\chi \lesssim m_s/2$ the dark matter can annihilate into $\gamma \gamma$, 
$gg$ and $aa$, as well as into $Z \gamma$ and $ZZ$, whether kinematically open 
(the last two modes are suppressed in the setup considered in this work).

The $\gamma \gamma$ and $gg$ channels mostly originate from s-channel exchange of the pseudoscalar.
The corresponding thermally averaged cross-sections, using eq.~(\ref{eq:braa}) and eq.~(\ref{Eq:siggg2}),
can be estimated as \cite{Mambrini:2015wyu,D'Eramo:2016mgv,Backovic:2015fnp}: 
%%%%%%%%%%%%%%%%%%%%%%%%%%%%%%%%%%%%%%%%%%%%%%%%%%%%%%%%%%%%%
\bea
\label{eq:sigmalight}
 \langle \sigma v \rangle_{gg}&&= \frac{16 \lambda_\Phi^2 C_{GG}^2 m_\chi^2}{\pi m_s^4} \, 
\approx 5 \times 10^{-30} 
\left( \frac{\sigma_{4 \gamma}}{16~\mrm{fb}}\right) \left( \frac{\Gamma_s/m_s}{0.05} \right) 
\left(\frac{m_\chi}{100 ~\mrm{GeV}} \right)^2 \left( \frac{750~\mrm{GeV}}{m_s} \right)^4,
\nonumber\\
\langle \sigma v \rangle_{\gamma \gamma}&&=\frac{2 \lambda_\Phi^2 C_{BB}^2 c_W^4 m_\chi^2}{\pi m_s^4}\nonumber\\
&&\approx \frac{9 \times 10^{-27}}{(1 +32 C_{GG}^2)^2} \left( \frac{C_{BB}}{0.1} \right)^2 
\left( \frac{\Gamma_s / m_s}{0.05} \right)^2 
\left( \frac{m_\chi}{100~\mrm{GeV}} \right)^2 \left( \frac{750 ~\mrm{GeV}}{m_s} \right)^4,
\eea
%%%%%%%%%%%%%%%%%%%%%%%%%%%%%%%%%%%%%%%%%%%%%%%%%%%%%%%%%%%%%

in the unit of $\mrm{cm^3 s^{-1}}$. We now clearly see the difficulty to achieve the correct 
relic abundance through the s-wave channel 
due to the constraints on $\sigma_{4 \gamma}$ and $C_{BB} \lesssim 0.1$ 
for the $gg$ and $\gamma \gamma$ final states, respectively.
Moreover, the gluon channel is also constrained by the di-jet constraint, eq.~(\ref{eq:dijet}), and, to a 
lesser extent and only for $m_a \geq 500\,\mbox{MeV}$, by the FERMI searches of gamma-rays 
from DSph~\cite{Ackermann:2015zua,Ackermann:2013yva}.
On the other hand, the contribution from the photon channel is, instead, substantially excluded by the 
FERMI searches of monochromatic gamma-ray signals ~\cite{Ackermann:2015lka}, which give a limit as strong as 
$10^{-(29 \div 30)}$ $\mrm{cm^3 s^{-1}}$ for $m_\chi \lsim 100$ GeV.

\begin{figure}[tbp]
\centering
 \includegraphics[width=0.5\linewidth]{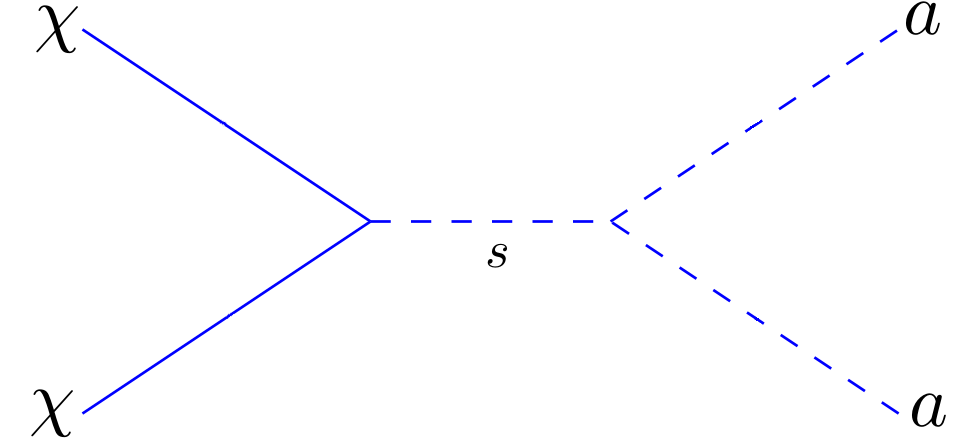}
 \caption{Main channel contributing to the relic abundance for $m_\chi \lesssim 375$ GeV.}
\label{Fig:feynman0}
\end{figure}

Even if the $aa$ channel gives rise to a p-wave velocity suppressed cross-section (by CP arguments), it is the
dominant annihilation process at the decoupling time. It proceeds mostly 
through the $s$-channel exchange of the scalar $s$ as depicted in the figure~\ref{Fig:feynman0}. 
The concerned thermally averaged cross-section can be estimated as: 
%%%%%%%%%%%%%%%%%%%%%%%%%%%%%%%%%%%%%%%%%%%%%%%%%%%%%%%%%%%%%
\bea
\langle \sigma v \rangle_{aa}&&=\frac{9 m_\chi^2 \lambda_\Phi^2}{384 \pi m_s^4}v^2 \nonumber\\
&& \approx  5.3 \times 10^{-27}\,{\mbox{cm}}^3 {\mbox{s}}^{-1}\frac{1}{(1+32 C_{GG}^2)^2}
{\left(\frac{\Gamma_s/m_s}{0.05}\right)}^2 
{\left(\frac{m_\chi}{100\,\mbox{GeV}}\right)}^{2}{\left(\frac{750\,\mbox{GeV}}{m_s}\right)}^4. 
\eea
%%%%%%%%%%%%%%%%%%%%%%%%%%%%%%%%%%%%%%%%%%%%%%%%%%%%%%%%%%%%%

In figure~\ref{Fig:mchigammas} we show the result of complete numerical analysis, 
performed precisely by determining the dark matter annihilation cross-sections and its relic density 
through the package {\tt{MicrOMEGAs}}~\cite{Belanger:2014vza}. 
Fixed values for the $C_{GG}$ and $C_{BB}$ couplings, 
as $0.001$ and $0.005$, respectively, are used
for this figure. Here we have plotted the contour of correct relic density
in the $(\,m_\chi,\,\Gamma_s)$ plane corresponding to different di-photon production
cross-sections. We have also used the same methods
and tools for figure~\ref{Fig:cgglphi}. 
We notice from figure~\ref{Fig:mchigammas} that, when $m_\chi\lesssim 375$ GeV, the 
correct relic density can be achieved, through the 
annihilation channel of figure~\ref{Fig:feynman0}, for
$m_\chi\sim \mathcal{O}(200-300$ GeV) and for $12~{\rm GeV}\leq\Gamma_s \leq 50$ GeV 
while still fitting the LHC di-photon data. We also notice the typical
pole effect, i.e., when $m_\chi$ approaches $m_s/2$, the width needs to 
be narrow to avoid the under-abundance due to thermal broadening effect \cite{Gondolo:1990dk}. 
On the other hand, being velocity suppressed, this channel cannot account for any indirect detection signal
as it will be discussed in sub-section~\ref{Sec:detection}.

%%%%%%%%%%%%%%%%%%%%%%%%%%%%%%%%%%%%%%%%%%%%%%%%%%%%%%%%%%%%%
%%%%%%%%%%%%%%%%%%%%%%%%%%%%%%%%%%%%%%%%%%%%%%%%%%%%%%%%%%%%%

%%%%%%%%%%%%%%%%%%%%%%%%%%%%%%%%%%%%%%%%%%%%%%%%%%%%%%%%%%%%%
\subsection{\boldmath $m_\chi > 375$ GeV}
%%%%%%%%%%%%%%%%%%%%%%%%%%%%%%%%%%%%%%%%%%%%%%%%%%%%%%%%%%%%%

\begin{figure}[tbp]
\centering
 \includegraphics[width=0.5\linewidth]{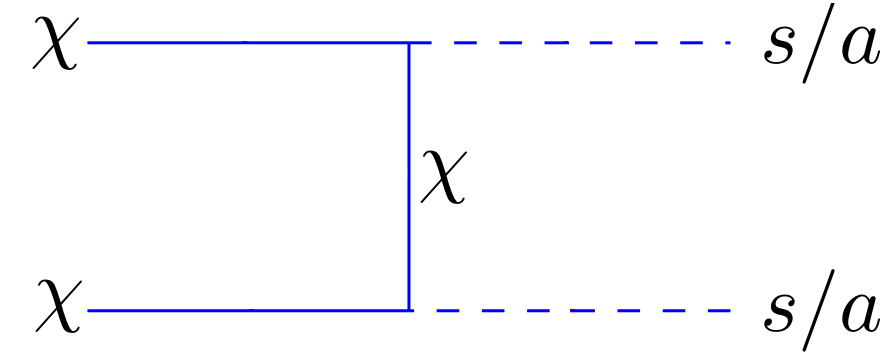}
 \caption{Main channel responsible for the relic abundance for $m_\chi > 375$ GeV.}
\label{Fig:feynman2}
\end{figure}

When $m_\chi > m_s/2$, the annihilation process $\chi \chi \rightarrow sa$, shown 
in figure~\ref{Fig:feynman2}, becomes kinematically allowed. This channel will dominate
the annihilation process compared to $aa$ (or $ss$ when $m_\chi > m_s$) as it is the only channel which is not 
velocity suppressed. The relevant thermally averaged cross-section is given by:

%%%%%%%%%%%%%%%%%%%%%%%%%%%%%%%%%%%%%%%%%%%%%%%%%%%%%%%%%%%%%
\begin{figure}[tbp]
\centering
\includegraphics[width=8cm]{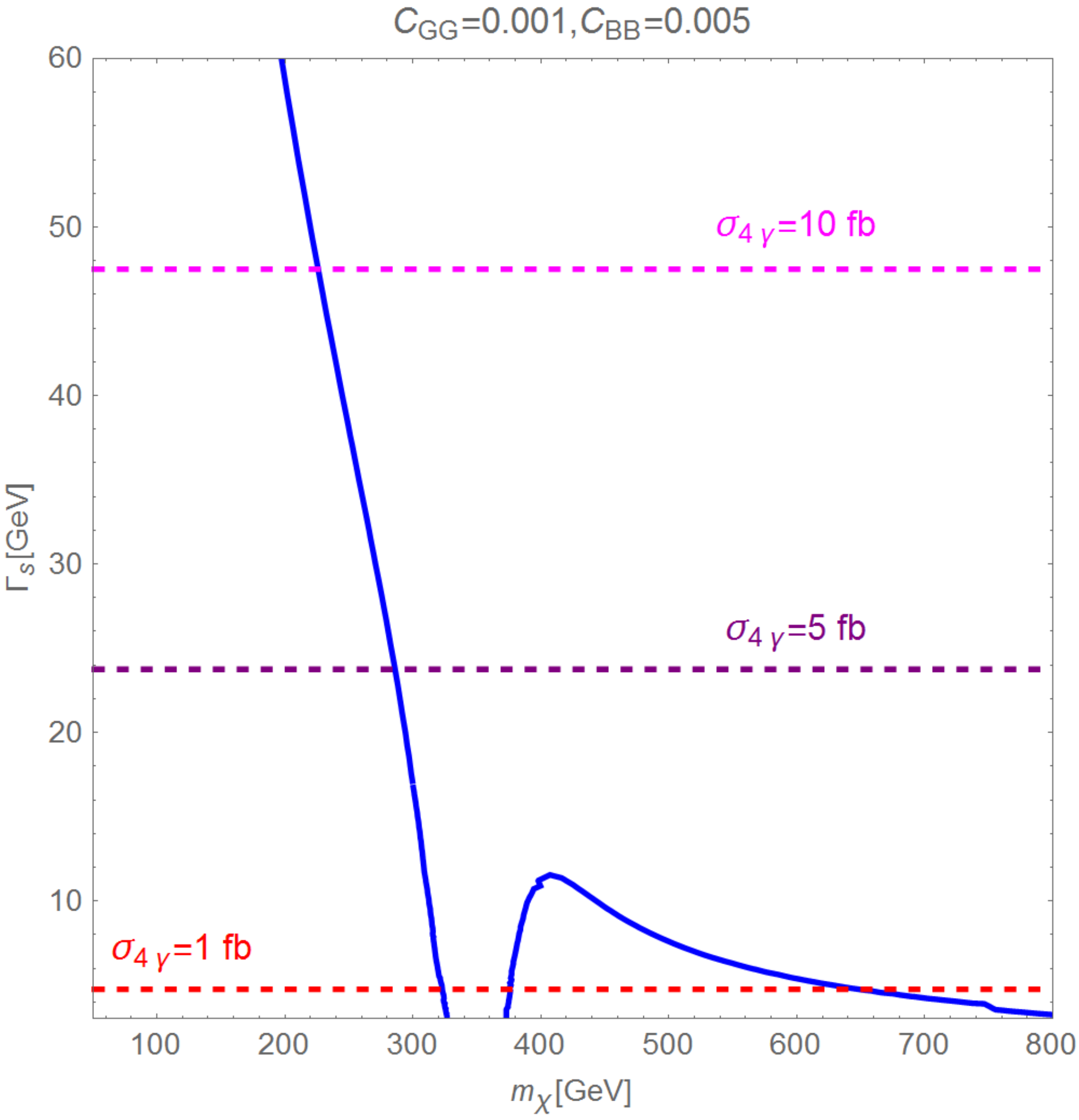}
\caption{The contour of correct relic density in the 
$(\,m_\chi,\,\Gamma_s)$ plane. The differently coloured horizontal lines
correspond to the various values of the di-photon production cross-sections.}
\label{Fig:mchigammas}
\end{figure}
%%%%%%%%%%%%%%%%%%%%%%%%%%%%%%%%%%%%%%%%%%%%%%%%%%%%%%%%%%%%%

%%%%%%%%%%%%%%%%%%%%%%%%%%%%%%%%%%%%%%%%%%%%%%%%%%%%%%%%%%%%%
\bea\label{Eq:sigvsa}
\langle \sigma v \rangle_{sa} &&\simeq \frac{\lambda^2_\Phi}{8 \pi} \frac{m_\chi^2}{m_s^4} \,
\simeq \frac{ 3 \times 10^{-25}~\mrm{cm^3 s^{-1}}}{(1+32 C_{GG}^2)^2} \left( \frac{\Gamma_s /m_s}{0.05} 
\right)^2 \left(\frac{m_\chi}{400~\mrm{GeV}}  \right)^2 {\left(\frac{750\,\mbox{GeV}}{m_s}\right)}^4.
\eea 
%%%%%%%%%%%%%%%%%%%%%%%%%%%%%%%%%%%%%%%%%%%%%%%%%%%%%%%%%%%%%

As evident from eq.~(\ref{Eq:sigvsa}) that in the case of $\Gamma_s/m_s \sim 5\%$,
the cross-section exceeds the thermally favoured 
value and leads, consequently, to an underabundant dark matter. 
This is also reflected from figure~\ref{Fig:mchigammas} that the 
correct relic density for $m_\chi >375\,\mbox{GeV}$ is achieved
only for $\Gamma_s < 12$ GeV. The latter,
for the chosen values of $C_{GG},\,C_{BB}$ couplings, corresponds
to a value of $\sigma_{4\gamma}$ on the lower side.
%
%%%%%%%%%%%%%%%%%%%%%%%%%%%%%%%%%%%%%%%%%%%%%%%%%%%%%%%%%%%%%
%%%%%%%%%%%%%%%%%%%%%%%%%%%%%%%%%%%%%%%%%%%%%%%%%%%%%%%%%%%%%

%%%%%%%%%%%%%%%%%%%%%%%%%%%%%%%%%%%%%%%%%%%%%%%%%%%%%%%%%%%%%
\subsection{Very light case : the freeze in  regime}
%%%%%%%%%%%%%%%%%%%%%%%%%%%%%%%%%%%%%%%%%%%%%%%%%%%%%%%%%%%%%

In this case the dominant contribution to the dark matter relic density is given by the 
annihilations into $gg$ and $\gamma \gamma$ final states through s-channel exchange of 
the pseudoscalar. The corresponding thermally averaged cross-sections are:
%%%%%%%%%%%%%%%%%%%%%%%%%%%%%%%%%%%%%%%%%%%%%%%%%%%%%%%%%%%%%
\bea\label{eq:fr1}
 \langle \sigma v \rangle_{gg}&&= \frac{1024}{\pi} \frac{m_\chi^6}{m_a^4 m_s^4} \lambda_\Phi^2 C_{GG}^2
 \nonumber\\
&&\approx 3 \times 10^{-33}\,{\mbox{cm}}^3 {\mbox{s}}^{-1} \lambda_{\Phi}^2 C_{GG}^2 
{\left(\frac{m_\chi}{0.1\,\mbox{GeV}}\right)}^6 
{\left(\frac{1\,\mbox{GeV}}{m_a}\right)}^{4}{\left(\frac{750\,\mbox{GeV}}{m_s}\right)}^{4}, \nonumber\\
 \langle \sigma v \rangle_{\gamma \gamma}&&= \frac{128}{\pi} \frac{m_\chi^6}{m_a^4 m_s^4} 
\lambda_\Phi^2 C_{BB}^2 c_W^4 \nonumber\\ 
&&\approx 2.2 \times 10^{-34}\,{\mbox{cm}}^3 {\mbox{s}}^{-1} \lambda_{\Phi}^2 C_{BB}^2 
{\left(\frac{m_\chi}{0.1\,\mbox{GeV}}\right)}^6 
{\left(\frac{1\,\mbox{GeV}}{m_a}\right)}^{4}{\left(\frac{750\,\mbox{GeV}}{m_s}\right)}^{4}.
\eea
%%%%%%%%%%%%%%%%%%%%%%%%%%%%%%%%%%%%%%%%%%%%%%%%%%%%%%%%%%%%%

As evident from eq.~(\ref{eq:fr1}) that these cross-sections are very far from the thermally favoured value, 
$\sim \mathcal{O}(10^{-26})\,{\mbox{cm}}^3 {\mbox{s}}^{-1}$, such that the validity 
of the WIMP paradigm itself becomes 
questionable. Thus, we apply the conventional rule-of-thumb of comparing the dark matter annihilation rate 
$\Gamma_{\rm ann}=\langle \sigma v \rangle n_{\chi,eq}$ with the Hubble expansion rate, both evaluated at
temperature $T\sim\mathcal{O}(m_\chi)$. Considering for reference the $\gamma \gamma$ annihilation 
channel~\footnote{As $m_\chi$ decreases, the relevant annihilation processes occur below the temperature 
of the QCD phase transition, $\sim\mathcal{O}(200~{\rm MeV}$), and then the hadronic annihilation 
channel remains inaccessible.},  one gets: 
%%%%%%%%%%%%%%%%%%%%%%%%%%%%%%%%%%%%%%%%%%%%%%%%%%%%%%%%%%%%%  
\beq
\frac{\Gamma_{\rm ann} (T=m_\chi)}{H(T=m_\chi)}  \approx \frac{3.2}{{\left(1+32 C_{GG}^2\right)}^2}
{\left(\frac{\Gamma_s/m_s}{0.05}\right)}^2 {\left(\frac{C_{BB}}{0.005}\right)}^2 
{\left(\frac{m_\chi}{0.1\,\mbox{GeV}}\right)}^7{\left(\frac{0.5\,
\mbox{GeV}}{m_a}\right)}^4 {\left(\frac{750\,\mbox{GeV}}{m_s}\right)}^4.
\eeq
%%%%%%%%%%%%%%%%%%%%%%%%%%%%%%%%%%%%%%%%%%%%%%%%%%%%%%%%%%%%%

For the chosen model parameters with $m_\chi\lsim 100$ MeV, compatible with the 
LHC di-photon signal, the dark matter annihilations become inefficient.
The dark matter, however, still remains relativistic and the ratio $\frac{\Gamma_{\rm ann}}{H}$ drops
very steeply as long as the dark matter mass is further reduced. The dark matter 
could nevertheless be kept into thermal equilibrium
by the decay process $\Gamma(a \rightarrow \bar{\chi} \chi)=\frac{\lambda_\Phi 
m_a m_\chi^2}{8 \pi m_s^2}$, with $a$ 
in turn, kept into equilibrium by its couplings with the gauge bosons, 
at least down to temperatures $\sim\mathcal{O}(m_a)$.
This requirement can be translated into the following condition:
%%%%%%%%%%%%%%%%%%%%%%%%%%%%%%%%%%%%%%%%%%%%%%%%%%%%%%%%%%%%%
\bea
&&\frac{\Gamma (a \rightarrow \bar{\chi} \chi)(T=m_a)}{H(T=m_a)} \geq 1, \nonumber\\
&& \rightarrow \,\,\,\, m_\chi \gtrsim 1.3\,\mbox{keV} \sqrt{1+32 C_{GG}^2} 
\sqrt{\frac{0.02}{\Gamma_s/m_s}} {\left(\frac{m_a}{1\,\mbox{GeV}}\right)}^{1/2} 
\left(\frac{m_s}{750\,\mbox{GeV}}\right).
\eea
%%%%%%%%%%%%%%%%%%%%%%%%%%%%%%%%%%%%%%%%%%%%%%%%%%%%%%%%%%%%%

We thus notice that the $a \leftrightarrow \bar{\chi} \chi$ process is rather efficient and 
keeps the dark matter in the equilibrium unless $m_\chi \sim\mathcal{O}(1 ~{\rm keV})$ (or even below) 
for $\Gamma_s/m_s >2 \%$. Remembering that $m_\chi$  below a keV is excluded by structure formation, we conclude 
that the light dark matter is coupled to the primordial thermal bath until the temperature
$T \sim m_a$. After this temperature relativistic decoupling happens with a relic density given by:
%%%%%%%%%%%%%%%%%%%%%%%%%%%%%%%%%%%%%%%%%%%%%%%%%%%%%%%%%%%%%
\beq
\Omega_{\chi,\rm rel}h^2 \approx 9.6\times 10^{-2} \frac{g_{\rm eff}}{g_{*,S} (T_d)}
\left(\frac{m_\chi}{1 \,\mbox{eV}}\right),
\eeq
%%%%%%%%%%%%%%%%%%%%%%%%%%%%%%%%%%%%%%%%%%%%%%%%%%%%%%%%%%%%%

where $g_{*,S}(T_d)$ represents the number of relativistic degrees of freedom at the temperature 
$T_d$ of decoupling of the dark matter while $g_{\rm eff}$ 
is the number of internal degrees of freedom of the dark matter itself.
The correct relic density can be achieved for $m_\chi \sim \,100~ \mbox{eV}$, 
inconsistently with the conditions stated above. 
For viable values of the mass, i.e., $\gtrsim \mathcal{O}(\mbox{keV})$, the dark matter is overabundant.
For $\Gamma_s/m_s < 2 \%$, on the contrary, there exists a small window of masses, 
$m_\chi \sim 1-10\,\mbox{keV}$, for which the dark matter cannot get into the thermal equilibrium 
in the Early Universe. In this case the dark matter is produced through freeze-in 
\cite{Chu:2013jja,Chu:2011be,Hall:2009bx}
by the decay of the pseudoscalar. The corresponding relic density is given by:
%%%%%%%%%%%%%%%%%%%%%%%%%%%%%%%%%%%%%%%%%%%%%%%%%%%%%%%%%%%%% 
\bea
\label{eq:freezein}
\Omega_{FI}h^2 &&= \frac{1.09 \times 10^{27} g_a}{g_{*,S}(T=m_a)^{3/2}}\frac{m_\chi 
\Gamma(a \rightarrow \bar{\chi} \chi)}{m_a^2}\nonumber\\
&&\approx 0.3 \frac{1}{(1+32 C_{GG}^2)} \left(\frac{\Gamma_s/m_s}{0.01}\right) {\left(\frac{m_\chi}{1\,
\mbox{keV}}\right)}^3 {\left(\frac{750\,\mbox{GeV}}{m_s}\right)}^2 \left(\frac{1\,\mbox{GeV}}{m_a}\right).   
\eea 
%%%%%%%%%%%%%%%%%%%%%%%%%%%%%%%%%%%%%%%%%%%%%%%%%%%%%%%%%%%%%

If the dark matter is not in the thermal equilibrium, 
the $s \rightarrow \bar{\chi} \chi$ decay process can also give rise 
to freeze-in production. However, its contribution is suppressed 
with respect to eq.~(\ref{eq:freezein}),
achieved with the replacement $m_a \rightarrow m_s$.
As can be seen, the correct dark matter relic density requires a $\Gamma_s$
on the smaller side, i.e., $\lesssim 7.5$ GeV giving $\Gamma_s/m_s \lesssim 1\%$
or $m_a \sim 2$ GeV.

The analytical expressions reported above have been validated by solving a system of 
coupled Boltzmann equations tracking the abundance of the dark matter itself as well as the ones of the 
scalar field $s$ and of the pseudoscalar field $a$. 
%%%%%%%%%%%%%%%%%%%%%%%%%%%%%%%%%%%%%%%%%%%%%%%%%%%%%%%%%%%%%
\begin{figure}[tbp]
\centering
\includegraphics[width=7.25cm]{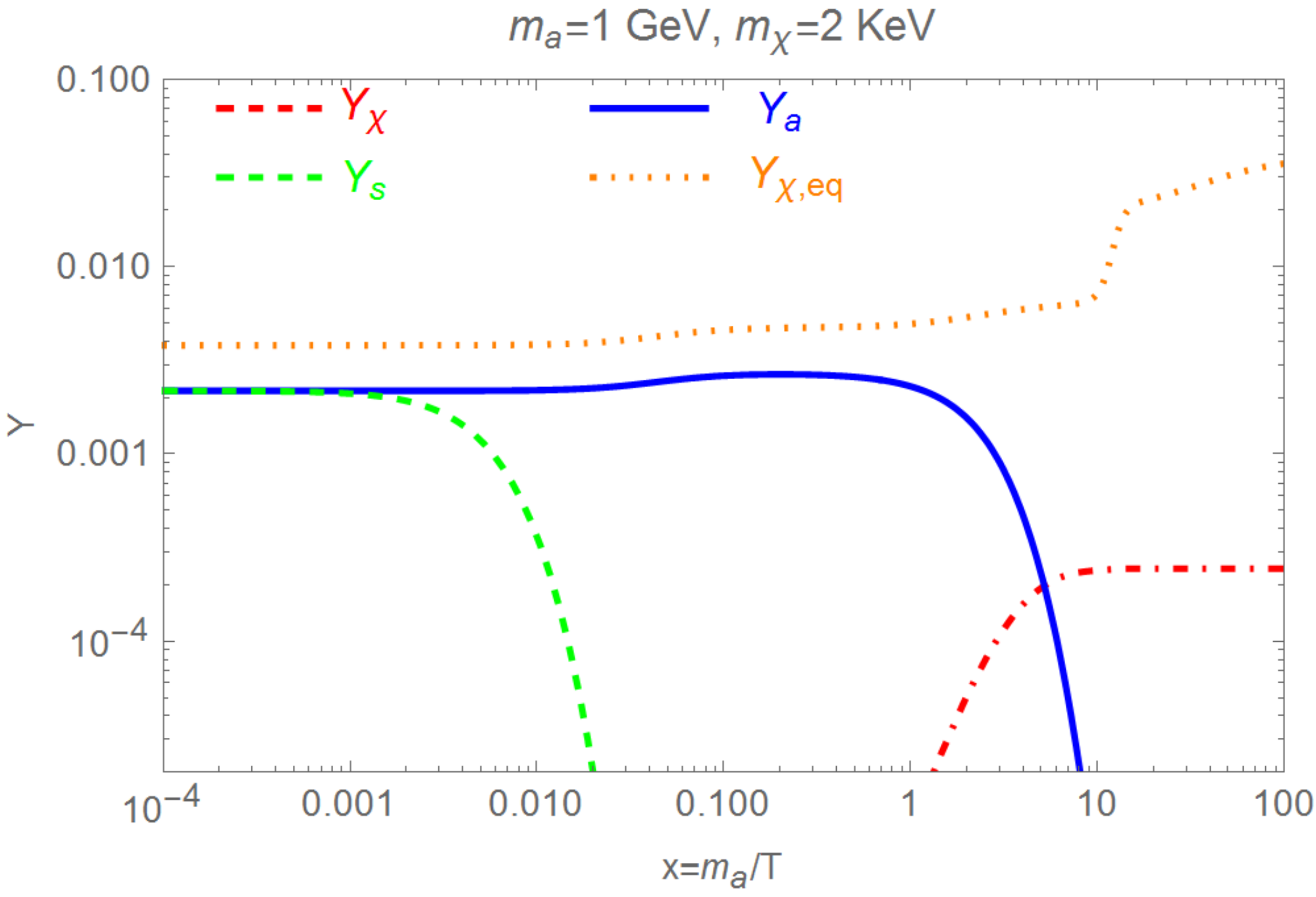}
\includegraphics[width=7.25cm]{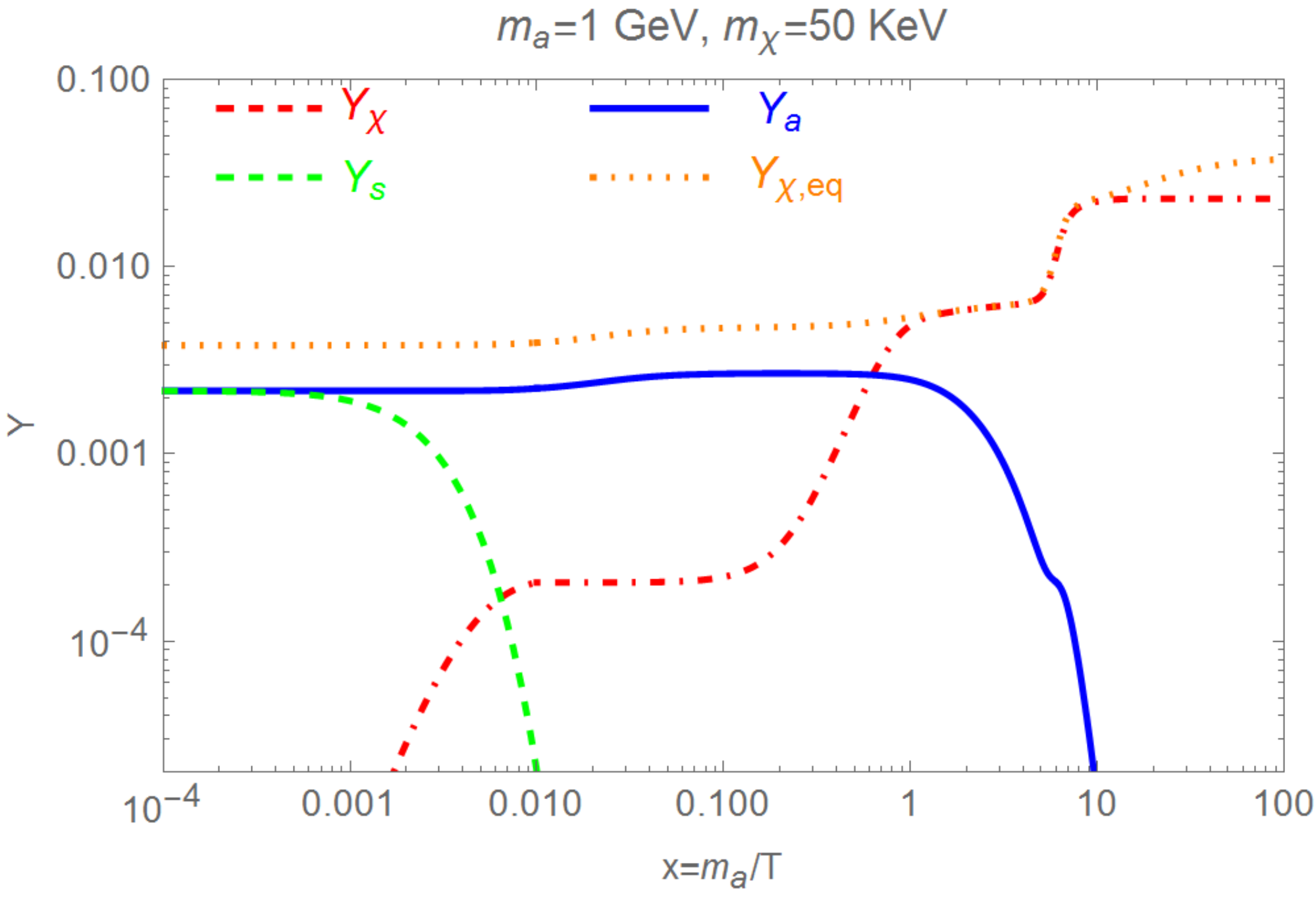}
\caption{Evolution of the abundance of the $s$, $a$ and $\chi$ fields, 
as determined by the numerical solution of a system of coupled Boltzmann equations 
in the very light dark matter regime. The two panels refer to two values of 
$m_\chi$, namely, $2$ keV (left) and $50$ keV (right).
As evident from the comparison with the dark matter equilibrium distribution 
(dotted orange coloured line), that for the lighter mass, the dark matter is not capable 
of getting into the thermal equilibrium and it is produced through freeze-in. 
In the case of $m_\chi=50$ keV, the dark matter can however go into the thermal 
equilibrium and get decoupled while still being relativistic.}
\label{fig:lightDM}
\end{figure}
%%%%%%%%%%%%%%%%%%%%%%%%%%%%%%%%%%%%%%%%%%%%%%%%%%%%%%%%%%%%%

Two examples of the numerical solution of this system are shown in figure~\ref{fig:lightDM}. 
Here two values of the dark matter masses have been considered, namely, $2$ keV (left) and
$50$ keV (right). In the lighter dark matter case, we notice that the dark matter abundance is always 
well below the equilibrium one, represented by the dotted orange coloured\footnote{The 
raise of the curve at high values of $x$ is due to the change of $g_{*,S}$,
coming from the QCD phase transition occurring at $T\sim\mathcal{O}(100~{\rm MeV})$.
The latter corresponds to $x\sim \mathcal{O}(10)$.} curve. It is produced 
at $T\sim\mathcal{O}(m_a)$, as expected for the freeze-in mechanism, and the 
numerical value of its abundance, as determined by the solution of the system of equations, 
is in very good agreement with the theoretical prediction stated in eq.~(\ref{eq:freezein}). 
We also notice that, for the chosen set of parameters, the dark matter abundance matches 
the experimental expectation. On the contrary, for the heavier value of the mass, i.e., $50$ keV, 
the dark matter abundance perfectly tracks the thermal equilibrium function until, again, 
$T\sim\mathcal{O}(m_a)$ and then remains constant confirming the prediction of a relativistic decoupling.
Interesting signatures of such a scenario would be the interpretation of the $3.5$ keV monochromatic signal 
as recently observed in different clusters of galaxies \cite{Dudas:2014ixa}.

%%%%%%%%%%%%%%%%%%%%%%%%%%%%%%%%%%%%%%%%%%%%%%%%%%%%%%%%%%%%%
\subsection{Dark Matter Detection}
\label{Sec:detection}
%%%%%%%%%%%%%%%%%%%%%%%%%%%%%%%%%%%%%%%%%%%%%%%%%%%%%%%%%%%%%

One of the main characteristics of the proposed scenario is the fact that here 
the observed di-photon production cross-section is obtained for relatively 
lower values of the $C_{GG}$ and $C_{BB}$ couplings, compared to the direct production 
from a $750$ GeV resonance. As already mentioned in the introduction, 
such values can potentially evade the constraints from the dark matter searches. 
Regarding direct detection of the dark matter the 
most relevant interactions are the Spin Independent (SI) ones,
produced by the exchange of the scalar field $s$,
as noted also in refs.~\cite{Mambrini:2015wyu,D'Eramo:2016mgv,Backovic:2015fnp,Barducci:2015gtd}.
The SI cross-section, in the studied framework, can be written as:
%%%%%%%%%%%%%%%%%%%%%%%%%%%%%%%%%%%%%%%%%%%%%%%%%%%%%%%%%%%%%
\beq
\sigma_{\chi p}^{\rm SI}\approx 6.7\times 10^{-48}\,{\mbox{cm}}^2 
\frac{1}{{\left(1+32 C_{GG}^2\right)}^2} {\left(\frac{\Gamma_s/m_s}{0.05}\right)}^2 
{\left(\frac{750\,\mbox{GeV}}{m_s}\right)}^8{\left(\frac{m_\chi}{100\,\mbox{GeV}}
\right)}^2 {\left(\frac{C_{GG}}{0.001}\right)}^2,
\eeq 
%%%%%%%%%%%%%%%%%%%%%%%%%%%%%%%%%%%%%%%%%%%%%%%%%%%%%%%%%%%%%

which remains well below the current limit from LUX~\cite{Akerib:2015rjg}, 
given the very small value of the $C_{GG}$ coupling. 
The latter also implies an very suppressed monojet production cross-section and 
thereby, easily evades the corresponding experimental constraint.

On the contrary, a potential impact is still retained by the indirect dark matter searches. At this moment, the 
most effective constraint is the one coming from the FERMI and HESS searches of the gamma-ray lines.
These  searches can trace $C_{BB}$ values $\sim\mathcal{O}(0.001)$ or even smaller. 
The latter, especially in the low $m_a$ regime, may appear incompatible with the 
requirement of a pseudoscalar decay length $\lesssim 1$ meter, as depicted
in figure~\ref{Fig:alptraum}. 
%%%%%%%%%%%%%%%%%%%%%%%%%%%%%%%%%%%%%%%%%%%%%%%%%%%%%%%%%%%%%
\begin{figure}[tbp]
\centering
\includegraphics[width=8cm]{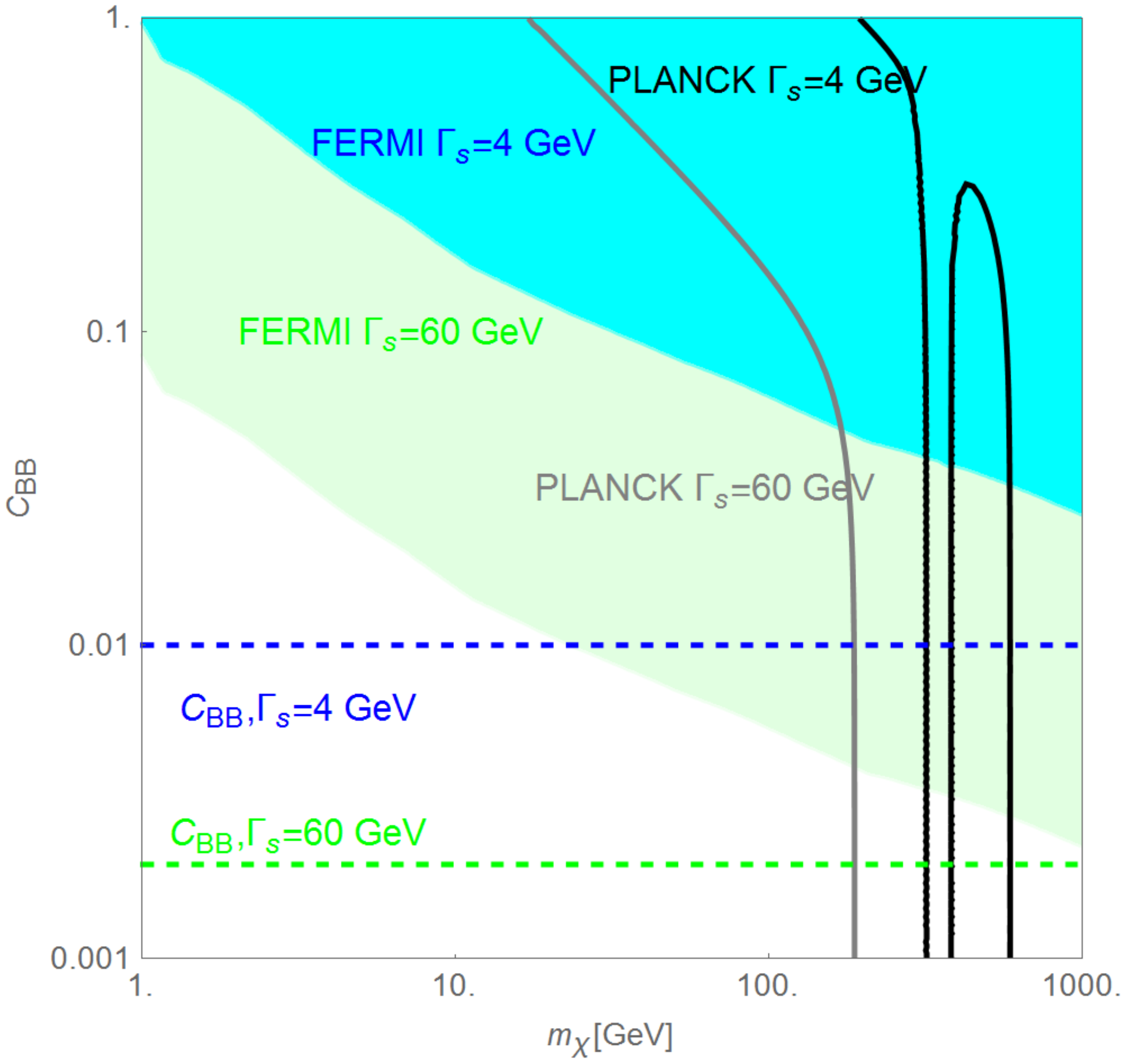}
\caption{Allowed region in the $(m_\chi,C_{BB})$ plane using the 
constraints from the FERMI/HESS searches of the gamma-ray lines. The 
cyan and light green coloured regimes, excluded from the FERMI/HESS searches,
correspond to $\Gamma_s=4$ and $60$ GeV, respectively.  
The black and gray coloured contours represent the correct relic density for the 
same two values of $\Gamma_s$. The blue and green coloured horizontal dashed 
lines represent two reference $C_{BB}$ values for the same 
two $\Gamma_s$ values (see figure~\ref{Fig:ma}), which assure a pseudoscalar 
decay length $\lesssim 1$ meter for $m_a\lesssim 500$ MeV.}
\label{fig:pFERMI} 
\end{figure}
%%%%%%%%%%%%%%%%%%%%%%%%%%%%%%%%%%%%%%%%%%%%%%%%%%%%%%%%%%%%%

The impact of the indirect detection on the relevant model parameters is represented in figure~\ref{fig:pFERMI}. 
Here we have shown the excluded regions in the $(m_\chi,C_{BB})$ plane from the  FERMI/HESS searches
for the minimum and maximum $\Gamma_s$ values considered in this work, i.e., $4$ GeV and $60$ GeV, respectively.
The regions in the $(m_\chi,C_{BB})$ plane excluded by the FERMI searches, corresponding to $\Gamma_s=4$ GeV 
and $60$ GeV, are represented with the cyan and light green colour, respectively.
The black and the gray coloured contours correspond to
the correct relic density, as suggested by the PLANCK. 
It is apparent from figure~\ref{fig:pFERMI} that one can extract a {\it{decreasing}} minimum
value for $C_{BB}$ depending on the {\it{increasing}} $m_\chi$ value. As a reference,
we also compare these limits with the ones obtained from figure~\ref{Fig:ma}. 
The latter are represented by the dashed blue coloured line (for $\Gamma_s=4$ GeV) and 
by dashed green coloured line (for $\Gamma_s=60$ GeV), respectively.
One should note that the $C_{BB}$ values considered for 
figure~\ref{fig:pFERMI} always give a pseudoscalar decay length $\lesssim 1$ meter
as observed in figure~\ref{Fig:ma}.
%%%%%%%%%%%%%%%%%%%%%%%%%%%%%%%%%%%%%%%%%%%%%%%%%%%%%%%%%%%%%
\begin{figure}[tbp]
\centering
\includegraphics[width=8cm]{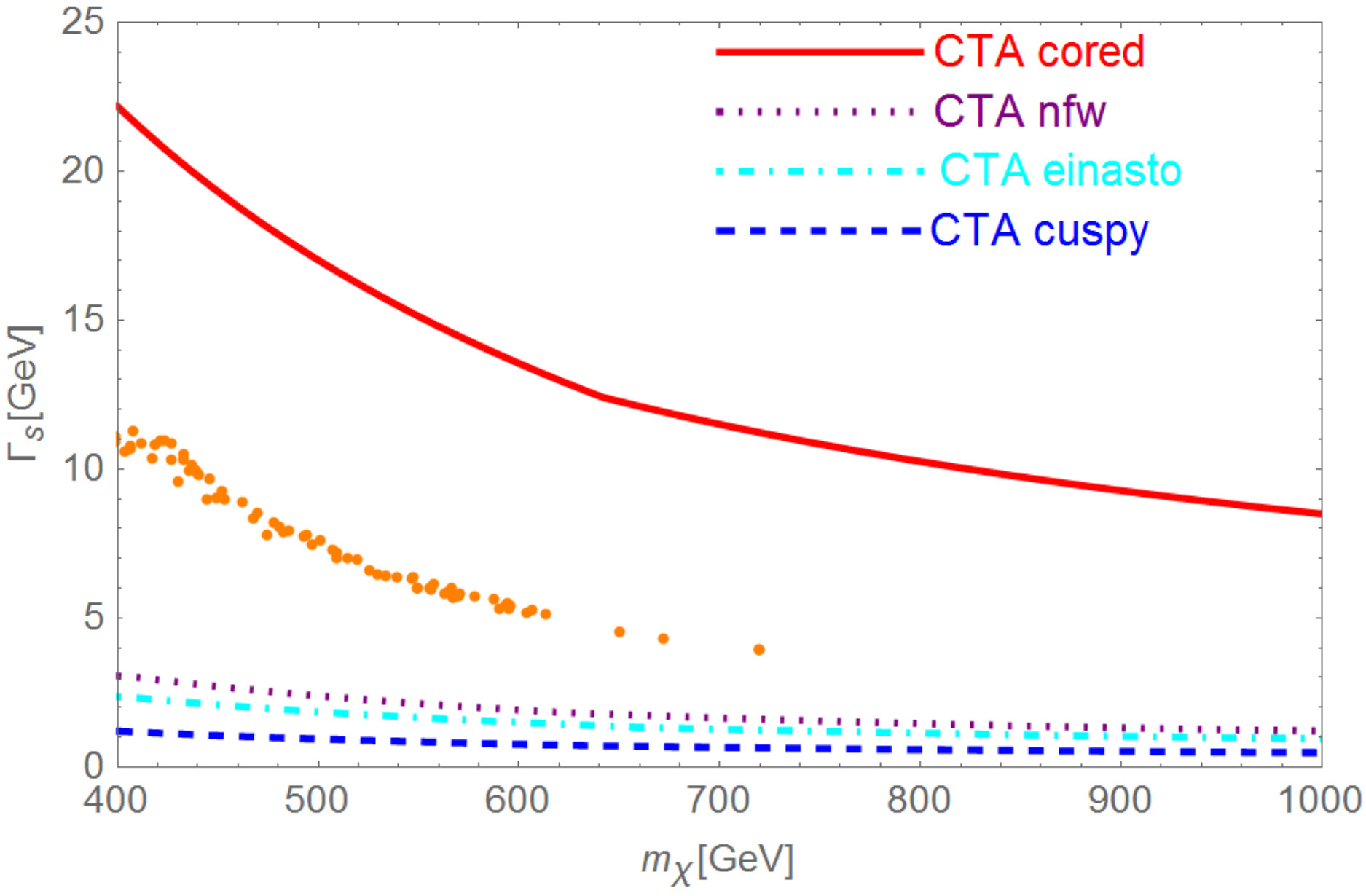}
\caption{Plot showing the CTA detection prospects in the 
$(m_\chi,\Gamma_s)$ plane for the gamma-ray boxes originating from 
the annihilation process $\bar{\chi} \chi \rightarrow sa$.
This plot is prepared for the four possible choices of the dark matter profile, as 
mentioned in the plot. The orange coloured points represent the 
set of viable model points which emerged from the parameter 
scan described in the sub-section~\ref{Sec:detection}.}
\label{fig:pCTA}
\end{figure}
%%%%%%%%%%%%%%%%%%%%%%%%%%%%%%%%%%%%%%%%%%%%%%%%%%%%%%%%%%%%%

Concerning detection of the dark matter, an interesting 
scenario appears for $m_\chi > m_s/2$. In this case the s-wave annihilation process $\bar{\chi} \chi 
\rightarrow sa$ gives a characteristic signal represented by a wide gamma-ray 
box~\cite{Ibarra:2013eda}\footnote{A wide gamma-ray box is actually achieved only 
when $m_\chi > 400$ GeV. The gamma-ray box is narrow when close to the kinematical threshold 
and becomes sensitive to the gamma-ray line searches. 
We have included the corresponding constraint in the parameter scan described in the 
sub-section~\ref{Sec:detection}.} which can be probed by the CTA~\cite{Ibarra:2015tya} in the near future. 
The detection prospects are shown in figure~\ref{fig:pCTA}. 
It is evident from figure~\ref{fig:pCTA} that the CTA can efficiently probe the studied
corner of the parameter space and thus, in the absence of any signal,
can also exclude the entire viable parameter space.

%%%%%%%%%%%%%%%%%%%%%%%%%%%%%%%%%%%%%%%%%%%%%%%%%%%%%%%%%%%%%
\subsection{Summary}
\label{Sec:summary}
%%%%%%%%%%%%%%%%%%%%%%%%%%%%%%%%%%%%%%%%%%%%%%%%%%%%%%%%%%%%%

\begin{figure}[tbp]
\centering
\includegraphics[width=8cm]{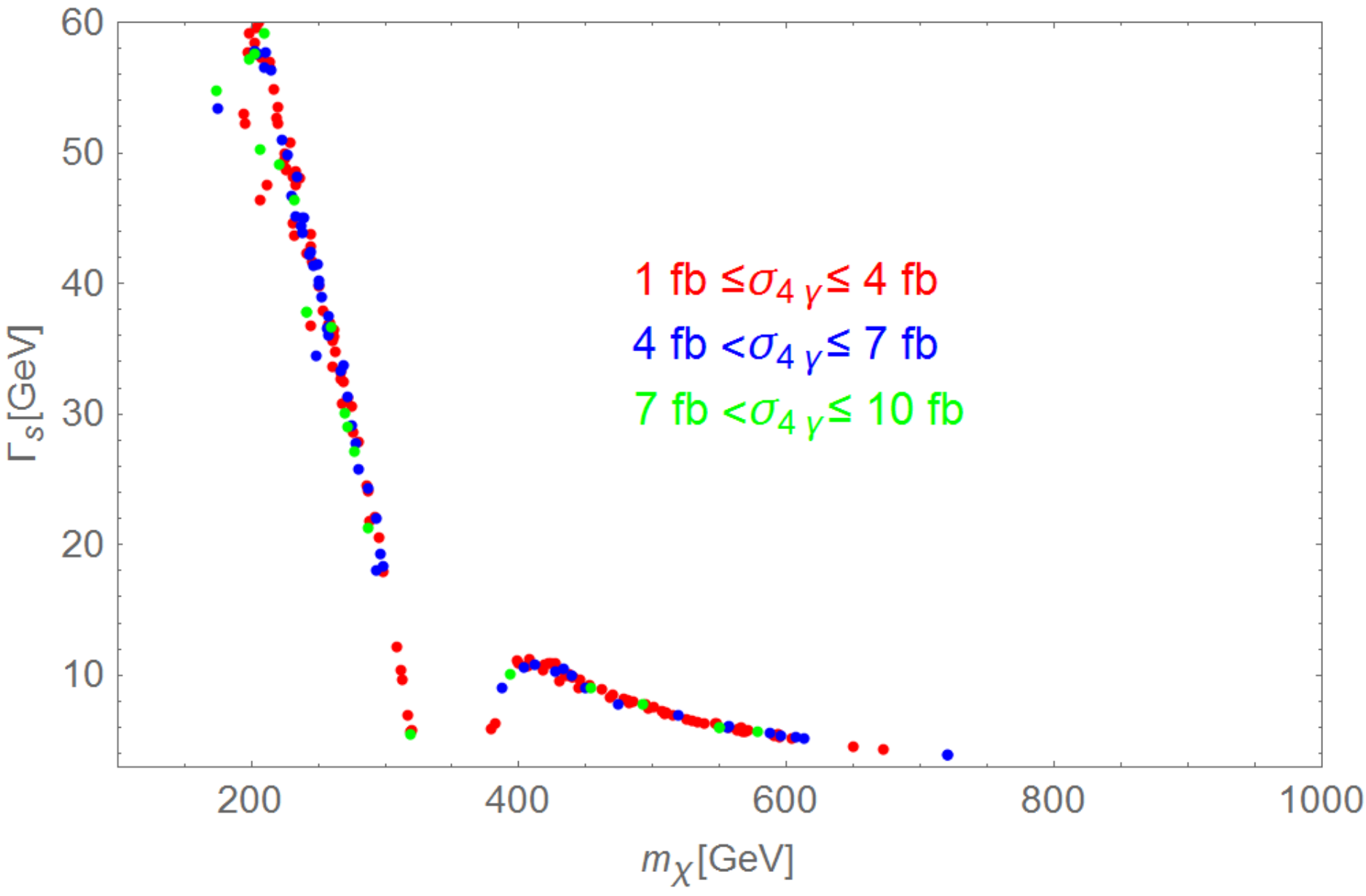}
\caption{Summary plot representing points in the ($m_\chi$, $\Gamma_s$) plane 
that simultaneously respect the LHC and cosmological constraints and correspond to
different di-photon production cross-sections, as depicted with the three
different colours.}
\label{Fig:scatter}
\end{figure}

We summarized our results in figure~\ref{Fig:scatter}. This plot
accommodates a set of points in the $(m_\chi,\,\Gamma_s)$ plane
possessing the correct relic density and corresponds to different
values of the di-photon production cross-sections, as represented
with the three different colours.
This result is obtained after a dedicated numerical scan on the set of parameters
$(C_{GG},C_{BB}, \lambda_\Phi,m_\chi, m_a)$ over the following ranges, 
keeping $m_s$ fixed at $750$ GeV:
%%%%%%%%%%%%%%%%%%%%%%%%%%%%%%%%%%%%%%%%%%%%%%%%%%%%%%%%%%%%%
\bea
&& C_{GG} \in \left[5\times 10^{-4},0.1\right], \, C_{BB} \in \left[5\times 10^{-4},0.1\right], \nonumber\\
&& m_\chi \in \left[10,1000\right]\,\mbox{GeV}, \, m_a \in \left[0.2,2\right]\,\mbox{GeV},\nonumber\\
&& \lambda_\Phi \in \left[0.25,4\right].
\eea
%%%%%%%%%%%%%%%%%%%%%%%%%%%%%%%%%%%%%%%%%%%%%%%%%%%%%%%%%%%%%

Finally, we kept only the points respecting
%%%%%%%%%%%%%%%%%%%%%%%%%%%%%%%%%%%%%%%%%%%%%%%%%%%%%%%%%%%%%
\beq\label{Eq:final}
1\,\mbox{fb} \lsim \sigma_{4\gamma}\lsim 10\,\mbox{fb},\,\,
4\,\mbox{GeV} \lsim \Gamma_s\lsim 60\,\mbox{GeV},\,\,\Omega
h^2 \approx 0.12,
\eeq
%%%%%%%%%%%%%%%%%%%%%%%%%%%%%%%%%%%%%%%%%%%%%%%%%%%%%%%%%%%%%

and remain compatible with the relevant accelerator constraints ($8$ TeV
searches of the di-jets, $Z\gamma$, $ZZ$, decay length of the pseudoscalar
$\lesssim 1$ meter etc.) as well as with the
dark matter detection limits discussed in the previous sub-section. 
We can clearly distinguish three regions
in figure~\ref{Fig:scatter}: 
(i) The low mass region
($m_\chi \lesssim 300$ GeV) in which the relic abundance is dominated by the
p-wave (velocity suppressed) annihilation
channel  $\bar{\chi} \chi \rightarrow s \rightarrow aa$
(see figure~\ref{Fig:feynman0}). In this case a large width ($\Gamma_s\gtrsim 20$ GeV)
is necessary to compensate the velocity suppression effect and to avoid
{\it{over}}abundance of the dark matter in the Universe. 
(ii) In the second region,
$300~{\rm GeV} \lesssim m_\chi \lesssim 400$ GeV, the correct abundance is obtained
through the pole region of the same diagram (figure~\ref{Fig:feynman0}).
In this case, a narrow width of $s$ ($\lsim 12$ GeV) is necessary to avoid
{\it{under}}abundance of the dark matter.
(iii) Finally, in the last region ($m_\chi \gtrsim$ 400 GeV), the t-channel
(s-wave) annihilation depicted in figure~\ref{Fig:feynman2}
determines the relic abundance of $\chi$. As the final state $sa$ has an
odd parity, there is no velocity suppression and a moderate width
($\sim\mathcal{O}(10~{\rm GeV})$) is needed to respect the WMAP/PLANCK
constraint.

It is also interesting to notice from figure~\ref{Fig:scatter} that if the
future ATLAS/CMS analysis determines the value of the width 
$\Gamma_s$ more precisely, it will become possible to estimate the
dark matter mass, consistent with the
cosmological constraint. If the large width scenario is confirmed, an
electroweak scale mass will be favoured.

%%%%%%%%%%%%%%%%%%%%%%%%%%%%%%%%%%%%%%%%%%%%%%%%%%%%%%%%%%%%%
%%%%%%%%%%%%%%%%%%%%%%%%%%%%%%%%%%%%%%%%%%%%%%%%%%%%%%%%%%%%%

Before moving to the conclusion we just briefly comment on some
theoretical aspects relative to our scenario. Our work is motivated by pure
phenomenological purposes and thus, we have 
not referred to any particular setup for the origin of the couplings
$C_{GG}$ and $C_{BB}$. As pointed out, for example, in refs.~\cite{Aparicio:2016iwr,
Bae:2016xni,Bertuzzo:2016fmv,Salvio:2016hnf,Son:2015vfl},
the extra matter needed at the loop level to generate 
the $C_{GG},C_{BB}$ couplings affects the running of
the gauge couplings, driving them towards the non-perturbative regime
already at energy scales of a few TeV. Thus, too much high values of these 
couplings remain disfavoured. Similar to
ref.~\cite{Aparicio:2016iwr}, we then remark that the low values 
of the $C_{GG},C_{BB}$ couplings needed to fit the observed LHC excess, compared to
the case of a resonance decaying into two isolated photons, relax
also the theoretical constraints, besides the experimental ones. In scenarios
like the one considered in this work, a strong impact is still retained by the renormalization group equation
running of the quartic coupling $\lambda_\Phi$. In view of this, the higher
values of $\lambda_\Phi$ considered here, corresponding to $\Gamma_s \sim 40-60\,\mbox{GeV}$, might result in
tension with constraints from the theoretical consistency. On the other hand,
we remark that except for $m_\chi \lesssim 300 ~{\rm GeV}$,
the phenomenological constraints favour values of $\Gamma_s$ below $12$ GeV. The latter corresponds to 
$\lambda_\Phi \lesssim 0.8$ which allows to
maintain a safe theoretical framework at least up to an energy scale relevant
for the phenomenological analysis.

%%%%%%%%%%%%%%%%%%%%%%%%%%%%%%%%%%%%%%%%%%%%%%%%%%%%%%%%%%%%%
\section{Conclusion}
%%%%%%%%%%%%%%%%%%%%%%%%%%%%%%%%%%%%%%%%%%%%%%%%%%%%%%%%%%%%%

We have considered in this note the possibility that the di-photon signal is in reality produced through 
the decay of two light pseudoscalars into two pairs of highly collimated photons, indistinguishable from
the isolated photons. We have shown that, in a spontaneous symmetry breaking framework, 
once one couples the scalar
sector to the dark matter, it appears feasible to re-open a large range of 
the dark matter masses compatible with all the present 
experimental constraints.

%%%%%%%%%%%%%%%%%%%%%%%%%%%%%%%%%%%%%%%%%%%%%%%%%%%%%%%%%%%%%
\acknowledgments
%%%%%%%%%%%%%%%%%%%%%%%%%%%%%%%%%%%%%%%%%%%%%%%%%%%%%%%%%%%%%

Y.M. wants to thank especially J.B de Vivie whose help was fundamental throughout our work.
The authors thanks Emilian Dudas and Ulrich Ellwanger for fruitful discussions. This work is also supported by the
Spanish MICINN's Consolider-Ingenio 2010 Programme under grant Multi-Dark {\bf CSD2009-00064}, the contract 
{\bf FPA2010-17747}, the France-US PICS no. 06482 and the LIA-TCAP of CNRS. 
Y.~M. and G. A. acknowledges partial support from the European
Union FP7 ITN INVISIBLES (Marie Curie Actions, {\bf PITN-GA-2011- 289442})
and the ERC advanced grants Higgs@LHC and MassTeV. 
G. A. thanks the CERN theory division for the hospitality during
part of the completion of this project.
P.G. acknowledges the support from P2IO Excellence Laboratory (LABEX). 
This research was also supported in part by the Research
Executive Agency (REA) of the European Union under
the Grant Agreement {\bf PITN-GA2012-316704} (``HiggsTools'').

%%%%%%%%%%%%%%%%%%%%%%%%%%%%%%%%%%%%%%%%%%%%%%%%%%%%%%%%%%%%%
\bibliography{AGMPref2}

\providecommand{\href}[2]{#2}\begingroup\raggedright\begin{thebibliography}{100}

\bibitem{ATLASdiph}
{\scshape ATLAS} collaboration, \emph{{Search for resonances decaying to photon
  pairs in 3.2 fb$^{-1}$ of $pp$ collisions at $\sqrt{s}$ = 13 TeV with the
  ATLAS detector}}, {\emph{ATLAS-CONF-2015-081} (2015) }.

\bibitem{CMS:2015dxe}
{\scshape CMS} collaboration, \emph{{Search for new physics in high mass
  diphoton events in proton-proton collisions at 13TeV}},
  {\emph{CMS-PAS-EXO-15-004} (2015) }.

\bibitem{CMS:2016owr}
{\scshape CMS} collaboration, \emph{{Search for new physics in high mass
  diphoton events in $3.3~\mathrm{fb}^{-1}$ of proton-proton collisions at
  $\sqrt{s}=13~\mathrm{TeV}$ and combined interpretation of searches at
  $8~\mathrm{TeV}$ and $13~\mathrm{TeV}$}}, {\emph{CMS-PAS-EXO-16-018} (2016)
  }.

\bibitem{Moriondatlas}
{\scshape ATLAS} collaboration, \emph{{Search for resonances in diphoton events
  with the ATLAS detector at $\sqrt{s}$ = 13 TeV}}, {\emph{ATLAS-CONF-2016-018}
  (2016) }.

\bibitem{Huong:2016kpa}
D.~T. Huong and P.~V. Dong, \emph{{Left-right asymmetry and 750 GeV diphoton
  excess}}, \href{http://dx.doi.org/10.1103/PhysRevD.93.095019}{\emph{Phys.
  Rev.} {\bf D93} (2016) 095019}, [\href{http://arxiv.org/abs/1603.05146}{{\tt
  1603.05146}}].

\bibitem{DeRomeri:2016xpb}
V.~De~Romeri, J.~S. Kim, V.~Martín-Lozano, K.~Rolbiecki and R.~R. de~Austri,
  \emph{{Confronting dark matter with the diphoton excess from a parent
  resonance decay}},
  \href{http://dx.doi.org/10.1140/epjc/s10052-016-4116-7}{\emph{Eur. Phys. J.}
  {\bf C76} (2016) 262}, [\href{http://arxiv.org/abs/1603.04479}{{\tt
  1603.04479}}].

\bibitem{Ahriche:2016mcx}
A.~Ahriche, G.~Faisel, S.~Nasri and J.~Tandean, \emph{{Addressing the LHC 750
  GeV diphoton excess without new colored states}},
  \href{http://arxiv.org/abs/1603.01606}{{\tt 1603.01606}}.

\bibitem{Tsai:2016lfg}
Y.~Tsai, L.-T. Wang and Y.~Zhao, \emph{{Faking The Diphoton Excess by Displaced
  Dark Photon Decays}},  \href{http://arxiv.org/abs/1603.00024}{{\tt
  1603.00024}}.

\bibitem{Li:2016tqf}
T.~Li, J.~A. Maxin, V.~E. Mayes and D.~V. Nanopoulos, \emph{{The $750$ GeV
  Diphoton Excesses in a Realistic D-brane Model}},
  \href{http://arxiv.org/abs/1602.09099}{{\tt 1602.09099}}.

\bibitem{Lazarides:2016ofd}
G.~Lazarides and Q.~Shafi, \emph{{Diphoton resonances in a U(1)$_{B-L}$
  extension of the minimal supersymmetric standard model}},
  \href{http://dx.doi.org/10.1103/PhysRevD.93.111702}{\emph{Phys. Rev.} {\bf
  D93} (2016) 111702}, [\href{http://arxiv.org/abs/1602.07866}{{\tt
  1602.07866}}].

\bibitem{Ren:2016gyg}
J.~Ren and J.-H. Yu, \emph{{$SU(2) \times SU(2) \times U(1)$ Interpretation on
  the 750 GeV Diphoton Excess}},  \href{http://arxiv.org/abs/1602.07708}{{\tt
  1602.07708}}.

\bibitem{Molinaro:2016oix}
E.~Molinaro, F.~Sannino and N.~Vignaroli, \emph{{Collider Tests of (Composite)
  Diphoton Resonances}},  \href{http://arxiv.org/abs/1602.07574}{{\tt
  1602.07574}}.

\bibitem{Redi:2016kip}
M.~Redi, A.~Strumia, A.~Tesi and E.~Vigiani, \emph{{Di-photon resonance and
  Dark Matter as heavy pions}},
  \href{http://dx.doi.org/10.1007/JHEP05(2016)078}{\emph{JHEP} {\bf 05} (2016)
  078}, [\href{http://arxiv.org/abs/1602.07297}{{\tt 1602.07297}}].

\bibitem{Ko:2016sxg}
P.~Ko, T.~Nomura, H.~Okada and Y.~Orikasa, \emph{{Confronting a New Three-loop
  Seesaw Model with the 750 GeV Diphoton Excess}},
  \href{http://arxiv.org/abs/1602.07214}{{\tt 1602.07214}}.

\bibitem{Baek:2016uqf}
S.~Baek and J.-h. Park, \emph{{LHC 750 GeV diphoton excess and muon $(g-2)$}},
  \href{http://dx.doi.org/10.1016/j.physletb.2016.05.040}{\emph{Phys. Lett.}
  {\bf B758} (2016) 416--422}, [\href{http://arxiv.org/abs/1602.05588}{{\tt
  1602.05588}}].

\bibitem{Staub:2016dxq}
F.~Staub et~al., \emph{{Precision tools and models to narrow in on the 750 GeV
  diphoton resonance}},  \href{http://arxiv.org/abs/1602.05581}{{\tt
  1602.05581}}.

\bibitem{Mantilla:2016sew}
S.~F. Mantilla, R.~Martinez, F.~Ochoa and C.~F. Sierra, \emph{{Diphoton decay
  for a 750 GeV scalar boson in a $SU(6)\otimes U(1)_{X}$ model}},
  \href{http://arxiv.org/abs/1602.05216}{{\tt 1602.05216}}.

\bibitem{Hamada:2016vwk}
Y.~Hamada, H.~Kawai, K.~Kawana and K.~Tsumura, \emph{{Models of LHC Diphoton
  Excesses Valid up to the Planck scale}},
  \href{http://arxiv.org/abs/1602.04170}{{\tt 1602.04170}}.

\bibitem{Gross:2016ioi}
C.~Gross, O.~Lebedev and J.~M. No, \emph{{Drell-Yan Constraints on New
  Electroweak States and the Di-photon Anomaly}},
  \href{http://arxiv.org/abs/1602.03877}{{\tt 1602.03877}}.

\bibitem{Bae:2016xni}
K.~J. Bae, M.~Endo, K.~Hamaguchi and T.~Moroi, \emph{{Diphoton Excess and
  Running Couplings}},
  \href{http://dx.doi.org/10.1016/j.physletb.2016.04.031}{\emph{Phys. Lett.}
  {\bf B757} (2016) 493--500}, [\href{http://arxiv.org/abs/1602.03653}{{\tt
  1602.03653}}].

\bibitem{Ge:2016xcq}
S.-F. Ge, H.-J. He, J.~Ren and Z.-Z. Xianyu, \emph{{Realizing Dark Matter and
  Higgs Inflation in Light of LHC Diphoton Excess}},
  \href{http://dx.doi.org/10.1016/j.physletb.2016.04.008}{\emph{Phys. Lett.}
  {\bf B757} (2016) 480--492}, [\href{http://arxiv.org/abs/1602.01801}{{\tt
  1602.01801}}].

\bibitem{Li:2016xcj}
T.~Li, J.~A. Maxin, V.~E. Mayes and D.~V. Nanopoulos, \emph{{A Flippon Related
  Singlet at the LHC II}},  \href{http://arxiv.org/abs/1602.01377}{{\tt
  1602.01377}}.

\bibitem{Ding:2016udc}
R.~Ding, Y.~Fan, L.~Huang, C.~Li, T.~Li, S.~Raza et~al., \emph{{Systematic
  Study of Diphoton Resonance at 750 GeV from Sgoldstino}},
  \href{http://arxiv.org/abs/1602.00977}{{\tt 1602.00977}}.

\bibitem{Hektor:2016uth}
A.~Hektor and L.~Marzola, \emph{{Di-photon excess at LHC and the gamma ray
  excess at the Galactic Centre}},  \href{http://arxiv.org/abs/1602.00004}{{\tt
  1602.00004}}.

\bibitem{Bertuzzo:2016fmv}
E.~Bertuzzo, P.~A.~N. Machado and M.~Taoso, \emph{{Di-Photon excess in the
  2HDM: hasting towards the instability and the non-perturbative regime}},
  \href{http://arxiv.org/abs/1601.07508}{{\tt 1601.07508}}.

\bibitem{Kawamura:2016idj}
J.~Kawamura and Y.~Omura, \emph{{Diphoton excess at 750 GeV and LHC constraints
  in models with vectorlike particles}},
  \href{http://dx.doi.org/10.1103/PhysRevD.93.115011}{\emph{Phys. Rev.} {\bf
  D93} (2016) 115011}, [\href{http://arxiv.org/abs/1601.07396}{{\tt
  1601.07396}}].

\bibitem{Geng:2016xin}
C.-Q. Geng and D.~Huang, \emph{{Note on Spin-2 Particle Interpretation of the
  750~GeV Diphoton Excess}},
  \href{http://dx.doi.org/10.1103/PhysRevD.93.115032}{\emph{Phys. Rev.} {\bf
  D93} (2016) 115032}, [\href{http://arxiv.org/abs/1601.07385}{{\tt
  1601.07385}}].

\bibitem{Nomura:2016rjf}
T.~Nomura and H.~Okada, \emph{{Generalized Zee-Babu model with $750$ GeV
  diphoton resonance}},
  \href{http://dx.doi.org/10.1016/j.physletb.2016.03.034}{\emph{Phys. Lett.}
  {\bf B756} (2016) 295--302}, [\href{http://arxiv.org/abs/1601.07339}{{\tt
  1601.07339}}].

\bibitem{King:2016wep}
S.~F. King and R.~Nevzorov, \emph{{750 GeV Diphoton Resonance from Singlets in
  an Exceptional Supersymmetric Standard Model}},
  \href{http://dx.doi.org/10.1007/JHEP03(2016)139}{\emph{JHEP} {\bf 03} (2016)
  139}, [\href{http://arxiv.org/abs/1601.07242}{{\tt 1601.07242}}].

\bibitem{Abel:2016pyc}
S.~Abel and V.~V. Khoze, \emph{{Photo-production of a 750 GeV di-photon
  resonance mediated by Kaluza-Klein leptons in the loop}},
  \href{http://dx.doi.org/10.1007/JHEP05(2016)063}{\emph{JHEP} {\bf 05} (2016)
  063}, [\href{http://arxiv.org/abs/1601.07167}{{\tt 1601.07167}}].

\bibitem{Aydemir:2016qqj}
U.~Aydemir and T.~Mandal, \emph{{Interpretation of the 750 GeV diphoton excess
  with colored scalars in $\mathbf{SO(10)}$ grand unification}},
  \href{http://arxiv.org/abs/1601.06761}{{\tt 1601.06761}}.

\bibitem{Chiang:2016ydx}
C.-W. Chiang and A.-L. Kuo, \emph{{Can the 750-GeV diphoton resonance be the
  singlet Higgs boson of custodial Higgs triplet model?}},
  \href{http://arxiv.org/abs/1601.06394}{{\tt 1601.06394}}.

\bibitem{Cao:2016cok}
Q.-H. Cao, Y.-Q. Gong, X.~Wang, B.~Yan and L.~L. Yang, \emph{{One bump or two
  peaks: The 750 GeV diphoton excess and dark matter with a complex mediator}},
  \href{http://dx.doi.org/10.1103/PhysRevD.93.075034}{\emph{Phys. Rev.} {\bf
  D93} (2016) 075034}, [\href{http://arxiv.org/abs/1601.06374}{{\tt
  1601.06374}}].

\bibitem{Han:2016bvl}
X.-F. Han, L.~Wang and J.~M. Yang, \emph{{An extension of two-Higgs-doublet
  model and the excesses of 750 GeV diphoton, muon g-2 and $h\to\mu\tau$}},
  \href{http://dx.doi.org/10.1016/j.physletb.2016.04.036}{\emph{Phys. Lett.}
  {\bf B757} (2016) 537--547}, [\href{http://arxiv.org/abs/1601.04954}{{\tt
  1601.04954}}].

\bibitem{Chao:2016aer}
W.~Chao, \emph{{The Diphoton Excess Inspired Electroweak Baryogenesis}},
  \href{http://arxiv.org/abs/1601.04678}{{\tt 1601.04678}}.

\bibitem{Nomura:2016seu}
T.~Nomura and H.~Okada, \emph{{Four-loop Radiative Seesaw Model with 750 GeV
  Diphoton Resonance}},  \href{http://arxiv.org/abs/1601.04516}{{\tt
  1601.04516}}.

\bibitem{Davis:2016hlw}
J.~H. Davis, M.~Fairbairn, J.~Heal and P.~Tunney, \emph{{The Significance of
  the 750 GeV Fluctuation in the ATLAS Run 2 Diphoton Data}},
  \href{http://arxiv.org/abs/1601.03153}{{\tt 1601.03153}}.

\bibitem{Ding:2016ldt}
R.~Ding, Z.-L. Han, Y.~Liao and X.-D. Ma, \emph{{Interpretation of 750 GeV
  Diphoton Excess at LHC in Singlet Extension of Color-octet Neutrino Mass
  Model}}, \href{http://dx.doi.org/10.1140/epjc/s10052-016-4052-6}{\emph{Eur.
  Phys. J.} {\bf C76} (2016) 204}, [\href{http://arxiv.org/abs/1601.02714}{{\tt
  1601.02714}}].

\bibitem{Yu:2016lof}
J.-H. Yu, \emph{{Hidden Gauged U(1) Model: Unifying Scotogenic Neutrino and
  Flavor Dark Matter}},
  \href{http://dx.doi.org/10.1103/PhysRevD.93.113007}{\emph{Phys. Rev.} {\bf
  D93} (2016) 113007}, [\href{http://arxiv.org/abs/1601.02609}{{\tt
  1601.02609}}].

\bibitem{Cao:2016udb}
J.~Cao, L.~Shang, W.~Su, Y.~Zhang and J.~Zhu, \emph{{Interpreting the 750 GeV
  diphoton excess in the Minimal Dilaton Model}},
  \href{http://dx.doi.org/10.1140/epjc/s10052-016-4098-5}{\emph{Eur. Phys. J.}
  {\bf C76} (2016) 239}, [\href{http://arxiv.org/abs/1601.02570}{{\tt
  1601.02570}}].

\bibitem{Ko:2016wce}
P.~Ko and T.~Nomura, \emph{{Dark sector shining through 750 GeV dark Higgs
  boson at the LHC}},
  \href{http://dx.doi.org/10.1016/j.physletb.2016.05.014}{\emph{Phys. Lett.}
  {\bf B758} (2016) 205--211}, [\href{http://arxiv.org/abs/1601.02490}{{\tt
  1601.02490}}].

\bibitem{Hati:2016thk}
C.~Hati, \emph{{Explaining the diphoton excess in Alternative Left-Right
  Symmetric Model}},
  \href{http://dx.doi.org/10.1103/PhysRevD.93.075002}{\emph{Phys. Rev.} {\bf
  D93} (2016) 075002}, [\href{http://arxiv.org/abs/1601.02457}{{\tt
  1601.02457}}].

\bibitem{Stolarski:2016dpa}
D.~Stolarski and R.~Vega-Morales, \emph{{Probing a Virtual Diphoton Excess}},
  \href{http://dx.doi.org/10.1103/PhysRevD.93.055008}{\emph{Phys. Rev.} {\bf
  D93} (2016) 055008}, [\href{http://arxiv.org/abs/1601.02004}{{\tt
  1601.02004}}].

\bibitem{Borah:2016uoi}
D.~Borah, S.~Patra and S.~Sahoo, \emph{{Subdominant Left-Right Scalar Dark
  Matter as Origin of the 750 GeV Di-photon Excess at LHC}},
  \href{http://arxiv.org/abs/1601.01828}{{\tt 1601.01828}}.

\bibitem{Fichet:2016pvq}
S.~Fichet, G.~von Gersdorff and C.~Royon, \emph{{Measuring the Diphoton
  Coupling of a 750 GeV Resonance}},
  \href{http://dx.doi.org/10.1103/PhysRevLett.116.231801}{\emph{Phys. Rev.
  Lett.} {\bf 116} (2016) 231801}, [\href{http://arxiv.org/abs/1601.01712}{{\tt
  1601.01712}}].

\bibitem{Sahin:2016lda}
I.~Sahin, \emph{{Semi-elastic cross section for a scalar resonance of mass 750
  GeV}},  \href{http://arxiv.org/abs/1601.01676}{{\tt 1601.01676}}.

\bibitem{D'Eramo:2016mgv}
F.~D'Eramo, J.~de~Vries and P.~Panci, \emph{{A 750 GeV Portal: LHC
  Phenomenology and Dark Matter Candidates}},
  \href{http://dx.doi.org/10.1007/JHEP05(2016)089}{\emph{JHEP} {\bf 05} (2016)
  089}, [\href{http://arxiv.org/abs/1601.01571}{{\tt 1601.01571}}].

\bibitem{Bhattacharya:2016lyg}
S.~Bhattacharya, S.~Patra, N.~Sahoo and N.~Sahu, \emph{{750 GeV diphoton excess
  at CERN LHC from a dark sector assisted scalar decay}},
  \href{http://dx.doi.org/10.1088/1475-7516/2016/06/010}{\emph{JCAP} {\bf 1606}
  (2016) 010}, [\href{http://arxiv.org/abs/1601.01569}{{\tt 1601.01569}}].

\bibitem{Zhang:2016pip}
H.~Zhang, \emph{{The 750GeV Diphoton Excess: Who Introduces It?}},
  \href{http://arxiv.org/abs/1601.01355}{{\tt 1601.01355}}.

\bibitem{Ito:2016zkz}
H.~Ito, T.~Moroi and Y.~Takaesu, \emph{{Studying 750 GeV di-photon resonance at
  photon-photon collider}},
  \href{http://dx.doi.org/10.1016/j.physletb.2016.03.008}{\emph{Phys. Lett.}
  {\bf B756} (2016) 147--152}, [\href{http://arxiv.org/abs/1601.01144}{{\tt
  1601.01144}}].

\bibitem{Deppisch:2016scs}
F.~F. Deppisch, C.~Hati, S.~Patra, P.~Pritimita and U.~Sarkar,
  \emph{{Implications of the diphoton excess on left–right models and gauge
  unification}},
  \href{http://dx.doi.org/10.1016/j.physletb.2016.03.081}{\emph{Phys. Lett.}
  {\bf B757} (2016) 223--230}, [\href{http://arxiv.org/abs/1601.00952}{{\tt
  1601.00952}}].

\bibitem{Dutta:2016jqn}
B.~Dutta, Y.~Gao, T.~Ghosh, I.~Gogoladze, T.~Li, Q.~Shafi et~al.,
  \emph{{Diphoton Excess in Consistent Supersymmetric SU(5) Models with
  Vector-like Particles}},  \href{http://arxiv.org/abs/1601.00866}{{\tt
  1601.00866}}.

\bibitem{Hernandez:2016rbi}
A.~E.~C. Hern\'andez, I.~d.~M. Varzielas and E.~Schumacher, \emph{{The
  $750\,\text{GeV}$ diphoton resonance in the light of a 2HDM with $S_3$
  flavour symmetry}},  \href{http://arxiv.org/abs/1601.00661}{{\tt
  1601.00661}}.

\bibitem{Karozas:2016hcp}
A.~Karozas, S.~F. King, G.~K. Leontaris and A.~K. Meadowcroft, \emph{{750 GeV
  diphoton excess from $E_6$ in F-theory GUTs}},
  \href{http://dx.doi.org/10.1016/j.physletb.2016.03.054}{\emph{Phys. Lett.}
  {\bf B757} (2016) 73--78}, [\href{http://arxiv.org/abs/1601.00640}{{\tt
  1601.00640}}].

\bibitem{Chao:2016mtn}
W.~Chao, \emph{{The Diphoton Excess from an Exceptional Supersymmetric Standard
  Model}},  \href{http://arxiv.org/abs/1601.00633}{{\tt 1601.00633}}.

\bibitem{Danielsson:2016nyy}
U.~Danielsson, R.~Enberg, G.~Ingelman and T.~Mandal, \emph{{The force awakens -
  the 750 GeV diphoton excess at the LHC from a varying electromagnetic
  coupling}},  \href{http://arxiv.org/abs/1601.00624}{{\tt 1601.00624}}.

\bibitem{Ghorbani:2016jdq}
K.~Ghorbani and H.~Ghorbani, \emph{{The 750 GeV Diphoton Excess from a
  Pseudoscalar in Fermionic Dark Matter Scenario}},
  \href{http://arxiv.org/abs/1601.00602}{{\tt 1601.00602}}.

\bibitem{Ko:2016lai}
P.~Ko, Y.~Omura and C.~Yu, \emph{{Diphoton Excess at 750 GeV in leptophobic
  U(1)$^\prime$ model inspired by $E_6$ GUT}},
  \href{http://dx.doi.org/10.1007/JHEP04(2016)098}{\emph{JHEP} {\bf 04} (2016)
  098}, [\href{http://arxiv.org/abs/1601.00586}{{\tt 1601.00586}}].

\bibitem{Han:2016bus}
X.-F. Han, L.~Wang, L.~Wu, J.~M. Yang and M.~Zhang, \emph{{Explaining 750 GeV
  diphoton excess from top/bottom partner cascade decay in two-Higgs-doublet
  model extension}},
  \href{http://dx.doi.org/10.1016/j.physletb.2016.03.035}{\emph{Phys. Lett.}
  {\bf B756} (2016) 309--316}, [\href{http://arxiv.org/abs/1601.00534}{{\tt
  1601.00534}}].

\bibitem{Nomura:2016fzs}
T.~Nomura and H.~Okada, \emph{{Four-loop Neutrino Model Inspired by Diphoton
  Excess at 750 GeV}},
  \href{http://dx.doi.org/10.1016/j.physletb.2016.02.022}{\emph{Phys. Lett.}
  {\bf B755} (2016) 306--311}, [\href{http://arxiv.org/abs/1601.00386}{{\tt
  1601.00386}}].

\bibitem{Palti:2016kew}
E.~Palti, \emph{{Vector-Like Exotics in F-Theory and 750 GeV Diphotons}},
  \href{http://dx.doi.org/10.1016/j.nuclphysb.2016.04.026}{\emph{Nucl. Phys.}
  {\bf B907} (2016) 597--616}, [\href{http://arxiv.org/abs/1601.00285}{{\tt
  1601.00285}}].

\bibitem{Dasgupta:2015pbr}
A.~Dasgupta, M.~Mitra and D.~Borah, \emph{{Minimal Left-Right Symmetry
  Confronted with the 750 GeV Di-photon Excess at LHC}},
  \href{http://arxiv.org/abs/1512.09202}{{\tt 1512.09202}}.

\bibitem{Kaneta:2015qpf}
K.~Kaneta, S.~Kang and H.-S. Lee, \emph{{Diphoton excess at the LHC Run 2 and
  its implications for a new heavy gauge boson}},
  \href{http://arxiv.org/abs/1512.09129}{{\tt 1512.09129}}.

\bibitem{Jiang:2015oms}
Y.~Jiang, Y.-Y. Li and T.~Liu, \emph{{750 GeV Resonance in the Gauged
  $U(1)'$-Extended MSSM}},
  \href{http://dx.doi.org/10.1016/j.physletb.2016.05.006}{\emph{Phys. Lett.}
  {\bf B759} (2016) 354--360}, [\href{http://arxiv.org/abs/1512.09127}{{\tt
  1512.09127}}].

\bibitem{Hernandez:2015hrt}
A.~E.~C. Hernández, \emph{{The 750 GeV diphoton resonance can cause the SM
  fermion mass and mixing pattern}},
  \href{http://arxiv.org/abs/1512.09092}{{\tt 1512.09092}}.

\bibitem{Kanemura:2015vcb}
S.~Kanemura, N.~Machida, S.~Odori and T.~Shindou, \emph{{Diphoton excess at 750
  GeV in an extended scalar sector}},
  \href{http://arxiv.org/abs/1512.09053}{{\tt 1512.09053}}.

\bibitem{Kanemura:2015bli}
S.~Kanemura, K.~Nishiwaki, H.~Okada, Y.~Orikasa, S.~C. Park and R.~Watanabe,
  \emph{{LHC 750 GeV Diphoton excess in a radiative seesaw model}},
  \href{http://arxiv.org/abs/1512.09048}{{\tt 1512.09048}}.

\bibitem{Huang:2015svl}
X.-J. Huang, W.-H. Zhang and Y.-F. Zhou, \emph{{A 750 GeV dark matter messenger
  at the Galactic Center}},
  \href{http://dx.doi.org/10.1103/PhysRevD.93.115006}{\emph{Phys. Rev.} {\bf
  D93} (2016) 115006}, [\href{http://arxiv.org/abs/1512.08992}{{\tt
  1512.08992}}].

\bibitem{Hamada:2015skp}
Y.~Hamada, T.~Noumi, S.~Sun and G.~Shiu, \emph{{An O(750) GeV Resonance and
  Inflation}}, \href{http://dx.doi.org/10.1103/PhysRevD.93.123514}{\emph{Phys.
  Rev.} {\bf D93} (2016) 123514}, [\href{http://arxiv.org/abs/1512.08984}{{\tt
  1512.08984}}].

\bibitem{Kang:2015roj}
S.~K. Kang and J.~Song, \emph{{Top-phobic heavy Higgs boson as the 750 GeV
  diphoton resonance}},
  \href{http://dx.doi.org/10.1103/PhysRevD.93.115012}{\emph{Phys. Rev.} {\bf
  D93} (2016) 115012}, [\href{http://arxiv.org/abs/1512.08963}{{\tt
  1512.08963}}].

\bibitem{Ibanez:2015uok}
L.~E. Ib\'a\~nez and V.~Mart\'in-Lozano, \emph{{A Megaxion at 750 GeV as a
  First Hint of Low Scale String Theory}},
  \href{http://arxiv.org/abs/1512.08777}{{\tt 1512.08777}}.

\bibitem{Bizot:2015qqo}
N.~Bizot, S.~Davidson, M.~Frigerio and J.~L. Kneur, \emph{{Two Higgs doublets
  to explain the excesses $pp\rightarrow \gamma\gamma(750\ {\rm GeV})$ and $h
  \to \tau^\pm \mu^\mp$}},
  \href{http://dx.doi.org/10.1007/JHEP03(2016)073}{\emph{JHEP} {\bf 03} (2016)
  073}, [\href{http://arxiv.org/abs/1512.08508}{{\tt 1512.08508}}].

\bibitem{Dev:2015vjd}
P.~S.~B. Dev, R.~N. Mohapatra and Y.~Zhang, \emph{{Quark Seesaw, Vectorlike
  Fermions and Diphoton Excess}},
  \href{http://dx.doi.org/10.1007/JHEP02(2016)186}{\emph{JHEP} {\bf 02} (2016)
  186}, [\href{http://arxiv.org/abs/1512.08507}{{\tt 1512.08507}}].

\bibitem{Goertz:2015nkp}
F.~Goertz, J.~F. Kamenik, A.~Katz and M.~Nardecchia, \emph{{Indirect
  Constraints on the Scalar Di-Photon Resonance at the LHC}},
  \href{http://dx.doi.org/10.1007/JHEP05(2016)187}{\emph{JHEP} {\bf 05} (2016)
  187}, [\href{http://arxiv.org/abs/1512.08500}{{\tt 1512.08500}}].

\bibitem{Chao:2015nac}
W.~Chao, \emph{{Neutrino Catalyzed Diphoton Excess}},
  \href{http://arxiv.org/abs/1512.08484}{{\tt 1512.08484}}.

\bibitem{Gao:2015igz}
J.~Gao, H.~Zhang and H.~X. Zhu, \emph{{Diphoton excess at 750 GeV: gluon-gluon
  fusion or quark-antiquark annihilation?}},
  \href{http://arxiv.org/abs/1512.08478}{{\tt 1512.08478}}.

\bibitem{Kim:2015xyn}
J.~E. Kim, \emph{{Is an axizilla possible for di-photon resonance?}},
  \href{http://dx.doi.org/10.1016/j.physletb.2016.02.016}{\emph{Phys. Lett.}
  {\bf B755} (2016) 190--195}, [\href{http://arxiv.org/abs/1512.08467}{{\tt
  1512.08467}}].

\bibitem{Cao:2015scs}
Q.-H. Cao, Y.~Liu, K.-P. Xie, B.~Yan and D.-M. Zhang, \emph{{Diphoton excess,
  low energy theorem, and the 331 model}},
  \href{http://dx.doi.org/10.1103/PhysRevD.93.075030}{\emph{Phys. Rev.} {\bf
  D93} (2016) 075030}, [\href{http://arxiv.org/abs/1512.08441}{{\tt
  1512.08441}}].

\bibitem{Cai:2015hzc}
C.~Cai, Z.-H. Yu and H.-H. Zhang, \emph{{750 GeV diphoton resonance as a
  singlet scalar in an extra dimensional model}},
  \href{http://dx.doi.org/10.1103/PhysRevD.93.075033}{\emph{Phys. Rev.} {\bf
  D93} (2016) 075033}, [\href{http://arxiv.org/abs/1512.08440}{{\tt
  1512.08440}}].

\bibitem{Wang:2015omi}
F.~Wang, W.~Wang, L.~Wu, J.~M. Yang and M.~Zhang, \emph{{Interpreting 750 GeV
  Diphoton Resonance in the NMSSM with Vector-like Particles}},
  \href{http://arxiv.org/abs/1512.08434}{{\tt 1512.08434}}.

\bibitem{Cao:2015apa}
J.~Cao, L.~Shang, W.~Su, F.~Wang and Y.~Zhang, \emph{{Interpreting The 750 GeV
  Diphoton Excess Within Topflavor Seesaw Model}},
  \href{http://arxiv.org/abs/1512.08392}{{\tt 1512.08392}}.

\bibitem{An:2015cgp}
H.~An, C.~Cheung and Y.~Zhang, \emph{{Broad Diphotons from Narrow States}},
  \href{http://arxiv.org/abs/1512.08378}{{\tt 1512.08378}}.

\bibitem{Tang:2015eko}
Y.-L. Tang and S.-h. Zhu, \emph{{NMSSM extended with vector-like particles and
  the diphoton excess on the LHC}},
  \href{http://arxiv.org/abs/1512.08323}{{\tt 1512.08323}}.

\bibitem{Li:2015jwd}
G.~Li, Y.-n. Mao, Y.-L. Tang, C.~Zhang, Y.~Zhou and S.-h. Zhu,
  \emph{{Pseudoscalar Decaying Only via Loops as an Explanation for the 750 GeV
  Diphoton Excess}},
  \href{http://dx.doi.org/10.1103/PhysRevLett.116.151803}{\emph{Phys. Rev.
  Lett.} {\bf 116} (2016) 151803}, [\href{http://arxiv.org/abs/1512.08255}{{\tt
  1512.08255}}].

\bibitem{Salvio:2015jgu}
A.~Salvio and A.~Mazumdar, \emph{{Higgs Stability and the 750 GeV Diphoton
  Excess}}, \href{http://dx.doi.org/10.1016/j.physletb.2016.02.057}{\emph{Phys.
  Lett.} {\bf B755} (2016) 469--474},
  [\href{http://arxiv.org/abs/1512.08184}{{\tt 1512.08184}}].

\bibitem{Park:2015ysf}
J.-C. Park and S.~C. Park, \emph{{Indirect signature of dark matter with the
  diphoton resonance at 750 GeV}},  \href{http://arxiv.org/abs/1512.08117}{{\tt
  1512.08117}}.

\bibitem{Han:2015yjk}
H.~Han, S.~Wang and S.~Zheng, \emph{{Dark Matter Theories in the Light of
  Diphoton Excess}},  \href{http://arxiv.org/abs/1512.07992}{{\tt 1512.07992}}.

\bibitem{Hall:2015xds}
L.~J. Hall, K.~Harigaya and Y.~Nomura, \emph{{750 GeV Diphotons: Implications
  for Supersymmetric Unification}},
  \href{http://dx.doi.org/10.1007/JHEP03(2016)017}{\emph{JHEP} {\bf 03} (2016)
  017}, [\href{http://arxiv.org/abs/1512.07904}{{\tt 1512.07904}}].

\bibitem{Zhang:2015uuo}
J.~Zhang and S.~Zhou, \emph{{Electroweak Vacuum Stability and Diphoton Excess
  at 750 GeV}},
  \href{http://dx.doi.org/10.1088/1674-1137/40/8/081001}{\emph{Chin. Phys.}
  {\bf C40} (2016) 081001}, [\href{http://arxiv.org/abs/1512.07889}{{\tt
  1512.07889}}].

\bibitem{Liu:2015yec}
J.~Liu, X.-P. Wang and W.~Xue, \emph{{LHC diphoton excess from colorful
  resonances}},  \href{http://arxiv.org/abs/1512.07885}{{\tt 1512.07885}}.

\bibitem{Cheung:2015cug}
K.~Cheung, P.~Ko, J.~S. Lee, J.~Park and P.-Y. Tseng, \emph{{A Higgcision study
  on the 750 GeV Di-photon Resonance and 125 GeV SM Higgs boson with the
  Higgs-Singlet Mixing}},  \href{http://arxiv.org/abs/1512.07853}{{\tt
  1512.07853}}.

\bibitem{Das:2015enc}
K.~Das and S.~K. Rai, \emph{{750 GeV diphoton excess in a $U(1)$ hidden
  symmetry model}},
  \href{http://dx.doi.org/10.1103/PhysRevD.93.095007}{\emph{Phys. Rev.} {\bf
  D93} (2016) 095007}, [\href{http://arxiv.org/abs/1512.07789}{{\tt
  1512.07789}}].

\bibitem{Davoudiasl:2015cuo}
H.~Davoudiasl and C.~Zhang, \emph{{750 GeV messenger of dark conformal symmetry
  breaking}}, \href{http://dx.doi.org/10.1103/PhysRevD.93.055006}{\emph{Phys.
  Rev.} {\bf D93} (2016) 055006}, [\href{http://arxiv.org/abs/1512.07672}{{\tt
  1512.07672}}].

\bibitem{Allanach:2015ixl}
B.~C. Allanach, P.~S.~B. Dev, S.~A. Renner and K.~Sakurai, \emph{{750 GeV
  diphoton excess explained by a resonant sneutrino in R-parity violating
  supersymmetry}},
  \href{http://dx.doi.org/10.1103/PhysRevD.93.115022}{\emph{Phys. Rev.} {\bf
  D93} (2016) 115022}, [\href{http://arxiv.org/abs/1512.07645}{{\tt
  1512.07645}}].

\bibitem{Gu:2015lxj}
J.~Gu and Z.~Liu, \emph{{Physics implications of the diphoton excess from the
  perspective of renormalization group flow}},
  \href{http://dx.doi.org/10.1103/PhysRevD.93.075006}{\emph{Phys. Rev.} {\bf
  D93} (2016) 075006}, [\href{http://arxiv.org/abs/1512.07624}{{\tt
  1512.07624}}].

\bibitem{Cvetic:2015vit}
M.~Cvetič, J.~Halverson and P.~Langacker, \emph{{String Consistency, Heavy
  Exotics, and the $750$ GeV Diphoton Excess at the LHC}},
  \href{http://arxiv.org/abs/1512.07622}{{\tt 1512.07622}}.

\bibitem{Altmannshofer:2015xfo}
W.~Altmannshofer, J.~Galloway, S.~Gori, A.~L. Kagan, A.~Martin and J.~Zupan,
  \emph{{750 GeV diphoton excess}},
  \href{http://dx.doi.org/10.1103/PhysRevD.93.095015}{\emph{Phys. Rev.} {\bf
  D93} (2016) 095015}, [\href{http://arxiv.org/abs/1512.07616}{{\tt
  1512.07616}}].

\bibitem{Cao:2015xjz}
Q.-H. Cao, S.-L. Chen and P.-H. Gu, \emph{{Strong CP Problem, Neutrino Masses
  and the 750 GeV Diphoton Resonance}},
  \href{http://arxiv.org/abs/1512.07541}{{\tt 1512.07541}}.

\bibitem{Chakraborty:2015gyj}
S.~Chakraborty, A.~Chakraborty and S.~Raychaudhuri, \emph{{Diphoton resonance
  at 750 GeV in the broken MRSSM}},
  \href{http://arxiv.org/abs/1512.07527}{{\tt 1512.07527}}.

\bibitem{Badziak:2015zez}
M.~Badziak, \emph{{Interpreting the 750 GeV diphoton excess in minimal
  extensions of Two-Higgs-Doublet models}},
  \href{http://dx.doi.org/10.1016/j.physletb.2016.06.003}{\emph{Phys. Lett.}
  {\bf B759} (2016) 464--470}, [\href{http://arxiv.org/abs/1512.07497}{{\tt
  1512.07497}}].

\bibitem{Patel:2015ulo}
K.~M. Patel and P.~Sharma, \emph{{Interpreting 750 GeV diphoton excess in SU(5)
  grand unified theory}},
  \href{http://dx.doi.org/10.1016/j.physletb.2016.04.006}{\emph{Phys. Lett.}
  {\bf B757} (2016) 282--288}, [\href{http://arxiv.org/abs/1512.07468}{{\tt
  1512.07468}}].

\bibitem{Moretti:2015pbj}
S.~Moretti and K.~Yagyu, \emph{{The 750 GeV diphoton excess and its explanation
  in 2-Higgs Doublet Models with a real inert scalar multiplet}},
  \href{http://dx.doi.org/10.1103/PhysRevD.93.055043}{\emph{Phys. Rev.} {\bf
  D93} (2016) 055043}, [\href{http://arxiv.org/abs/1512.07462}{{\tt
  1512.07462}}].

\bibitem{Huang:2015rkj}
W.-C. Huang, Y.-L.~S. Tsai and T.-C. Yuan, \emph{{Gauged Two Higgs Doublet
  Model confronts the LHC 750 GeV diphoton anomaly}},
  \href{http://dx.doi.org/10.1016/j.nuclphysb.2016.05.002}{\emph{Nucl. Phys.}
  {\bf B909} (2016) 122--134}, [\href{http://arxiv.org/abs/1512.07268}{{\tt
  1512.07268}}].

\bibitem{Dev:2015isx}
P.~S.~B. Dev and D.~Teresi, \emph{{Asymmetric Dark Matter in the Sun and the
  Diphoton Excess at the LHC}},  \href{http://arxiv.org/abs/1512.07243}{{\tt
  1512.07243}}.

\bibitem{deBlas:2015hlv}
J.~de~Blas, J.~Santiago and R.~Vega-Morales, \emph{{New vector bosons and the
  diphoton excess}},
  \href{http://dx.doi.org/10.1016/j.physletb.2016.05.078}{\emph{Phys. Lett.}
  {\bf B759} (2016) 247--252}, [\href{http://arxiv.org/abs/1512.07229}{{\tt
  1512.07229}}].

\bibitem{Dey:2015bur}
U.~K. Dey, S.~Mohanty and G.~Tomar, \emph{{750 GeV Resonance in the Dark
  Left-Right Model}},
  \href{http://dx.doi.org/10.1016/j.physletb.2016.03.048}{\emph{Phys. Lett.}
  {\bf B756} (2016) 384--389}, [\href{http://arxiv.org/abs/1512.07212}{{\tt
  1512.07212}}].

\bibitem{Hernandez:2015ywg}
A.~E.~C. Hernández and I.~Nisandzic, \emph{{LHC diphoton 750 GeV resonance as
  an indication of $SU(3)_c\times SU(3)_L\times U(1)_X$ gauge symmetry}},
  \href{http://arxiv.org/abs/1512.07165}{{\tt 1512.07165}}.

\bibitem{Murphy:2015kag}
C.~W. Murphy, \emph{{Vector Leptoquarks and the 750 GeV Diphoton Resonance at
  the LHC}},
  \href{http://dx.doi.org/10.1016/j.physletb.2016.03.076}{\emph{Phys. Lett.}
  {\bf B757} (2016) 192--198}, [\href{http://arxiv.org/abs/1512.06976}{{\tt
  1512.06976}}].

\bibitem{Boucenna:2015pav}
S.~M. Boucenna, S.~Morisi and A.~Vicente, \emph{{The LHC diphoton resonance
  from gauge symmetry}},
  \href{http://dx.doi.org/10.1103/PhysRevD.93.115008}{\emph{Phys. Rev.} {\bf
  D93} (2016) 115008}, [\href{http://arxiv.org/abs/1512.06878}{{\tt
  1512.06878}}].

\bibitem{Bauer:2015boy}
M.~Bauer and M.~Neubert, \emph{{Flavor anomalies, the 750 GeV diphoton excess,
  and a dark matter candidate}},
  \href{http://dx.doi.org/10.1103/PhysRevD.93.115030}{\emph{Phys. Rev.} {\bf
  D93} (2016) 115030}, [\href{http://arxiv.org/abs/1512.06828}{{\tt
  1512.06828}}].

\bibitem{Cline:2015msi}
J.~M. Cline and Z.~Liu, \emph{{LHC diphotons from electroweakly pair-produced
  composite pseudoscalars}},  \href{http://arxiv.org/abs/1512.06827}{{\tt
  1512.06827}}.

\bibitem{Cho:2015nxy}
W.~S. Cho, D.~Kim, K.~Kong, S.~H. Lim, K.~T. Matchev, J.-C. Park et~al.,
  \emph{{750 GeV Diphoton Excess May Not Imply a 750 GeV Resonance}},
  \href{http://dx.doi.org/10.1103/PhysRevLett.116.151805}{\emph{Phys. Rev.
  Lett.} {\bf 116} (2016) 151805}, [\href{http://arxiv.org/abs/1512.06824}{{\tt
  1512.06824}}].

\bibitem{Berthier:2015vbb}
L.~Berthier, J.~M. Cline, W.~Shepherd and M.~Trott, \emph{{Effective
  interpretations of a diphoton excess}},
  \href{http://dx.doi.org/10.1007/JHEP04(2016)084}{\emph{JHEP} {\bf 04} (2016)
  084}, [\href{http://arxiv.org/abs/1512.06799}{{\tt 1512.06799}}].

\bibitem{Kim:2015ksf}
J.~S. Kim, K.~Rolbiecki and R.~Ruiz~de Austri, \emph{{Model-independent
  combination of diphoton constraints at 750 GeV}},
  \href{http://dx.doi.org/10.1140/epjc/s10052-016-4102-0}{\emph{Eur. Phys. J.}
  {\bf C76} (2016) 251}, [\href{http://arxiv.org/abs/1512.06797}{{\tt
  1512.06797}}].

\bibitem{Bi:2015uqd}
X.-J. Bi, Q.-F. Xiang, P.-F. Yin and Z.-H. Yu, \emph{{The 750 GeV diphoton
  excess at the LHC and dark matter constraints}},
  \href{http://dx.doi.org/10.1016/j.nuclphysb.2016.04.042}{\emph{Nucl. Phys.}
  {\bf B909} (2016) 43--64}, [\href{http://arxiv.org/abs/1512.06787}{{\tt
  1512.06787}}].

\bibitem{Dhuria:2015ufo}
M.~Dhuria and G.~Goswami, \emph{{Perturbativity, vacuum stability and inflation
  in the light of 750 GeV diphoton excess}},
  \href{http://arxiv.org/abs/1512.06782}{{\tt 1512.06782}}.

\bibitem{Heckman:2015kqk}
J.~J. Heckman, \emph{{750 GeV Diphotons from a D3-brane}},
  \href{http://dx.doi.org/10.1016/j.nuclphysb.2016.02.031}{\emph{Nucl. Phys.}
  {\bf B906} (2016) 231--240}, [\href{http://arxiv.org/abs/1512.06773}{{\tt
  1512.06773}}].

\bibitem{Huang:2015evq}
F.~P. Huang, C.~S. Li, Z.~L. Liu and Y.~Wang, \emph{{750 GeV Diphoton Excess
  from Cascade Decay}},  \href{http://arxiv.org/abs/1512.06732}{{\tt
  1512.06732}}.

\bibitem{Cao:2015twy}
J.~Cao, C.~Han, L.~Shang, W.~Su, J.~M. Yang and Y.~Zhang, \emph{{Interpreting
  the 750 GeV diphoton excess by the singlet extension of the Manohar-Wise
  model}}, \href{http://dx.doi.org/10.1016/j.physletb.2016.02.045}{\emph{Phys.
  Lett.} {\bf B755} (2016) 456--463},
  [\href{http://arxiv.org/abs/1512.06728}{{\tt 1512.06728}}].

\bibitem{Wang:2015kuj}
F.~Wang, L.~Wu, J.~M. Yang and M.~Zhang, \emph{{750 GeV diphoton resonance, 125
  GeV Higgs and muon $g-2$ anomaly in deflected anomaly mediation SUSY breaking
  scenarios}},
  \href{http://dx.doi.org/10.1016/j.physletb.2016.05.071}{\emph{Phys. Lett.}
  {\bf B759} (2016) 191--199}, [\href{http://arxiv.org/abs/1512.06715}{{\tt
  1512.06715}}].

\bibitem{Feng:2015wil}
T.-F. Feng, X.-Q. Li, H.-B. Zhang and S.-M. Zhao, \emph{{The LHC 750 GeV
  diphoton excess in supersymmetry with gauged baryon and lepton numbers}},
  \href{http://arxiv.org/abs/1512.06696}{{\tt 1512.06696}}.

\bibitem{Bardhan:2015hcr}
D.~Bardhan, D.~Bhatia, A.~Chakraborty, U.~Maitra, S.~Raychaudhuri and T.~Samui,
  \emph{{Radion Candidate for the LHC Diphoton Resonance}},
  \href{http://arxiv.org/abs/1512.06674}{{\tt 1512.06674}}.

\bibitem{Chang:2015sdy}
J.~Chang, K.~Cheung and C.-T. Lu, \emph{{Interpreting the 750 GeV diphoton
  resonance using photon jets in hidden-valley-like models}},
  \href{http://dx.doi.org/10.1103/PhysRevD.93.075013}{\emph{Phys. Rev.} {\bf
  D93} (2016) 075013}, [\href{http://arxiv.org/abs/1512.06671}{{\tt
  1512.06671}}].

\bibitem{Luo:2015yio}
M.-x. Luo, K.~Wang, T.~Xu, L.~Zhang and G.~Zhu, \emph{{Squarkonium,
  diquarkonium and octetonium at the LHC and their di-photon decays}},
  \href{http://dx.doi.org/10.1103/PhysRevD.93.055042}{\emph{Phys. Rev.} {\bf
  D93} (2016) 055042}, [\href{http://arxiv.org/abs/1512.06670}{{\tt
  1512.06670}}].

\bibitem{Han:2015qqj}
X.-F. Han and L.~Wang, \emph{{Implication of the 750 GeV diphoton resonance on
  two-Higgs-doublet model and its extensions with Higgs field}},
  \href{http://dx.doi.org/10.1103/PhysRevD.93.055027}{\emph{Phys. Rev.} {\bf
  D93} (2016) 055027}, [\href{http://arxiv.org/abs/1512.06587}{{\tt
  1512.06587}}].

\bibitem{Han:2015dlp}
H.~Han, S.~Wang and S.~Zheng, \emph{{Scalar Explanation of Diphoton Excess at
  LHC}}, \href{http://dx.doi.org/10.1016/j.nuclphysb.2016.04.002}{\emph{Nucl.
  Phys.} {\bf B907} (2016) 180--186},
  [\href{http://arxiv.org/abs/1512.06562}{{\tt 1512.06562}}].

\bibitem{Ding:2015rxx}
R.~Ding, L.~Huang, T.~Li and B.~Zhu, \emph{{Interpreting $750$ GeV Diphoton
  Excess with R-parity Violation Supersymmetry}},
  \href{http://arxiv.org/abs/1512.06560}{{\tt 1512.06560}}.

\bibitem{Chakraborty:2015jvs}
I.~Chakraborty and A.~Kundu, \emph{{Diphoton excess at 750 GeV: Singlet scalars
  confront triviality}},
  \href{http://dx.doi.org/10.1103/PhysRevD.93.055003}{\emph{Phys. Rev.} {\bf
  D93} (2016) 055003}, [\href{http://arxiv.org/abs/1512.06508}{{\tt
  1512.06508}}].

\bibitem{Chang:2015bzc}
S.~Chang, \emph{{A Simple $U(1)$ Gauge Theory Explanation of the Diphoton
  Excess}}, \href{http://dx.doi.org/10.1103/PhysRevD.93.055016}{\emph{Phys.
  Rev.} {\bf D93} (2016) 055016}, [\href{http://arxiv.org/abs/1512.06426}{{\tt
  1512.06426}}].

\bibitem{Han:2015cty}
C.~Han, H.~M. Lee, M.~Park and V.~Sanz, \emph{{The diphoton resonance as a
  gravity mediator of dark matter}},
  \href{http://dx.doi.org/10.1016/j.physletb.2016.02.040}{\emph{Phys. Lett.}
  {\bf B755} (2016) 371--379}, [\href{http://arxiv.org/abs/1512.06376}{{\tt
  1512.06376}}].

\bibitem{Chao:2015nsm}
W.~Chao, \emph{{Symmetries behind the 750 GeV diphoton excess}},
  \href{http://dx.doi.org/10.1103/PhysRevD.93.115013}{\emph{Phys. Rev.} {\bf
  D93} (2016) 115013}, [\href{http://arxiv.org/abs/1512.06297}{{\tt
  1512.06297}}].

\bibitem{Bernon:2015abk}
J.~Bernon and C.~Smith, \emph{{Could the width of the diphoton anomaly signal a
  three-body decay?}},
  \href{http://dx.doi.org/10.1016/j.physletb.2016.03.068}{\emph{Phys. Lett.}
  {\bf B757} (2016) 148--153}, [\href{http://arxiv.org/abs/1512.06113}{{\tt
  1512.06113}}].

\bibitem{Carpenter:2015ucu}
L.~M. Carpenter, R.~Colburn and J.~Goodman, \emph{{Supersoft SUSY Models and
  the 750 GeV Diphoton Excess, Beyond Effective Operators}},
  \href{http://arxiv.org/abs/1512.06107}{{\tt 1512.06107}}.

\bibitem{Alves:2015jgx}
A.~Alves, A.~G. Dias and K.~Sinha, \emph{{The 750 GeV $S$-cion: Where else
  should we look for it?}},
  \href{http://dx.doi.org/10.1016/j.physletb.2016.03.052}{\emph{Phys. Lett.}
  {\bf B757} (2016) 39--46}, [\href{http://arxiv.org/abs/1512.06091}{{\tt
  1512.06091}}].

\bibitem{Kim:2015ron}
J.~S. Kim, J.~Reuter, K.~Rolbiecki and R.~Ruiz~de Austri, \emph{{A resonance
  without resonance: scrutinizing the diphoton excess at 750 GeV}},
  \href{http://dx.doi.org/10.1016/j.physletb.2016.02.041}{\emph{Phys. Lett.}
  {\bf B755} (2016) 403--408}, [\href{http://arxiv.org/abs/1512.06083}{{\tt
  1512.06083}}].

\bibitem{Benbrik:2015fyz}
R.~Benbrik, C.-H. Chen and T.~Nomura, \emph{{Higgs singlet boson as a diphoton
  resonance in a vectorlike quark model}},
  \href{http://dx.doi.org/10.1103/PhysRevD.93.055034}{\emph{Phys. Rev.} {\bf
  D93} (2016) 055034}, [\href{http://arxiv.org/abs/1512.06028}{{\tt
  1512.06028}}].

\bibitem{Gabrielli:2015dhk}
E.~Gabrielli, K.~Kannike, B.~Mele, M.~Raidal, C.~Spethmann and H.~Veermäe,
  \emph{{A SUSY Inspired Simplified Model for the 750 GeV Diphoton Excess}},
  \href{http://dx.doi.org/10.1016/j.physletb.2016.02.069}{\emph{Phys. Lett.}
  {\bf B756} (2016) 36--41}, [\href{http://arxiv.org/abs/1512.05961}{{\tt
  1512.05961}}].

\bibitem{Bai:2015nbs}
Y.~Bai, J.~Berger and R.~Lu, \emph{{750 GeV dark pion: Cousin of a dark G
  -parity odd WIMP}},
  \href{http://dx.doi.org/10.1103/PhysRevD.93.076009}{\emph{Phys. Rev.} {\bf
  D93} (2016) 076009}, [\href{http://arxiv.org/abs/1512.05779}{{\tt
  1512.05779}}].

\bibitem{Csaki:2015vek}
C.~Cs\'aki, J.~Hubisz and J.~Terning, \emph{{Minimal model of a diphoton
  resonance: Production without gluon couplings}},
  \href{http://dx.doi.org/10.1103/PhysRevD.93.035002}{\emph{Phys. Rev.} {\bf
  D93} (2016) 035002}, [\href{http://arxiv.org/abs/1512.05776}{{\tt
  1512.05776}}].

\bibitem{Chakrabortty:2015hff}
J.~Chakrabortty, A.~Choudhury, P.~Ghosh, S.~Mondal and T.~Srivastava,
  \emph{{Di-photon resonance around 750 GeV: shedding light on the theory
  underneath}},  \href{http://arxiv.org/abs/1512.05767}{{\tt 1512.05767}}.

\bibitem{Bian:2015kjt}
L.~Bian, N.~Chen, D.~Liu and J.~Shu, \emph{{Hidden confining world on the 750
  GeV diphoton excess}},
  \href{http://dx.doi.org/10.1103/PhysRevD.93.095011}{\emph{Phys. Rev.} {\bf
  D93} (2016) 095011}, [\href{http://arxiv.org/abs/1512.05759}{{\tt
  1512.05759}}].

\bibitem{Curtin:2015jcv}
D.~Curtin and C.~B. Verhaaren, \emph{{Quirky Explanations for the Diphoton
  Excess}}, \href{http://dx.doi.org/10.1103/PhysRevD.93.055011}{\emph{Phys.
  Rev.} {\bf D93} (2016) 055011}, [\href{http://arxiv.org/abs/1512.05753}{{\tt
  1512.05753}}].

\bibitem{Chao:2015ttq}
W.~Chao, R.~Huo and J.-H. Yu, \emph{{The Minimal Scalar-Stealth Top
  Interpretation of the Diphoton Excess}},
  \href{http://arxiv.org/abs/1512.05738}{{\tt 1512.05738}}.

\bibitem{Demidov:2015zqn}
S.~V. Demidov and D.~S. Gorbunov, \emph{{On the sgoldstino interpretation of
  the diphoton excess}},
  \href{http://dx.doi.org/10.1134/S0021364016040044}{\emph{JETP Lett.} {\bf
  103} (2016) 219--222}, [\href{http://arxiv.org/abs/1512.05723}{{\tt
  1512.05723}}].

\bibitem{No:2015bsn}
J.~M. No, V.~Sanz and J.~Setford, \emph{{See-saw composite Higgs model at the
  LHC: Linking naturalness to the 750 GeV diphoton resonance}},
  \href{http://dx.doi.org/10.1103/PhysRevD.93.095010}{\emph{Phys. Rev.} {\bf
  D93} (2016) 095010}, [\href{http://arxiv.org/abs/1512.05700}{{\tt
  1512.05700}}].

\bibitem{Becirevic:2015fmu}
D.~Bečirević, E.~Bertuzzo, O.~Sumensari and R.~Zukanovich~Funchal, \emph{{Can
  the new resonance at LHC be a CP-Odd Higgs boson?}},
  \href{http://dx.doi.org/10.1016/j.physletb.2016.03.073}{\emph{Phys. Lett.}
  {\bf B757} (2016) 261--267}, [\href{http://arxiv.org/abs/1512.05623}{{\tt
  1512.05623}}].

\bibitem{Cox:2015ckc}
P.~Cox, A.~D. Medina, T.~S. Ray and A.~Spray, \emph{{Diphoton Excess at 750 GeV
  from a Radion in the Bulk-Higgs Scenario}},
  \href{http://arxiv.org/abs/1512.05618}{{\tt 1512.05618}}.

\bibitem{Martinez:2015kmn}
R.~Martinez, F.~Ochoa and C.~F. Sierra, \emph{{Diphoton decay for a $750$ GeV
  scalar boson in an $U(1)'$ model}},
  \href{http://arxiv.org/abs/1512.05617}{{\tt 1512.05617}}.

\bibitem{Kobakhidze:2015ldh}
A.~Kobakhidze, F.~Wang, L.~Wu, J.~M. Yang and M.~Zhang, \emph{{750 GeV diphoton
  resonance in a top and bottom seesaw model}},
  \href{http://dx.doi.org/10.1016/j.physletb.2016.03.067}{\emph{Phys. Lett.}
  {\bf B757} (2016) 92--96}, [\href{http://arxiv.org/abs/1512.05585}{{\tt
  1512.05585}}].

\bibitem{Cao:2015pto}
Q.-H. Cao, Y.~Liu, K.-P. Xie, B.~Yan and D.-M. Zhang, \emph{{A Boost Test of
  Anomalous Diphoton Resonance at the LHC}},
  \href{http://arxiv.org/abs/1512.05542}{{\tt 1512.05542}}.

\bibitem{Molinaro:2015cwg}
E.~Molinaro, F.~Sannino and N.~Vignaroli, \emph{{Minimal Composite Dynamics
  versus Axion Origin of the Diphoton excess}},
  \href{http://arxiv.org/abs/1512.05334}{{\tt 1512.05334}}.

\bibitem{Petersson:2015mkr}
C.~Petersson and R.~Torre, \emph{{750 GeV Diphoton Excess from the Goldstino
  Superpartner}},
  \href{http://dx.doi.org/10.1103/PhysRevLett.116.151804}{\emph{Phys. Rev.
  Lett.} {\bf 116} (2016) 151804}, [\href{http://arxiv.org/abs/1512.05333}{{\tt
  1512.05333}}].

\bibitem{Gupta:2015zzs}
R.~S. Gupta, S.~Jäger, Y.~Kats, G.~Perez and E.~Stamou, \emph{{Interpreting a
  750 GeV Diphoton Resonance}},  \href{http://arxiv.org/abs/1512.05332}{{\tt
  1512.05332}}.

\bibitem{Bellazzini:2015nxw}
B.~Bellazzini, R.~Franceschini, F.~Sala and J.~Serra, \emph{{Goldstones in
  Diphotons}}, \href{http://dx.doi.org/10.1007/JHEP04(2016)072}{\emph{JHEP}
  {\bf 04} (2016) 072}, [\href{http://arxiv.org/abs/1512.05330}{{\tt
  1512.05330}}].

\bibitem{Higaki:2015jag}
T.~Higaki, K.~S. Jeong, N.~Kitajima and F.~Takahashi, \emph{{The QCD Axion from
  Aligned Axions and Diphoton Excess}},
  \href{http://dx.doi.org/10.1016/j.physletb.2016.01.055}{\emph{Phys. Lett.}
  {\bf B755} (2016) 13--16}, [\href{http://arxiv.org/abs/1512.05295}{{\tt
  1512.05295}}].

\bibitem{DiChiara:2015vdm}
S.~Di~Chiara, L.~Marzola and M.~Raidal, \emph{{First interpretation of the 750
  GeV diphoton resonance at the LHC}},
  \href{http://dx.doi.org/10.1103/PhysRevD.93.095018}{\emph{Phys. Rev.} {\bf
  D93} (2016) 095018}, [\href{http://arxiv.org/abs/1512.04939}{{\tt
  1512.04939}}].

\bibitem{Pilaftsis:2015ycr}
A.~Pilaftsis, \emph{{Diphoton Signatures from Heavy Axion Decays at the CERN
  Large Hadron Collider}},
  \href{http://dx.doi.org/10.1103/PhysRevD.93.015017}{\emph{Phys. Rev.} {\bf
  D93} (2016) 015017}, [\href{http://arxiv.org/abs/1512.04931}{{\tt
  1512.04931}}].

\bibitem{Buttazzo:2015txu}
D.~Buttazzo, A.~Greljo and D.~Marzocca, \emph{{Knocking on new physics’ door
  with a scalar resonance}},
  \href{http://dx.doi.org/10.1140/epjc/s10052-016-3970-7}{\emph{Eur. Phys. J.}
  {\bf C76} (2016) 116}, [\href{http://arxiv.org/abs/1512.04929}{{\tt
  1512.04929}}].

\bibitem{Nakai:2015ptz}
Y.~Nakai, R.~Sato and K.~Tobioka, \emph{{Footprints of New Strong Dynamics via
  Anomaly and the 750 GeV Diphoton}},
  \href{http://dx.doi.org/10.1103/PhysRevLett.116.151802}{\emph{Phys. Rev.
  Lett.} {\bf 116} (2016) 151802}, [\href{http://arxiv.org/abs/1512.04924}{{\tt
  1512.04924}}].

\bibitem{Harigaya:2015ezk}
K.~Harigaya and Y.~Nomura, \emph{{Composite Models for the 750 GeV Diphoton
  Excess}}, \href{http://dx.doi.org/10.1016/j.physletb.2016.01.026}{\emph{Phys.
  Lett.} {\bf B754} (2016) 151--156},
  [\href{http://arxiv.org/abs/1512.04850}{{\tt 1512.04850}}].

\bibitem{Bharucha:2016jyr}
A.~Bharucha, A.~Djouadi and A.~Goudelis, \emph{{Threshold enhancement of
  diphoton resonances}},  \href{http://arxiv.org/abs/1603.04464}{{\tt
  1603.04464}}.

\bibitem{Djouadi:2016eyy}
A.~Djouadi, J.~Ellis, R.~Godbole and J.~Quevillon, \emph{{Future Collider
  Signatures of the Possible 750 GeV State}},
  \href{http://dx.doi.org/10.1007/JHEP03(2016)205}{\emph{JHEP} {\bf 03} (2016)
  205}, [\href{http://arxiv.org/abs/1601.03696}{{\tt 1601.03696}}].

\bibitem{Falkowski:2015swt}
A.~Falkowski, O.~Slone and T.~Volansky, \emph{{Phenomenology of a 750 GeV
  Singlet}}, \href{http://dx.doi.org/10.1007/JHEP02(2016)152}{\emph{JHEP} {\bf
  02} (2016) 152}, [\href{http://arxiv.org/abs/1512.05777}{{\tt 1512.05777}}].

\bibitem{Dutta:2015wqh}
B.~Dutta, Y.~Gao, T.~Ghosh, I.~Gogoladze and T.~Li, \emph{{Interpretation of
  the diphoton excess at CMS and ATLAS}},
  \href{http://dx.doi.org/10.1103/PhysRevD.93.055032}{\emph{Phys. Rev.} {\bf
  D93} (2016) 055032}, [\href{http://arxiv.org/abs/1512.05439}{{\tt
  1512.05439}}].

\bibitem{Low:2015qep}
M.~Low, A.~Tesi and L.-T. Wang, \emph{{A pseudoscalar decaying to photon pairs
  in the early LHC Run 2 data}},
  \href{http://dx.doi.org/10.1007/JHEP03(2016)108}{\emph{JHEP} {\bf 03} (2016)
  108}, [\href{http://arxiv.org/abs/1512.05328}{{\tt 1512.05328}}].

\bibitem{Ellis:2015oso}
J.~Ellis, S.~A.~R. Ellis, J.~Quevillon, V.~Sanz and T.~You, \emph{{On the
  Interpretation of a Possible $\sim 750$ GeV Particle Decaying into $\gamma
  \gamma$}}, \href{http://dx.doi.org/10.1007/JHEP03(2016)176}{\emph{JHEP} {\bf
  03} (2016) 176}, [\href{http://arxiv.org/abs/1512.05327}{{\tt 1512.05327}}].

\bibitem{McDermott:2015sck}
S.~D. McDermott, P.~Meade and H.~Ramani, \emph{{Singlet Scalar Resonances and
  the Diphoton Excess}},
  \href{http://dx.doi.org/10.1016/j.physletb.2016.02.033}{\emph{Phys. Lett.}
  {\bf B755} (2016) 353--357}, [\href{http://arxiv.org/abs/1512.05326}{{\tt
  1512.05326}}].

\bibitem{Angelescu:2015uiz}
A.~Angelescu, A.~Djouadi and G.~Moreau, \emph{{Scenarii for interpretations of
  the LHC diphoton excess: two Higgs doublets and vector-like quarks and
  leptons}},
  \href{http://dx.doi.org/10.1016/j.physletb.2016.02.064}{\emph{Phys. Lett.}
  {\bf B756} (2016) 126--132}, [\href{http://arxiv.org/abs/1512.04921}{{\tt
  1512.04921}}].

\bibitem{Franceschini:2015kwy}
R.~Franceschini, G.~F. Giudice, J.~F. Kamenik, M.~McCullough, A.~Pomarol,
  R.~Rattazzi et~al., \emph{{What is the $\gamma \gamma$ resonance at 750
  GeV?}}, \href{http://dx.doi.org/10.1007/JHEP03(2016)144}{\emph{JHEP} {\bf 03}
  (2016) 144}, [\href{http://arxiv.org/abs/1512.04933}{{\tt 1512.04933}}].

\bibitem{CMS:2015neg}
{\scshape CMS} collaboration, \emph{{Search for Resonances Decaying to Dijet
  Final States at $\sqrt{s} = 8$ TeV with Scouting Data}},
  {\emph{CMS-PAS-EXO-14-005} (2015) }.

\bibitem{Aad:2014aqa}
{\scshape ATLAS} collaboration, G.~Aad et~al., \emph{{Search for new phenomena
  in the dijet mass distribution using $p-p$ collision data at $\sqrt{s}=8$ TeV
  with the ATLAS detector}},
  \href{http://dx.doi.org/10.1103/PhysRevD.91.052007}{\emph{Phys. Rev.} {\bf
  D91} (2015) 052007}, [\href{http://arxiv.org/abs/1407.1376}{{\tt
  1407.1376}}].

\bibitem{Fichet:2015vvy}
S.~Fichet, G.~von Gersdorff and C.~Royon, \emph{{Scattering light by light at
  750 GeV at the LHC}},
  \href{http://dx.doi.org/10.1103/PhysRevD.93.075031}{\emph{Phys. Rev.} {\bf
  D93} (2016) 075031}, [\href{http://arxiv.org/abs/1512.05751}{{\tt
  1512.05751}}].

\bibitem{Mambrini:2015wyu}
Y.~Mambrini, G.~Arcadi and A.~Djouadi, \emph{{The LHC diphoton resonance and
  dark matter}},
  \href{http://dx.doi.org/10.1016/j.physletb.2016.02.049}{\emph{Phys. Lett.}
  {\bf B755} (2016) 426--432}, [\href{http://arxiv.org/abs/1512.04913}{{\tt
  1512.04913}}].

\bibitem{Backovic:2015fnp}
M.~Backovic, A.~Mariotti and D.~Redigolo, \emph{{Di-photon excess illuminates
  Dark Matter}}, \href{http://dx.doi.org/10.1007/JHEP03(2016)157}{\emph{JHEP}
  {\bf 03} (2016) 157}, [\href{http://arxiv.org/abs/1512.04917}{{\tt
  1512.04917}}].

\bibitem{Barducci:2015gtd}
D.~Barducci, A.~Goudelis, S.~Kulkarni and D.~Sengupta, \emph{{One jet to rule
  them all: monojet constraints and invisible decays of a 750 GeV diphoton
  resonance}}, \href{http://dx.doi.org/10.1007/JHEP05(2016)154}{\emph{JHEP}
  {\bf 05} (2016) 154}, [\href{http://arxiv.org/abs/1512.06842}{{\tt
  1512.06842}}].

\bibitem{Chu:2012qy}
X.~Chu, T.~Hambye, T.~Scarna and M.~H.~G. Tytgat, \emph{{What if Dark Matter
  Gamma-Ray Lines come with Gluon Lines?}},
  \href{http://dx.doi.org/10.1103/PhysRevD.86.083521}{\emph{Phys. Rev.} {\bf
  D86} (2012) 083521}, [\href{http://arxiv.org/abs/1206.2279}{{\tt
  1206.2279}}].

\bibitem{Khachatryan:2014rra}
{\scshape CMS} collaboration, V.~Khachatryan et~al., \emph{{Search for dark
  matter, extra dimensions, and unparticles in monojet events in proton -
  proton collisions at $\sqrt{s} = 8$ TeV}},
  \href{http://dx.doi.org/10.1140/epjc/s10052-015-3451-4}{\emph{Eur. Phys. J.}
  {\bf C75} (2015) 235}, [\href{http://arxiv.org/abs/1408.3583}{{\tt
  1408.3583}}].

\bibitem{Aad:2014nra}
{\scshape ATLAS} collaboration, G.~Aad et~al., \emph{{Search for pair-produced
  third-generation squarks decaying via charm quarks or in compressed
  supersymmetric scenarios in $pp$ collisions at $\sqrt{s}=8~$TeV with the
  ATLAS detector}},
  \href{http://dx.doi.org/10.1103/PhysRevD.90.052008}{\emph{Phys. Rev.} {\bf
  D90} (2014) 052008}, [\href{http://arxiv.org/abs/1407.0608}{{\tt
  1407.0608}}].

\bibitem{Dasgupta:2016wxw}
B.~Dasgupta, J.~Kopp and P.~Schwaller, \emph{{Photons, Photon Jets and Dark
  Photons at 750 GeV and Beyond}},
  \href{http://dx.doi.org/10.1140/epjc/s10052-016-4127-4}{\emph{Eur. Phys. J.}
  {\bf C76} (2016) 277}, [\href{http://arxiv.org/abs/1602.04692}{{\tt
  1602.04692}}].

\bibitem{Bi:2015lcf}
X.-J. Bi, R.~Ding, Y.~Fan, L.~Huang, C.~Li, T.~Li et~al., \emph{{A Promising
  Interpretation of Diphoton Resonance at 750 GeV}},
  \href{http://arxiv.org/abs/1512.08497}{{\tt 1512.08497}}.

\bibitem{Knapen:2015dap}
S.~Knapen, T.~Melia, M.~Papucci and K.~Zurek, \emph{{Rays of light from the
  LHC}}, \href{http://dx.doi.org/10.1103/PhysRevD.93.075020}{\emph{Phys. Rev.}
  {\bf D93} (2016) 075020}, [\href{http://arxiv.org/abs/1512.04928}{{\tt
  1512.04928}}].

\bibitem{Aparicio:2016iwr}
L.~Aparicio, A.~Azatov, E.~Hardy and A.~Romanino, \emph{{Diphotons from
  Diaxions}}, \href{http://dx.doi.org/10.1007/JHEP05(2016)077}{\emph{JHEP} {\bf
  05} (2016) 077}, [\href{http://arxiv.org/abs/1602.00949}{{\tt 1602.00949}}].

\bibitem{Badziak:2016cfd}
M.~Badziak, M.~Olechowski, S.~Pokorski and K.~Sakurai, \emph{{Interpreting 750
  GeV Diphoton Excess in Plain NMSSM}},
  \href{http://arxiv.org/abs/1603.02203}{{\tt 1603.02203}}.

\bibitem{Domingo:2016unq}
F.~Domingo, S.~Heinemeyer, J.~S. Kim and K.~Rolbiecki, \emph{{The NMSSM lives:
  with the 750 GeV diphoton excess}},
  \href{http://dx.doi.org/10.1140/epjc/s10052-016-4080-2}{\emph{Eur. Phys. J.}
  {\bf C76} (2016) 249}, [\href{http://arxiv.org/abs/1602.07691}{{\tt
  1602.07691}}].

\bibitem{Ellwanger:2016qax}
U.~Ellwanger and C.~Hugonie, \emph{{A 750 GeV Diphoton Signal from a Very Light
  Pseudoscalar in the NMSSM}},
  \href{http://dx.doi.org/10.1007/JHEP05(2016)114}{\emph{JHEP} {\bf 05} (2016)
  114}, [\href{http://arxiv.org/abs/1602.03344}{{\tt 1602.03344}}].

\bibitem{Dobrescu:2000jt}
B.~A. Dobrescu, G.~L. Landsberg and K.~T. Matchev, \emph{{Higgs boson decays to
  CP odd scalars at the Tevatron and beyond}},
  \href{http://dx.doi.org/10.1103/PhysRevD.63.075003}{\emph{Phys. Rev.} {\bf
  D63} (2001) 075003}, [\href{http://arxiv.org/abs/hep-ph/0005308}{{\tt
  hep-ph/0005308}}].

\bibitem{Dobrescu:2000yn}
B.~A. Dobrescu and K.~T. Matchev, \emph{{Light axion within the next-to-minimal
  supersymmetric standard model}},
  \href{http://dx.doi.org/10.1088/1126-6708/2000/09/031}{\emph{JHEP} {\bf 09}
  (2000) 031}, [\href{http://arxiv.org/abs/hep-ph/0008192}{{\tt
  hep-ph/0008192}}].

\bibitem{Hinshaw:2012aka}
{\scshape WMAP} collaboration, G.~Hinshaw et~al., \emph{{Nine-Year Wilkinson
  Microwave Anisotropy Probe (WMAP) Observations: Cosmological Parameter
  Results}},
  \href{http://dx.doi.org/10.1088/0067-0049/208/2/19}{\emph{Astrophys. J.
  Suppl.} {\bf 208} (2013) 19}, [\href{http://arxiv.org/abs/1212.5226}{{\tt
  1212.5226}}].

\bibitem{Ade:2013zuv}
{\scshape Planck} collaboration, P.~A.~R. Ade et~al., \emph{{Planck 2013
  results. XVI. Cosmological parameters}},
  \href{http://dx.doi.org/10.1051/0004-6361/201321591}{\emph{Astron.
  Astrophys.} {\bf 571} (2014) A16},
  [\href{http://arxiv.org/abs/1303.5076}{{\tt 1303.5076}}].

\bibitem{Chatrchyan:2013dga}
{\scshape CMS} collaboration, S.~Chatrchyan et~al., \emph{{Energy Calibration
  and Resolution of the CMS Electromagnetic Calorimeter in $pp$ Collisions at
  $\sqrt{s} = 7$ TeV}},
  \href{http://dx.doi.org/10.1088/1748-0221/8/09/P09009}{\emph{JINST} {\bf 8}
  (2013) P09009}, [\href{http://arxiv.org/abs/1306.2016}{{\tt 1306.2016}}].

\bibitem{ATLAS:2012ana}
{\scshape ATLAS} collaboration, \emph{{Measurements of the photon
  identification efficiency with the ATLAS detector using 4.9 fb$^{-1}$ of $pp$
  collision data collected in 2011}}, {\emph{ATLAS-CONF-2012-123} (2012) }.

\bibitem{Agrawal:2015dbf}
P.~Agrawal, J.~Fan, B.~Heidenreich, M.~Reece and M.~Strassler,
  \emph{{Experimental Considerations Motivated by the Diphoton Excess at the
  LHC}}, \href{http://dx.doi.org/10.1007/JHEP06(2016)082}{\emph{JHEP} {\bf 06}
  (2016) 082}, [\href{http://arxiv.org/abs/1512.05775}{{\tt 1512.05775}}].

\bibitem{Chala:2015cev}
M.~Chala, M.~Duerr, F.~Kahlhoefer and K.~Schmidt-Hoberg, \emph{{Tricking
  Landau-Yang: How to obtain the diphoton excess from a vector resonance}},
  \href{http://dx.doi.org/10.1016/j.physletb.2016.02.006}{\emph{Phys. Lett.}
  {\bf B755} (2016) 145--149}, [\href{http://arxiv.org/abs/1512.06833}{{\tt
  1512.06833}}].

\bibitem{Dobrich:2015jyk}
B.~Döbrich, J.~Jaeckel, F.~Kahlhoefer, A.~Ringwald and K.~Schmidt-Hoberg,
  \emph{{ALPtraum: ALP production in proton beam dump experiments}},
  \href{http://dx.doi.org/10.1007/JHEP02(2016)018}{\emph{JHEP} {\bf 02} (2016)
  018}, [\href{http://arxiv.org/abs/1512.03069}{{\tt 1512.03069}}].

\bibitem{CMS:2016all}
{\scshape CMS} collaboration, \emph{{Search for scalar resonances in the
  200--1200 GeV mass range decaying into a Z and a photon in pp collisions at
  $\sqrt{s}=8~\mathrm{TeV}$}}, {\emph{CMS-PAS-HIG-16-014} (2016) }.

\bibitem{Aad:2015kna}
{\scshape ATLAS} collaboration, G.~Aad et~al., \emph{{Search for an additional,
  heavy Higgs boson in the $H\rightarrow ZZ$ decay channel at $\sqrt{s} =
  8\;\text{ TeV }$ in $pp$ collision data with the ATLAS detector}},
  \href{http://dx.doi.org/10.1140/epjc/s10052-015-3820-z}{\emph{Eur. Phys. J.}
  {\bf C76} (2016) 45}, [\href{http://arxiv.org/abs/1507.05930}{{\tt
  1507.05930}}].

\bibitem{Khachatryan:2015cwa}
{\scshape CMS} collaboration, V.~Khachatryan et~al., \emph{{Search for a Higgs
  Boson in the Mass Range from 145 to 1000 GeV Decaying to a Pair of W or Z
  Bosons}}, \href{http://dx.doi.org/10.1007/JHEP10(2015)144}{\emph{JHEP} {\bf
  10} (2015) 144}, [\href{http://arxiv.org/abs/1504.00936}{{\tt 1504.00936}}].

\bibitem{Aad:2014fha}
{\scshape ATLAS} collaboration, G.~Aad et~al., \emph{{Search for new resonances
  in $W\gamma$ and $Z\gamma$ final states in $pp$ collisions at $\sqrt s=8$ TeV
  with the ATLAS detector}},
  \href{http://dx.doi.org/10.1016/j.physletb.2014.10.002}{\emph{Phys. Lett.}
  {\bf B738} (2014) 428--447}, [\href{http://arxiv.org/abs/1407.8150}{{\tt
  1407.8150}}].

\bibitem{CMS:2015lza}
{\scshape CMS} collaboration, \emph{{Search for scalar resonances in the
  200-500 GeV mass range decaying into a $Z$ and a photon in pp collisions at
  $\sqrt{s}=8$ TeV}}, {\emph{CMS-PAS-HIG-14-031} (2015) }.

\bibitem{Draper:2012xt}
P.~Draper and D.~McKeen, \emph{{Diphotons from Tetraphotons in the Decay of a
  125 GeV Higgs at the LHC}},
  \href{http://dx.doi.org/10.1103/PhysRevD.85.115023}{\emph{Phys. Rev.} {\bf
  D85} (2012) 115023}, [\href{http://arxiv.org/abs/1204.1061}{{\tt
  1204.1061}}].

\bibitem{ATLAS:2015nsi}
{\scshape ATLAS} collaboration, G.~Aad et~al., \emph{{Search for new phenomena
  in dijet mass and angular distributions from $pp$ collisions at $\sqrt{s}=
  13$ TeV with the ATLAS detector}},
  \href{http://dx.doi.org/10.1016/j.physletb.2016.01.032}{\emph{Phys. Lett.}
  {\bf B754} (2016) 302--322}, [\href{http://arxiv.org/abs/1512.01530}{{\tt
  1512.01530}}].

\bibitem{Khachatryan:2015dcf}
{\scshape CMS} collaboration, V.~Khachatryan et~al., \emph{{Search for narrow
  resonances decaying to dijets in proton-proton collisions at $\sqrt(s) = 13$
  TeV}}, \href{http://dx.doi.org/10.1103/PhysRevLett.116.071801}{\emph{Phys.
  Rev. Lett.} {\bf 116} (2016) 071801},
  [\href{http://arxiv.org/abs/1512.01224}{{\tt 1512.01224}}].

\bibitem{Martin:2010db}
A.~D. Martin, W.~J. Stirling, R.~S. Thorne and G.~Watt, \emph{{Heavy-quark mass
  dependence in global PDF analyses and 3- and 4-flavour parton
  distributions}},
  \href{http://dx.doi.org/10.1140/epjc/s10052-010-1462-8}{\emph{Eur. Phys. J.}
  {\bf C70} (2010) 51--72}, [\href{http://arxiv.org/abs/1007.2624}{{\tt
  1007.2624}}].

\bibitem{Martin:2009bu}
A.~D. Martin, W.~J. Stirling, R.~S. Thorne and G.~Watt, \emph{{Uncertainties on
  alpha(S) in global PDF analyses and implications for predicted hadronic cross
  sections}},
  \href{http://dx.doi.org/10.1140/epjc/s10052-009-1164-2}{\emph{Eur. Phys. J.}
  {\bf C64} (2009) 653--680}, [\href{http://arxiv.org/abs/0905.3531}{{\tt
  0905.3531}}].

\bibitem{Martin:2009iq}
A.~D. Martin, W.~J. Stirling, R.~S. Thorne and G.~Watt, \emph{{Parton
  distributions for the LHC}},
  \href{http://dx.doi.org/10.1140/epjc/s10052-009-1072-5}{\emph{Eur. Phys. J.}
  {\bf C63} (2009) 189--285}, [\href{http://arxiv.org/abs/0901.0002}{{\tt
  0901.0002}}].

\bibitem{Harland-Lang:2016qjy}
L.~A. Harland-Lang, V.~A. Khoze and M.~G. Ryskin, \emph{{The production of a
  diphoton resonance via photon-photon fusion}},
  \href{http://dx.doi.org/10.1007/JHEP03(2016)182}{\emph{JHEP} {\bf 03} (2016)
  182}, [\href{http://arxiv.org/abs/1601.07187}{{\tt 1601.07187}}].

\bibitem{Ackermann:2015zua}
{\scshape Fermi-LAT} collaboration, M.~Ackermann et~al., \emph{{Searching for
  Dark Matter Annihilation from Milky Way Dwarf Spheroidal Galaxies with Six
  Years of Fermi Large Area Telescope Data}},
  \href{http://dx.doi.org/10.1103/PhysRevLett.115.231301}{\emph{Phys. Rev.
  Lett.} {\bf 115} (2015) 231301}, [\href{http://arxiv.org/abs/1503.02641}{{\tt
  1503.02641}}].

\bibitem{Ackermann:2013yva}
{\scshape Fermi-LAT} collaboration, M.~Ackermann et~al., \emph{{Dark matter
  constraints from observations of 25 Milky Way satellite galaxies with the
  Fermi Large Area Telescope}},
  \href{http://dx.doi.org/10.1103/PhysRevD.89.042001}{\emph{Phys. Rev.} {\bf
  D89} (2014) 042001}, [\href{http://arxiv.org/abs/1310.0828}{{\tt
  1310.0828}}].

\bibitem{Ackermann:2015lka}
{\scshape Fermi-LAT} collaboration, M.~Ackermann et~al., \emph{{Updated search
  for spectral lines from Galactic dark matter interactions with pass 8 data
  from the Fermi Large Area Telescope}},
  \href{http://dx.doi.org/10.1103/PhysRevD.91.122002}{\emph{Phys. Rev.} {\bf
  D91} (2015) 122002}, [\href{http://arxiv.org/abs/1506.00013}{{\tt
  1506.00013}}].

\bibitem{Belanger:2014vza}
G.~B\'elanger, F.~Boudjema, A.~Pukhov and A.~Semenov, \emph{{micrOMEGAs4.1: two
  dark matter candidates}},
  \href{http://dx.doi.org/10.1016/j.cpc.2015.03.003}{\emph{Comput. Phys.
  Commun.} {\bf 192} (2015) 322--329},
  [\href{http://arxiv.org/abs/1407.6129}{{\tt 1407.6129}}].

\bibitem{Gondolo:1990dk}
P.~Gondolo and G.~Gelmini, \emph{{Cosmic abundances of stable particles:
  Improved analysis}},
  \href{http://dx.doi.org/10.1016/0550-3213(91)90438-4}{\emph{Nucl. Phys.} {\bf
  B360} (1991) 145--179}.

\bibitem{Chu:2013jja}
X.~Chu, Y.~Mambrini, J.~Quevillon and B.~Zaldivar, \emph{{Thermal and
  non-thermal production of dark matter via $Z'$-portal(s)}},
  \href{http://dx.doi.org/10.1088/1475-7516/2014/01/034}{\emph{JCAP} {\bf 1401}
  (2014) 034}, [\href{http://arxiv.org/abs/1306.4677}{{\tt 1306.4677}}].

\bibitem{Chu:2011be}
X.~Chu, T.~Hambye and M.~H.~G. Tytgat, \emph{{The Four Basic Ways of Creating
  Dark Matter Through a Portal}},
  \href{http://dx.doi.org/10.1088/1475-7516/2012/05/034}{\emph{JCAP} {\bf 1205}
  (2012) 034}, [\href{http://arxiv.org/abs/1112.0493}{{\tt 1112.0493}}].

\bibitem{Hall:2009bx}
L.~J. Hall, K.~Jedamzik, J.~March-Russell and S.~M. West, \emph{{Freeze-In
  Production of FIMP Dark Matter}},
  \href{http://dx.doi.org/10.1007/JHEP03(2010)080}{\emph{JHEP} {\bf 03} (2010)
  080}, [\href{http://arxiv.org/abs/0911.1120}{{\tt 0911.1120}}].

\bibitem{Dudas:2014ixa}
E.~Dudas, L.~Heurtier and Y.~Mambrini, \emph{{Generating X-ray lines from
  annihilating dark matter}},
  \href{http://dx.doi.org/10.1103/PhysRevD.90.035002}{\emph{Phys. Rev.} {\bf
  D90} (2014) 035002}, [\href{http://arxiv.org/abs/1404.1927}{{\tt
  1404.1927}}].

\bibitem{Akerib:2015rjg}
{\scshape LUX} collaboration, D.~S. Akerib et~al., \emph{{Improved Limits on
  Scattering of Weakly Interacting Massive Particles from Reanalysis of 2013
  LUX Data}},
  \href{http://dx.doi.org/10.1103/PhysRevLett.116.161301}{\emph{Phys. Rev.
  Lett.} {\bf 116} (2016) 161301}, [\href{http://arxiv.org/abs/1512.03506}{{\tt
  1512.03506}}].

\bibitem{Ibarra:2013eda}
A.~Ibarra, H.~M. Lee, S.~L\'opez~Gehler, W.-I. Park and M.~Pato,
  \emph{{Gamma-ray boxes from axion-mediated dark matter}},
  \href{http://dx.doi.org/10.1088/1475-7516/2016/03/E01,
  10.1088/1475-7516/2013/05/016}{\emph{JCAP} {\bf 1305} (2013) 016, {\bf{1603}}
  (2016) E01}, [\href{http://arxiv.org/abs/1303.6632}{{\tt 1303.6632}}].

\bibitem{Ibarra:2015tya}
A.~Ibarra, A.~S. Lamperstorfer, S.~L\'opez-Gehler, M.~Pato and G.~Bertone,
  \emph{{On the sensitivity of CTA to gamma-ray boxes from multi-TeV dark
  matter}}, \href{http://dx.doi.org/10.1088/1475-7516/2015/09/048}{\emph{JCAP}
  {\bf 1509} (2015) 048}, [\href{http://arxiv.org/abs/1503.06797}{{\tt
  1503.06797}}].

\bibitem{Salvio:2016hnf}
A.~Salvio, F.~Staub, A.~Strumia and A.~Urbano, \emph{{On the maximal diphoton
  width}}, \href{http://dx.doi.org/10.1007/JHEP03(2016)214}{\emph{JHEP} {\bf
  03} (2016) 214}, [\href{http://arxiv.org/abs/1602.01460}{{\tt 1602.01460}}].

\bibitem{Son:2015vfl}
M.~Son and A.~Urbano, \emph{{A new scalar resonance at 750 GeV: Towards a proof
  of concept in favor of strongly interacting theories}},
  \href{http://dx.doi.org/10.1007/JHEP05(2016)181}{\emph{JHEP} {\bf 05} (2016)
  181}, [\href{http://arxiv.org/abs/1512.08307}{{\tt 1512.08307}}].

\end{thebibliography}\endgroup

\end{document}